\documentclass[numberedappendix]{emulateapj}

\usepackage{amsmath}
\usepackage{graphicx}
\usepackage{psfig}
\usepackage{apjfonts}
\usepackage{multirow}

\def\cm{\textrm{cm}}
\def\meter{\textrm{m}}
\def\km{\textrm{km}}
\def\gram{\textrm{g}}
\def\erg{\textrm{erg}}
\def\kpc{\textrm{kpc}}
\def\pc{\textrm{pc}}

\def\Kelv{\textrm{K}}

\def\ergps{\textrm{erg}~\textrm{s}^{-1}}
\def\gcm2{\textrm{g}~\textrm{cm}^{-2}}

\def\kms{\textrm{km}~\textrm{s}^{-1}}

\def\phcm2s1{\textrm{photons}~\textrm{cm}^{-2}~\textrm{s}^{-1}}

\def\DiffTableUnits{10^{27}\ \cm^2\ \sec^{-1}}
\def\eV{\textrm{eV}}
\def\keV{\textrm{keV}}

\def\GeV{\textrm{GeV}}
\def\TeV{\textrm{TeV}}

\def\MHz{\textrm{MHz}}
\def\GHz{\textrm{GHz}}

\def\yr{\textrm{yr}}
\def\Myr{\textrm{Myr}}

\def\kyr{\textrm{kyr}}
\def\muGauss{\mu\textrm{G}}
\def\mGauss{\textrm{mG}}

\def\Msun{\textrm{M}_{\sun}}
\def\Lsun{\textrm{L}_{\sun}}
\def\radm2{{\rm rad}\ {\rm m}^{-2}}
\newcommand{\mean}[1]{\ensuremath{\langle #1 \rangle}}

\begin{document}

\title{\emph{Sturm und Drang}:\\ Supernova-Driven Turbulence, Magnetic Fields, and Cosmic Rays in the Chaotic Starburst Interstellar Medium}
\author{Brian C. Lacki\altaffilmark{1,2}}
\shorttitle{TURBULENCE AND MAGNETIC FIELDS IN THE STARBURST ISM}
\shortauthors{LACKI}
\altaffiltext{1}{Jansky Fellow of the National Radio Astronomy Observatory}
\altaffiltext{2}{Institute for Advanced Study, Einstein Drive, Princeton, NJ 08540, USA, brianlacki@ias.edu}

\begin{abstract}
Frequent supernova explosions in compact starburst regions are a main shaper of these regions' interstellar media (ISM).  In most starbursts, the supernova remnants blast open a hot phase that fills the regions and launches a superwind.  Denser starbursts are too overpressured for hot wind formation, but supernovae still stir up the ISM.   I argue that supernovae power ubiquitous turbulence through each of the starburst ISM phases, including the hot wind, and that a fluctuation dynamo amplifies magnetic fields until they are in equipartition with the turbulence.  Supernovae can drive turbulence with speeds of $\sim 1000\ \kms$ in the hot wind and $\sim 20\ \kms$ in the cold molecular gas, depending on the outer scale.  I predict magnetic field strengths of $70\ \muGauss$ in the Galactic Center starburst, $300\ \muGauss$ in M82 and NGC 253, and $2\ \mGauss$ in Arp 220's nuclei.  The mean magnetic field strengths are a few times stronger in molecular gas than in hot winds, but do not vary strongly with density within a starburst.  I explain how the dominance of supernova-driven turbulence leads to near equipartition between the components of starburst ISM.  I also consider implications for cosmic ray (CR) diffusion in starbursts.  The high amounts of power cascading to small scales could confine CRs very effectively in starbursts, so much that CR transport is dominated by the flow of gas rather than diffusion through the gas.  In addition, I discuss the effects of turbulence on X-ray line width, the far-infrared--radio correlation, observed radio polarization, and Faraday rotation measures.  Finally, I discuss the many questions raised regarding the physics of turbulence in starbursts.
\end{abstract}


\section{Introduction}
\label{sec:Introduction}

Starbursts are regions of galaxies that convert gas to stars quickly ($\tau_{\rm gas} \approx 10$ -- 100 Myr), resulting in high star-formation surface densities ($\Sigma_{\rm SFR} \ga 0.1\ \Msun\ \yr^{-1}\ \kpc^{-2}$).  They host up to 10\% of the cosmic star formation rate at all redshifts \citep{Hopkins10,Sargent12}.  In the present day Universe, they are the most extreme conditions for star formation.  They also serve as models for the conditions of star formation in high redshift main sequence galaxies, the birthplaces of most of the Universe's stars \citep[e.g.,][]{Chary01,LeFloch05,Noeske07,Daddi07,Magnelli09,Sargent12}.  Those galaxies typically had much larger $\Sigma_{\rm SFR}$  and gas densities and somewhat smaller $\tau_{\rm gas}$ than present-day normal galaxies \citep{Daddi10-BzK,Tacconi10,Genzel10}.

The intense conditions of a starburst drive mass and energy transfer through the interstellar medium (ISM), providing feedback that regulates star-formation.  The starburst ISM is extremely dynamic and chaotic, with several phases.  These phases include relatively cold ($\sim 100\ \Kelv$) and dense molecular gas that makes up most of the mass of starbursts, and a rarefied supernova-heated ($\sim 10^8\ \Kelv$) superwind phase that is thought to fill most of the volume of starbursts.  The molecular gas sets the physical conditions of how gas is converted into stars, while the superwind expels gas before it forms stars and enriches the metallicity of the surrounding intergalactic medium.  An important feature of the starburst ISM is turbulence, a phenomenon spanning many decades in eddy size.  Strong supersonic turbulence (with Mach number ${\cal M} \approx 100$) is present in the molecular gas of starbursts, creating huge density fluctuations and providing high pressures \citep{Downes98,MacLow04}.  

But aside from the thermalized gas of the starburst, star formation also propels the nonthermal phases of starbursts, cosmic rays (CRs) and magnetic fields, which thread through the other phases.  The presence of cosmic rays in starbursts is demonstrated by recent gamma-ray detections of M82 and NGC 253 at GeV and TeV energies \citep{Acero09,Acciari09,Abdo10-Starburst}.  The presence of CR electrons and positrons ($e^{\pm}$) and magnetic fields is demonstrated by the synchrotron radio emission of starburst galaxies \citep{Condon92}.  The existence of the FIR-radio correlation (FRC), a constant ratio of star-formation rate and continuum GHz radio emission for star-forming galaxies \citep{Helou85,Yun01}, indicates that the amount of star-formation in a galaxy or starburst is tied to the strength of the magnetic field and amount of cosmic rays \citep{Lacki10-FRC1}.  The nonthermal phases are not just passive ingredients of starburst ISM, but are active participants.  Magnetic fields provide strong pressure in starbursts \citep[e.g.,][]{Thompson06} and may halt gas collapse into stars \citep{Mouschovias76}, while CRs are another source of pressure \citep{Socrates08} and produce ionization in the molecular gas for the magnetic fields to hold onto \citep{Suchkov93,Papadopoulos10-CRDRs}.  

Turbulence links the energy flows in the nonthermal ISM with those in the nonrelativistic gas.  Enough energy is stored in the turbulent random motions of the ISM to dominate the fluid pressure and support the ISM from gravitational collapse.  On the other hand, some gas in a supersonic turbulent flow is compressed enough to collapse gravitationally and form stars and stellar clusters \citep[e.g.,][]{Krumholz05,Padoan11,Hopkins13-Frag}.  Turbulence mixes the ISM, distributing metals quickly \citep{Scalo04}.  A turbulent dynamo can amplify a magnetic field until it approaches equipartition with the turbulent energy density (\citealt{Cho00,Groves03}; I will discuss its efficiency in detail in Section~\ref{sec:Dynamoes}).  Turbulence in the magnetic fields in turn deflects CRs, confining them within a galaxy or starburst \citep{Schlickeiser02}.  More speculatively, strong turbulence can reaccelerate CRs to high energy \citep{Amano11,Melia11}.  To understand the gas of a starburst, and how magnetic fields and cosmic rays interact with it, we must therefore understand the role of turbulence in starburst galaxies.

\subsection{Fluctuation dynamoes and equipartition between turbulence and magnetic fields}
\label{sec:Dynamoes}
One of the basic premises of this work is that turbulence is roughly in equipartition with magnetic fields in starbursts \citep[c.f.,][]{Thompson06}, because of the fluctuation dynamo.  The fluctuation dynamo is thought to be a ubiquitous phenomenon.  It relies on the freezing of magnetic fields in most astrophysical plasmas.  In plasmas with random motions, the field lines get tangled, and the magnetic field strength increases.  

The characteristic residence time of turbulence is an eddy-crossing time, $\ell_{\rm outer} / \sigma$, where $\ell_{\rm outer}$ is the outer scale and $\sigma$ is the turbulent speed.  $\ell_{\rm outer}$ is the scale at which turbulence is driven; turbulent energy cascades to smaller scales.   The key result of the energy cascade is a spectrum of velocity fluctuations reaching from $\ell_{\rm outer}$ down to an inner scale $\ell_{\rm inner}$ where the energy dissipates.  Typically, the velocity fluctuation spectrum is thought to be some kind of power law \citep{Tennekes72}.

As the velocity spectrum develops, the chaotic motions of the plasma deforms the frozen-in magnetic fields, resulting in magnetic fluctuations.  A magnetic fluctuation spectrum grows in parallel to the velocity fluctuation spectrum, possibly from small scales to large scales.   On some timescale of order the eddy-crossing time, the magnetic fields approach equipartition with turbulence and alter the dynamics of the fluid motions.  At this point, the fluctuation dynamo enters a nonlinear regime; it ultimately ``saturates'' and the magnetic field growth stops \citep{Cho00,Beresnyak12}.  The fluctuation dynamo does not by itself generate regular fields, as observed in spiral galaxies -- these must arise from some other magnetic dynamo \citep{Beck96}.  

At least this is the basic picture, but there are many disagreements on the details.  The nature of the spectrum of magnetic fluctuations remains unclear.  For hydrodynamical turbulence, most power is at large scales (see the discussion in section~\ref{sec:TurbSpectrum}).  But whether the magnetic fluctuations are in equipartition with these velocity fluctuations at each scale is a matter of debate.  According to \citet{Schekochihin04}, in a flucatuation dynamo, the magnetic fluctuation spectrum only traces the velocity fluctuation spectrum at scales below the resistive scale ($\ell_{\eta} = \ell_{\rm outer} / \sqrt{{\cal R}_m}$, where ${\cal R}_m$ is the magnetic Reynolds number; \citealt{Subramanian06}), which is tiny in astrophysical systems.  As a result, the ratio of saturated  magnetic field energy density to turbulent energy density, $\epsilon_B$, is very small.  But \citet{Haugen04} instead argued that $\epsilon_B \approx 1/3$, with most of the power on large scales.

Galaxy clusters are a proving ground for fluctuation dynamo theories; they are filled with hot, turbulent plasmas and are known to have magnetic fields from Faraday rotation and synchrotron emission measurements \citep{Ferrari08}.  It's been known for some time that the magnetic fluctuations cannot peak at the resistive scale, because then there would be no Faraday rotation signal \citep{Goldshmidt93}.  \citet{Subramanian06} and \citet{Ensslin06} argued that the magnetic power peaks at a scale $\ell_{\rm peak} \approx \ell_{\rm outer} / \sqrt{{\cal R}_m^{\rm crit}}$, where ${\cal R}_m^{\rm crit} \approx 35$ is the minimum magnetic Reynolds number for driving magnetic turbulence \citep[see also][]{Brandenburg05}.  They then find that $\epsilon_B \approx 1/4$.  This gives results that are compatible with galaxy cluster rotation measures.  The simulations by \citet{Beresnyak12} imply that only $\sim 5\%$ of the turbulent energy is transferred into magnetic fields per eddy-crossing time; in starburst galaxies where the wind is advected away in a few eddy-crossing times, this suggests that $\epsilon_B$ is not much greater than $\sim 0.1$.

Even if $\epsilon_B \approx 1/6$ in galaxy clusters, it is not clear whether this applies throughout the starburst ISM.  \citet{Stone98} simulated magnetohydrodynamical (MHD) turbulence in molecular clouds and found $\epsilon_B \approx 1/2$.  In actual Galactic molecular clouds, magnetic fields and turbulence are in rough equipartition, implying $\epsilon_B \approx 1/3$--$1$ \citep{Crutcher99,Troland08,Falgarone08}.  Likewise, magnetic fields in the Galactic cold neutral medium are in equipartition with turbulence \citep{Heiles05}.  \citet{Balsara04} argued that this indicated that the magnetic fields arose from a fast magnetic dynamo, powered by supernova driven turbulence, although in their own simulations, $\epsilon_B$ only reached 0.01.   For that matter, the saturated state of the fluctuation dynamo may also depend on the kind of turbulent forcing.  \citet{Federrath11} found that solenoidal forcing led to much higher $\epsilon_B$ than compressive forcing.  Supernova explosions push the ISM in a compressive manner, yet they can amplify solenoidal motions dramatically in cold and warm gas \citep{Balsara04}.  A ``shear dynamo'' operating in sheared, turbulent flows leads to not only turbulent magnetic field growth, but a field stronger along one direction than others (\citealt{Yousef08}; c.f., \citealt{Laing80}).

For the rest of the paper, I simply scale all of my results to $\epsilon_B = 1$.  Remember, though, that the actual value of $\epsilon_B$ is plausibly as low as 0.1 or smaller.  I also ignore the time for the turbulent dynamo to reach saturation.  The growth time could be important in starbursts if the wind removes the turbulent gas from the star-forming region before the dynamo saturates.

Finally, magnetic fields driven by a turbulent dynamo are unlikely to be homogeneous.  Turbulence is ``intermittent'', with large fluctuations in strength from place to place and time to time.  Instead, the magnetic field structure is thought to include a low-strength volume-filling field, with much higher magnetic field strengths in sheets, ribbons, and filaments \citep[e.g.,][]{Schekochihin04,Subramanian06,Bhat13}.  These features are roughly $\ell_{\rm outer}$ in length, but only $\ell_{\rm peak}$ in thickness \citep{Subramanian06}.  I generally ignore this complication, but it may prove important in understanding starburst magnetic fields.

\subsection{Outline of the Paper}
This paper explores the role of turbulence in starbursts and how it determines the characteristics of the pervasive magnetic fields in these environments.  I begin in Section~\ref{sec:BroadView} by considering the basic structure of starburst ISM as it is shaped by supernova explosions.  The bulk characteristics of supernova-driven turbulence and the equipartition magnetic fields in the hot superwind are calculated in Section~\ref{sec:HotTurbulence}.  Similar calculations are performed for turbulence and magnetic fields in cold molecular gas in Section~\ref{sec:ColdTurbulence}.  I demonstrate that energy input from supernovae naturally leads to equipartition between many energy densities in the ISM of starbursts in Section~\ref{sec:Equipartition}.  Section~\ref{sec:Cascade} explores the cascade of turbulence to small scales, and its implications for CR propagation.  The observable effects of turbulence and the magnetic fields, including X-ray line widths, Faraday rotation measures, supernova remnant magnetic fields, and the infrared-radio correlation, are considered in Section~\ref{sec:Implications}.  Finally, I summarize these results in Section~\ref{sec:Conclusion}.

Throughout the paper, I use a Salpeter initial mass function (IMF) from 0.1 to 100$\ \Msun$.

\section{A broad view of the starburst ISM: Physical conditions and phases}
\label{sec:BroadView}
Throughout this work, I compare my results to four prototypical starburst regions: the Galactic Center Central Molecular Zone (GCCMZ), the starburst cores of NGC 253 and M82, and the starburst nuclei of Arp 220.  The conditions in these regions span those of $z = 0$ starburst regions.  In Table~\ref{table:BasicProperties}, I list the basic properties that I use in computations for these regions.  These regions are relatively well studied.  Models of the nonthermal emission from each has constrained the magnetic field strength and CR energy densities, and all but Arp 220 have gamma-ray detections.  All have star-formation surface densities above the ``Heckman threshold" of $0.1\ \Msun\ \yr^{-1}\ \kpc^{-2}$ \citep[c.f.,][]{Heckman90}.  The gas consumption times of NGC 253, M82, and Arp 220 are all of the order 10 -- 20 Myr, values typical of true starbursts throughout the Universe \citep{Tacconi06,Daddi09,Genzel10,Riechers11}.  The GCCMZ, by contrast, has a far slower gas consumption, more typical of $z \approx 2$ main sequence galaxies \citep{Genzel10}, but since it is much more easily resolved and displays several of the same phenomena as true starbursts, I include it.  

\begin{deluxetable*}{llccccc}
\tablecaption{Fundamental properties of prototypical starbursts}
\tablehead{ & \colhead{Units} & \colhead{GCCMZ} & \colhead{NGC 253} & \colhead{M82} & \colhead{Arp 220 Nuclei} & \colhead{Notes}}
\startdata
SFR                   & $\Msun\ \yr^{-1}$             & 0.07              & 3                 & 10                & 100  & a\\
R                     & pc                            & 100               & 150               & 300               & 100  & b\\
h                     & pc                            & 50                & 50                & 50                & 50   & c\\
$\Sigma_{\rm SFR}$    & $\Msun\ \yr^{-1}\ \kpc^{-2}$  & 2.2               & 42                & 35                & 3200 & d\\
$\Gamma_{\rm SN}$     & $\yr^{-1}$                    & 0.00052           & 0.022             & 0.074             & 0.74 & e\\
$M_{\rm gas}$         & $\Msun$                       & $3 \times 10^7$   & $3 \times 10^7$   & $2 \times 10^8$   & $10^9$ & f\\
$\tau_{\rm gas}$      & $\Myr$                        & 430               & 10                & 20                & 10   & g\\
$\Sigma_g$            & $\gcm2$                       & 0.20              & 0.087             & 0.15              & 6.7  & h\\
$\mean{n_H}_{\rm SB}$ & $\cm^{-3}$                    & 390               & 170               & 300               & 13000 & i\\
$B_{\rm true}$        & $\muGauss$                    & 50 - 100          & 100 - 300         & 100 - 300         & 2000 - 6000 & j\\
$L_{\rm bol}$         & $\Lsun$                       & $3.9 \times 10^8$ & $1.7 \times 10^{10}$ & $5.6 \times 10^{10}$ & $5.6 \times 10^{11}$ & k\\
$\dot{E}_{\rm mech}$  & $\Lsun$                       & $4.6 \times 10^6$ & $1.9 \times 10^8$ & $6.6 \times 10^8$ & $6.6 \times 10^9$ & l \\
$\dot{\varepsilon}_{\rm mech}$   & $\erg\ \cm^{-3}\ \sec^{-1}$   & $1.9 \times 10^{-22}$ & $3.7 \times 10^{-21}$ & $3.1 \times 10^{-21}$ & $2.7 \times 10^{-19}$ & m \\
$\dot{\varepsilon}_{\rm mech} / n_H$ & $\erg\ {\rm H}^{-1}\ \sec^{-1}$ & $4.9 \times 10^{-25}$ & $2.2 \times 10^{-23}$ & $1.0 \times 10^{-23}$ & $2.1 \times 10^{-23}$ & 
\enddata
\tablenotetext{a}{{\bf GCCMZ:} \citet{YusefZadeh09}.  {\bf NGC 253:} from bolometric luminosity \citep{Sanders03}, taking into account that half is from surrounding host galaxy \citep{Melo02}.  {\bf M82:} from bolometric luminosity \citep{Sanders03}.  {\bf Arp 220:} \citet{Downes98}.}
\tablenotetext{b}{{\bf GCCMZ:} \citet{Molinari11}.  {\bf NGC 253:} molecular gas disk radius from \citet{Sakamoto11}.  {\bf M82:} scale of radio disk from \citet{Williams10}.  {\bf Arp 220:} \citet{Downes98}.}
\tablenotetext{c}{Midplane-to-edge scale height.  While the exact values of $h$ are hard to measure because of projection effects, a value of $\sim 50\ \pc$ is reasonable for starbursts \citep{Crocker11-Wild,Sakamoto11,Downes98}.}
\tablenotetext{d}{$\Sigma_{\rm SFR} = {\rm SFR} / (\pi R^2)$.}
\tablenotetext{e}{Calculated assuming a Salpeter IMF from 0.1 to 100 $\Msun$, and that stars with masses $\ge 8\ \Msun$ go supernova: $\Gamma_{\rm SN} = 0.0074 ({\rm SFR} / \Msun)$.}
\tablenotetext{f}{{\bf GCCMZ:} \citet{Molinari11}.  {\bf NGC 253:} Rough value from \citet{Mauersberger96,Harrison99,Bradford03}.  {\bf M82:} \citet{Weiss01}.  {\bf Arp 220 nuclei:} \citet{Downes98}.}
\tablenotetext{g}{$\tau_{\rm gas} = M_{\rm gas} / {\rm SFR}$.}
\tablenotetext{h}{$\Sigma_g = M_{\rm gas} / (\pi R^2)$.  Note $1\ \gcm2 = 4790\ \Msun\ \pc^{-2}$.}
\tablenotetext{i}{$\mean{n_H}_{\rm SB} = M_{\rm gas} / (2 \pi R^2 h m_H)$.}
\tablenotetext{j}{I use estimates derived from one-zone modeling of radio and, if detected, gamma-ray emission.  These $B$ depend on gas density \citep{Paglione12}.{\bf GCCMZ:} \citet{Crocker11-Wild}.  {\bf NGC 253}: \citet{Domingo05,Rephaeli10}.  {\bf M82:} \citet{Persic08,deCeaDelPozo09-M82,YoastHull13}.   {\bf Arp 220:} \citet{Torres04}.  \citet{Robishaw08} found $\ga 3\ \mGauss$ magnetic fields in ULIRGs with Zeeman splitting.  Firm lower limits come from upper bound on observed Inverse Compton emission; these are $\ga 50\ \muGauss$ in the GCCMZ \citep{Crocker10}, NGC 253, and M82, and $\ga 250\ \muGauss$ in Arp 220 \citep{Lacki13-XRay}.}
\tablenotetext{k}{Calculated using the Kennicutt conversion factor between $L_{\rm IR}$ and SFR \citep{Kennicutt98}.}
\tablenotetext{l}{Calculated with equation~\ref{eqn:EDot}.}
\tablenotetext{m}{$\dot{\varepsilon} = \dot{E}/(2 \pi R^2 h)$ is the mechanical energy injected per unit volume.}
\label{table:BasicProperties}
\end{deluxetable*}

\subsection{Energy densities in starburst ISM}
\label{sec:UStarburst}
The flows of energy within galaxies are reflected in the energy densities present in the ISM.  The fact of equipartition between the Galactic ISM's energy fields (thermal, turbulent, radiation, magnetic, CRs) is well known.  It is often attributed to couplings between the components that drive them to equal pressures.  Equipartition between various ISM components in external galaxies is a frequent assumption, generally to estimate the strength of relatively unknown quantities like the magnetic field or CR pressure.  Appeals to pressure balance are also a common way to make a statement about the plausible existence of an ISM phase, such as the hot superwind, or a volume-filling WIM \citep{Murray10,Lacki13-LowNu}.  In this paper, I argue that the magnetic field and turbulent energy densities are roughly equal.  But what actually is our present knowledge of the energy densities in the starburst ISM?

While the pressures have large uncertainties, it is clear from Table~\ref{table:Pressures} that many of the energy densities in starburst regions are of the same order for each starburst region.  The typical pressure $P/k_B$ in the GCCMZ is a few $\times 10^6\ \cm^{-3}$ \citep[see also][]{Law10}, a few $\times 10^7\ \cm^{-3}$ in M82 and NGC 253, and a few $\times 10^9\ \cm^{-3}$ in the intense conditions of Arp 220's nuclei.  Equipartition seems to hold between turbulent energy density in molecular gas, the thermal pressure in superwinds, radiation pressure, and magnetic energy density.  

\emph{Molecular gas turbulence} -- Although they are in line with the other energy densities, the molecular gas turbulent pressures are actually poorly known.  I compute these as $\mean{\rho}_{\rm SB} \sigma^2 / 2$, where $\sigma$ is the turbulent speed.  The turbulent speeds of NGC 253 and M82's molecular gas are not even given in the literature.  For M82, $\sigma$ has an upper limit of $70 - 110\ \kms$ from the molecular line widths \citep{Muehle07}.  I assume $\sigma = 50\ \kms$, between the values in the GCCMZ ($\sim 15-30\ \kms$; \citealt{Shetty12}) and Arp 220 ($\sim 80\ \kms$; \citealt{Downes98}).  That speed also matches the turbulent speed in the H II region surrounding one of M82's super star clusters \citep{Smith06}.  However, measured turbulent speeds may overestimate the actual level of turbulence, in studies where the observing beam cannot resolve out larger-scale laminar flows \citep{Ostriker11}.

The other missing data are the density of the gas in the molecular clouds.  While the mean densities averaged over the entire starburst volume $\mean{\rho}_{\rm SB}$ are fairly well measured, the molecular clouds may only fill a small fraction of that volume, at least in the three weaker starbursts (in ULIRGs, the entire volume may be filled with molecular gas).  A filling fraction $\sim 0.1$, entirely plausible given our current state of knowledge, would raise the average density and turbulent energy densities of molecular clouds by $\sim 10$.  Thus the molecular gas turbulent pressures should be considered uncertain at the order of magnitude level.

At this point, I should differentiate the true molecular filling factor with the filling factor of the dense molecular clumps inferred by various photodissociation region models of (for example) M82.  These clumps are described as clouds with densities $\sim 10^3$ -- $10^5\ \cm^{-3}$, sometimes even denser, filling $\sim 0.1 - 1\%$ of the starburst volume \citep[e.g.,][]{Wolfire90,Wild92,Lord96,Mao00,Weiss01,Ward03,Naylor10}.  In an isothermal turbulent medium with Mach factor ${\cal M}$, most of the mass is concentrated into regions overdense by a factor $\sim \sqrt{1 + b^2 {\cal M}^2}$, where $b \sim 0.3 - 1$.  In the canonical lognormal density distribution for a medium with supersonic turbulence, most of the volume is underdense by the same factor $\sqrt{1 + b^2 {\cal M}^2}$ \citep{Padoan97,Ostriker01}, although the actual volume-filling density may in fact be near the volume-average density \citep{Hopkins13-rhoDist}.  If M82's molecular gas has a volume-averaged density $300 f_{\rm fill}^{-1}\ \cm^{-3}$ and ${\cal M} \approx 30$ turbulence, most of the mass is predicted to be in clumps of density $\ga 9000 f_{\rm fill}^{-1}\ \cm^{-3}$ for a lognormal distribution.  When the isothermal assumption is relaxed, though, the density distribution is not lognormal anymore, but tends to a power-law form at high densities \citep{Scalo98}.

Therefore the observed ``clouds'' may not be permanent objects separated by another phase, but rather, temporary structures of a much larger volume molecular medium.  The true filling factor of the molecular medium is a factor $\sim {\cal M}$ greater than the inferred filling factor of the clumps -- perhaps of the order $\sim 10\%$.  

But the thermal pressure in the molecular medium is far below equipartition with the turbulent energy density.  The sound speed is a mere $\sqrt{(5/6) k_B T/m_H} = 0.8\ \kms (T/100\ \Kelv)^{1/2}$, so the Mach factors in starburst molecular media are ${\cal M} \approx 10$ -- $100$.  The ratio of turbulent to thermal pressure scales as ${\cal M}^2$, and is of order $\sim 1000$ in starbursts.  Even in the clumps that contain most of the molecular mass, the density is enhanced by only a factor $\sim {\cal M}$, so the thermal pressure inferred for these clumps is still a factor $\sim {\cal M}$ below the mean turbulent pressure.

\begin{deluxetable*}{lrrrrc}
\tablecaption{Observed Mean Energy Densities in Prototypical Starbursts}
\tablehead{ & \colhead{GCCMZ} & \colhead{NGC 253} & \colhead{M82} & \colhead{Arp 220 Nuclei} &  \colhead{Notes} \\ & \colhead{$(K\ \cm^{-3})$} & \colhead{$(K\ \cm^{-3})$} & \colhead{$(K\ \cm^{-3})$} & \colhead{$(K\ \cm^{-3})$}}
\startdata
Molecular gas turbulence          & $\sim 14 \times 10^6$        & $\sim 3 \times 10^7$           & $\sim 5 \times 10^7$         & $\sim 5 \times 10^9$          & (1)\\
Superwind thermal (CC85)          & $1.1 \times 10^6$            & $2.4 \times 10^7$              & $2.5 \times 10^7$            & $1.5 \times 10^9$             & (2)\\
Superwind turbulence (this paper) & $(1 - 2) \times 10^6$        & $(1 - 3) \times 10^7$          & $(1 - 3) \times 10^7$        & $(0.4 - 2) \times 10^9$             & (3)\\
H II region thermal               & $(3 - 24) \times 10^6$       & $(0.2 - 2) \times 10^7$        & $(0.3 - 2) \times 10^7$      & $(0.02 - 4) \times 10^9$      & (4)\\
H II region turbulence            & \nodata                      & \nodata                        & $30 \times 10^7$             & \nodata                       & (5)\\
\multirow{2}{*}{Radiation, $\tau_{\rm IR}^{\rm eff} = \left\{ \begin{array}{l} 0 \\ \tau_{\rm IR} \end{array} \right.$} & $0.7 \times 10^6$ & $1.4 \times 10^7$ & $1.5 \times 10^7$ & $0.9 \times 10^9$ & \multirow{2}{*}{(6)}\\
                                  & $0.8 \times 10^6$            & $2.1 \times 10^7$              & $2.8 \times 10^7$            & $44 \times 10^9$              & \\
Magnetic fields                   & $\sim (0.7 - 3) \times 10^6$ & $\sim (0.3 - 3) \times 10^7$   & $\sim (0.3 - 3) \times 10^7$ & $\sim (1.2 - 10) \times 10^9$ & (7)\\
Cosmic rays                       & $\sim 0.2 \times 10^6$       & $\sim (0.3 - 0.4) \times 10^7$ & $\sim (0.3 - 0.4) \times 10^7$       & $\sim (0.03 - 1) \times 10^9$ & (8)\\
\hline
\multirow{2}{*}{Hydrostatic, $\Sigma_{\rm tot} = \left\{ \begin{array}{l} \Sigma_g \\ \Sigma_g / f_{\rm gas} \end{array} \right.$} & $60 \times 10^6$ & $1 \times 10^7$ & $3 \times 10^7$ & $70 \times 10^9$ & \multirow{2}{*}{(9)}\\
                                  & $1700 \times 10^6$           & $6 \times 10^7$                & $17 \times 10^7$             & $130 \times 10^9$ & \\
``Orbital''                       & $150 \times 10^6$            & $10 \times 10^7$               & $18 \times 10^7$             & $80 \times 10^9$  & (10)
\enddata
\label{table:Pressures}
\tablenotetext{1}{These energy densities are very uncertain because of the unknown filling factor and poorly constrained turbulent speeds.  These estimates assume $f_{\rm fill} = 1$ for cold gas; smaller $f_{\rm fill}$ lead to higher turbulent pressuers.  For the GCCMZ, I use the average density $390\ \cm^{-3}$ from \citet{Molinari11} and a velocity dispersion of $25\ \kms$ \citep{Shetty12}.  I assume turbulent speeds of $\sigma = 50\ \kms$ in NGC 253 and M82.  The molecular line widths of each ``lobe'' of molecular gas in M82 is $\sim 100 - 150\ \kms$; if it is all due to turbulence, the velocity dispersion is smaller by a factor $1.4$ \citep{Downes98}, giving turbulent velocities of $70 - 110\ \kms$ \citep{Muehle07}.  For Arp 220's nuclei, I use $\sigma = 100\ \kms$ \citep{Downes98}.}
\tablenotetext{2}{Calculated using the CC85 wind solution with $\epsilon_{\rm therm} = 0.75$, $\zeta = 1$, and $\beta = 2$.  I assume the hot superwind exists and is correctly described by the CC85 solution.}
\tablenotetext{3}{Calculated in this paper in Section~\ref{sec:HotTurbulence}.}
\tablenotetext{4}{{\bf GCCMZ:} From radio recombination line measurements of the Galactic Center radio lobe, \citet{Law09} determine a pressure of $\sim (4$ -- $7) \times 10^6\ \Kelv\ \cm^{-3}$.  \cite{Law10} compiles pressures in the GCCMZ region, finding values of order a few $10^6\ \Kelv\ \cm^{-3}$.  In individual H II regions, \citet{Zhao93} finds this range, again using radio recombination lines.  Most of the Galactic Center H II regions have pressures of $\sim 2.5 \times 10^6\ \Kelv\ \cm^{-3}$, with the region H2 alone accounting for the upper end of the range.  {\bf NGC 253:} The NGC 253 pressures are from \citet{Mohan05} and \citet{RodriguezRico06}.  {\bf M82:} According to \citet{Smith06}, the thermal pressures in the H II region around the super star cluster M82 A-1 is $(1 - 2) \times 10^7\ \Kelv\ \cm^{-3}$.  For other H II regions, the thermal pressures are $10^7\ \Kelv\ \cm^{-3}$ or smaller \citep{Lord96,Westmoquette09}.  {\bf Arp 220:} For Arp 220's nuclei, I used \citet{RodriguezRico05}.  In the multicomponent fit to radio-recombination line data, there is a low pressure component with $n_e \approx 1000\ \cm^{-3}$ that recieves most of the ionizing photons and a high pressure component with $n_e \approx 2.5 \times 10^5\ \cm^{-3}$ \citep{Anantharamaiah00}.  The low pressure component is more spatially extended, so it may not be representative of the nuclei themselves \citep{RodriguezRico05}.}
\tablenotetext{5}{Found for the H II region surrounding the super star cluster M82 A-1 by \citet{Smith06}.  It is denser than other H II regions in M82, and may either be in an unrepresentative region of M82 or still expanding.  Using the turbulent speed $\sigma = 45\ \kms$ for other M82 H II regions give turbulent pressures close to equipartition with the rest of the ISM.}
\tablenotetext{6}{The radiation optical depth is calculated as $\tau_{\rm IR} = \Sigma_g \times (5\ \cm^2\ \gram^{-1})$; the value of the opacity is appropriate for far-infrared radiation.}
\tablenotetext{7}{Equal to $B^2 / (8 \pi)$.}
\tablenotetext{8}{The Galactic Center CR pressure comes from \citet{Crocker11-Wild}.  NGC 253's CR pressure comes from \citet{Abramowski12}.   M82's CR pressure is from \citet{Acciari09} and \citet{Lacki11-Obs}.  For Arp 220, I calculate it by assuming a SFR of $100\ \Msun\ \yr^{-1}$ and $10^{50}$ ergs injected in CRs per supernova.  The lower limit is for a CR residence time of $\sim 5000\ \yr$, as is typically assumed in models \citep{Torres04}; the upper limit is the absolute bound by assuming a residence time from advection of $\sim h / v_{\rm wind} = 160\ \kyr$.}
\tablenotetext{9}{The hydrostatic pressure needed to support the gas disk is $\pi G \Sigma_g \Sigma_{\rm tot}$.  I find $f_{\rm gas} = 0.036$ for the GCCMZ, from \citet{Launhardt02}'s estimate of the stellar mass in the ``inner nuclear bulge'', but this implies an implausibly high hydrostatic pressure.  I assume $f_{\rm gas} = 0.2$ for NGC 253 and M82 \citep{Murray10}.  For comparison, \citet{ForsterSchreiber01} argued $M = 8 \times 10^8\ \Msun$ in M82's starburst, so that $f_{\rm gas} \approx 0.25$ (see also references therein).  For Arp 220's nuclei, $f_{\rm gas} = 0.5$ \citep{Downes98}.}
\tablenotetext{10}{I use $v_{\rm circ} \approx 80\ \kms$ for the GCCMZ, although the kinematics are complex in the region \citep{Molinari11}.  I take NGC 253's rotational speed to be $\sim 100\ \kms$ \citep{Sorai00}.  For M82, I use $\sim 100\ \kms$ from the CO measurements of \citet{Weiss01} and the stellar rotation curve at $R = 300\ \pc$ from \citet{Greco12}.  I use a rotational speed of $325\ \kms$ for Arp 220's nuclei, which is the average of the values for Arp 220 West and East in \citet{Downes98}.}
\end{deluxetable*}

\emph{Cosmic rays} -- The most glaring outlier in Table~\ref{table:BasicProperties} is the CR energy density, which is admittedly poorly constrained but appears to be an order of magnitude lower than the other densities.  In the GCCMZ, M82, and NGC 253, the CR energy density is constrained by gamma-ray detections.  But deriving these constraints requires knowledge of the gas density the CRs interact with; while typically the mean gas density is used, whether the CR lifetime is actually determined by the mean gas density is not clear.  Imposing equipartition between magnetic and CR energy density gives energy densities that are a bit lower than ISM pressure in M82 and NGC 253, $\sim 10^7\ \Kelv\ \cm^{-3}$ \citep{Lacki13-Equip}.  

While the CR energy density is perhaps an order of magnitude too low in M82 and NGC 253, the disparity is much more spectacular in Arp 220's nuclei.  The equipartition estimate between magnetic and CR energy densities is a mere $10^8\ \Kelv\ \cm^{-3}$ -- roughly 30 -- 100 times lower than the pressure in the molecular phase and H II regions \citep{Thompson06,Lacki13-Equip}.  But even this may be too high.  Models of the CR population in Arp 220's nuclei imply higher magnetic field strengths (roughly $\sim 2$ -- $6\ \muGauss$) but lower CR energy densities (a few times $10^7\ \Kelv\ \cm^{-3}$; \citealt{Torres04,Lacki10-FRC1,Lacki13-XRay}).  In that case, the CRs are a factor of $\sim 100$ below equipartition with the ISM pressure, while the magnetic field energy density is comparable.  As with the gamma-ray estimates, these models require knowledge of a density.

\emph{H II region pressure} -- Measured H II region thermal pressures show a wide spread, but on average are somewhat high in the GCCMZ and somewhat low in NGC 253, M82, and Arp 220.  Either way, it is a very different situation than in the Milky Way, where H II regions are massively overpressured bubbles expanding into the ISM (see section~\ref{sec:WarmGas}).  In addition to the thermal pressures, there is probably turbulent pressure in the H II regions \citep{Smith06}.  

\emph{The hydrostatic pressure, the radiation pressure, and rotation} -- A starburst's gas must be in hydrostatic equilibrium, otherwise it will either collapse or be blown apart.  For a thin, homogeneous disk with surface density $\Sigma_g$ and gas fraction $f_{\rm gas}$, the necessary midplane hydrostatic pressure is $\pi G \Sigma_g \Sigma_{\rm tot} = \pi G \Sigma_g^2 / f_{\rm gas}$.  When compared to the other pressures in Table~\ref{table:Pressures}, $\pi G \Sigma_g^2$ seems roughly in line with the other pressures in NGC 253, and M82, especially when considering that $f_{\rm gas}$ is probably of order a few tenths.  But it fails spectacularly at predicting the characteristic pressures in Arp 220.

What is the cause of this discrepancy?  The radiation pressure in Arp 220 could be high enough to reach $\pi G \Sigma_g^2$, but only if it is an extreme scattering atmosphere with $\tau_{\rm IR} \approx 100$, so that the energy density accumulates as photons diffuse out of the starburst.  Those high levels motivate radiation-pressure supported models of starbursts \citep[e.g.,][]{Thompson05,Thompson06,Murray10}.  Yet an odd fact is clear from Table~\ref{table:Pressures}: $\pi G \Sigma_g^2$ is much greater than the mean turbulent energy density.  Thus, if radiation pressure pressurizes ULIRGs that amount, it apparently does so \emph{without driving turbulence to equipartition} (Table~\ref{table:Pressures}), in defiance of the arguments of \citet{Thompson06}.  An additional problem is that actual starbursts are inhomogeneous, and may have low density chimneys where IR photons can escape.  A recent simulation by \citet{Krumholz12} indicates that instabilities reduce $\tau_{\rm IR}$ by a factor $\sim 5$ in actual starbursts.  That implies a radiation energy density $\sim 10^{10}\ \Kelv\ \cm^{-3}$ in Arp 220, comparable to the turbulent energy density.  

If the radiation pressure does not overwhelmingly dominate the pressures in Arp 220, why does $\pi G \Sigma_g^2$ fail as an estimate?  One possibility is that the surface density in Arp 220 is overestimated by a factor $\sim 10$, reducing both $\pi G \Sigma_g^2$ and $U_{\rm turb}$ to $\sim 10^9\ \Kelv\ \cm^{-3}$.  That would require substantial errors in our ability to measure gas masses, though.  A second possibility is that the ``nuclei'' of Arp 220 are in fact transient features and will collapse in a free-fall time $\sim 1 /\sqrt{G\rho} \approx 10^6\ \yr$.  Indeed, the two nuclei appear to be concentrations within a larger gas disk with $\mean{\Sigma_g} \approx 1\ \gcm2$ \citep{Downes98}.  Yet the concentration of radio supernovae in the nuclei requires that the nuclei have been stable for at least a few million years \citep[e.g.,][]{Smith98,HerreroIllana12}.  

The discrepancy could indicate that Arp 220's nuclei are very inhomogeneous.  This is supported by the relatively low Toomre ${\cal Q} \approx \sqrt{2} \sigma v_{\rm circ} / (\pi G \Sigma_g R)$ values of $\sim 0.3$ -- 1 for the nuclei \citep[c.f.,][]{Kruijssen12}.  Gas under such conditions fragments into small clumps.  The clumps' gravity holds down any interclump gas, but they do not themselves need to be supported by interclump gas.  The clumps orbit around the nucleus; the nucleus as a whole is supported by the orbital motions of these clumps rather than the turbulent motions of gas within the clumps (just as the orbital speeds of stars in the Milky Way's disk are much greater than their local velocity dispersion).  Indeed, I find that the orbital kinetic energy density in Arp 220's nuclei is comparable to $\pi G \Sigma_g^2$ (see Table~\ref{table:Pressures}), when I use the orbital speed of $325\ \kms$ \citep{Downes98}.

Similar considerations seem to apply to the GCCMZ, where the clumpiness of the gas can be directly observed \citep[e.g.,][]{Molinari11}.  Note that if $f_{\rm gas}$ is very small, as suggested by \citet{Launhardt02}, then the hydrostatic pressure apparently is much greater than even the orbital kinetic energy (Table~\ref{table:Pressures}).  This may mean that $f_{\rm gas}$ is actually of order a few tenths in this region, as in M82 and NGC 253.

\emph{Basic conclusions} -- Loose equipartition, at least, seems to hold between the various gas phases of the ISM in starbursts.  The major outliers are the CR energy density and the hydrostatic pressure estimate.  The hot superwind phase, if it exists, is in equipartition with the cold molecular gas; thus pressure arguments alone do not rule it out.  The magnetic energy density is comparable to the turbulent energy density.

\subsection{Cold clouds in a hot wind or hot bubbles in a cold sea: Which phase is volume-filling?}
\label{sec:VolumeFillingPhase}
With these estimates of the central pressure of starburst regions in mind, which phase fills most of the volume in starbursts?  Do supernova remnants (SNRs) expand until they largely overlap, or are they stopped and form little bubbles in the ISM?  The volume overlap factor is essentially given by:
\begin{equation}
Q_{\rm SNR} = \rho_{\rm SN} \times \frac{4}{3} \pi R_{\rm max}^3 \times t_{\rm SNR},
\end{equation}
the product of the volumetric rate of supernova explosions ($\rho_{\rm SN}$), the maximum volume of the SNR (set by the maximum radius of the remnant, $R_{\rm max}$), and the time the SNR survives before being washed away by the surrounding ISM ($t_{\rm SNR}$).  The filling fraction of the SNRs is then $f_{\rm fill} = 1 - \exp(-Q_{\rm SNR})$ \citep{McKee77,Heckman90}.  The survival time is generally assumed to be of order the ISM sound-crossing time of the SNR, $R_{\rm max} / \sqrt{P_{\rm ISM} / \rho_{\rm ISM}}$ \citep{McKee77}.

Suppose the ISM originally consists of cold molecular gas.  Then, the evolution of SNRs is most likely to be set by radiative losses in the dense ISM.  The maximum radius of a radiative SNR is 
\begin{equation}
R_{\rm max} = 6.6\ \pc\ E_{51}^{0.32} \left(\frac{n_{\rm ext}}{100\ \cm^{-3}}\right)^{-0.16} \left(\frac{P/k_B}{10^7\ \Kelv\ \cm^{-3}}\right)^{-0.2}
\end{equation}
from \citet{McKee77}.  In these conditions, and according to \citet{McKee77} and \citet{Heckman90}, the overlap factor is 
\begin{eqnarray}
\nonumber Q_{\rm SNR} & = & 3.4\ E_{51}^{1.28} \left(\frac{\rho_{\rm SN}}{10\ \kpc^{-3}\ \yr^{-1}}\right) \left(\frac{n_{\rm ext}}{100\ \cm^{-3}}\right)^{-0.14} \\
& & \times \left(\frac{P/k_B}{10^7\ \Kelv\ \cm^{-3}}\right)^{-1.3},
\end{eqnarray}
In this equation, $E_{51}$ is the mechanical energy of the supernova in units of $10^{51}\ \erg$, $\rho_{\rm SN}$ is the volumetric supernova rate, $n_{\rm ext}$ is the density of the surrounding ISM, and $P_{\rm ISM}$ is the ISM pressure.  

We need two ingredients: the density of the surrounding ISM, and the pressure of the surrounding ISM.  It is possible to relate the density to $\Sigma_{\rm SFR}$ with the Schmidt law \citep{Kennicutt98}, but the recent literature suggests that the gas masses in some starbursts (e.g., M82 and NGC 253) are lower than used in \citet{Kennicutt98} \citep{Weiss01}; furthermore, true merger-driven starbursts appear to lie on a different Schmidt law than main sequence galaxies \citep{Daddi10-Schmidt}.  I instead relate $\Sigma_g$ and $\Sigma_{\rm SFR}$ through the gas consumption time $\tau_{\rm gas}$:
\begin{equation}
%Exact value: 239.355 Msun/yr/kpc^2
\left(\frac{\Sigma_{\rm SFR}}{\Msun\ \yr^{-1}\ \kpc^{-2}}\right) = 240\ \left(\frac{\Sigma_g}{\gcm2}\right) \left(\frac{\tau_{\rm gas}}{20\ \Myr}\right)^{-1}.
\end{equation}
I then set $n_{\rm ext} = \delta (\Sigma_g / (2 h m_H)) = \delta \mean{\rho}_{\rm SB} / m_H$, where $\delta$ accounts for the possibility that SNe generally go off in an underdense medium.  

As noted in the previous subsection, there are several different estimates we can use for the ISM pressure.  The first is the turbulent energy density of the gas, $\mean{\rho}_{\rm ext} \sigma^2 / 2$.  In this case, the SNR overlap fraction is:
\begin{eqnarray}
\nonumber Q_{\rm SNR}^{\rm turb} & = & 0.55\ \left(\frac{\Sigma_{\rm SFR}}{\Msun\ \yr^{-1}\ \kpc^{-2}}\right)^{-0.44} \left(\frac{\tau_{\rm gas}}{20\ \Myr}\right)^{-1.44}\\
& & \times \left[E_{51}^{1.28} h_{50}^{0.44} \delta^{-0.14} \sigma_{50}^{-2.6}\right].
\end{eqnarray}
The auxiliary variables parameterize the mechanical energy per supernova as $10^{51} E_{51}\ \erg$, the starburst scale height as $50 h_{50}\ \pc$, and the turbulent dispersion as $50 \sigma_{50}\ \kms$.  The second option is to suppose the pressure everywhere in the ISM is equal to the CC85 wind pressure (given in Appendix~\ref{sec:CC85Properties}), giving:
\begin{align}
\nonumber Q_{\rm SNR}^{\rm CC85} & = 0.77\ \left(\frac{\Sigma_{\rm SFR}}{\Msun\ \yr^{-1}\ \kpc^{-2}}\right)^{-0.44} \left(\frac{\tau_{\rm gas}}{20\ \Myr}\right)^{-0.14}\\
& \times \left[E_{51}^{1.28} f_{\rm geom}^{2.6} \delta^{-0.14} \zeta^{-1.3} h_{50}^{-0.86} \left(\frac{\beta}{2}\right)^{-0.65} \left(\frac{\epsilon_{\rm therm}}{0.75}\right)^{-0.65}\right].
\end{align}
Finally, we can use the homogeneous hydrostatic pressure, $\pi G \Sigma_g^2 / f_{\rm gas}$:
\begin{eqnarray}
%Exact value: 67.2511
\nonumber Q_{\rm SNR}^{\rm hydro} & = & 80\ \left(\frac{\Sigma_{\rm SFR}}{\Msun\ \yr^{-1}\ \kpc^{-2}}\right)^{-1.74} \left(\frac{\tau_{\rm gas}}{20\ \Myr}\right)^{-2.74}\\
& & \times \left[E_{51}^{1.28} f_{\rm gas}^{1.3} \delta^{-0.14} h_{50}^{-0.86}\right].
\end{eqnarray}

\begin{figure}
\centerline{\includegraphics[width=8cm]{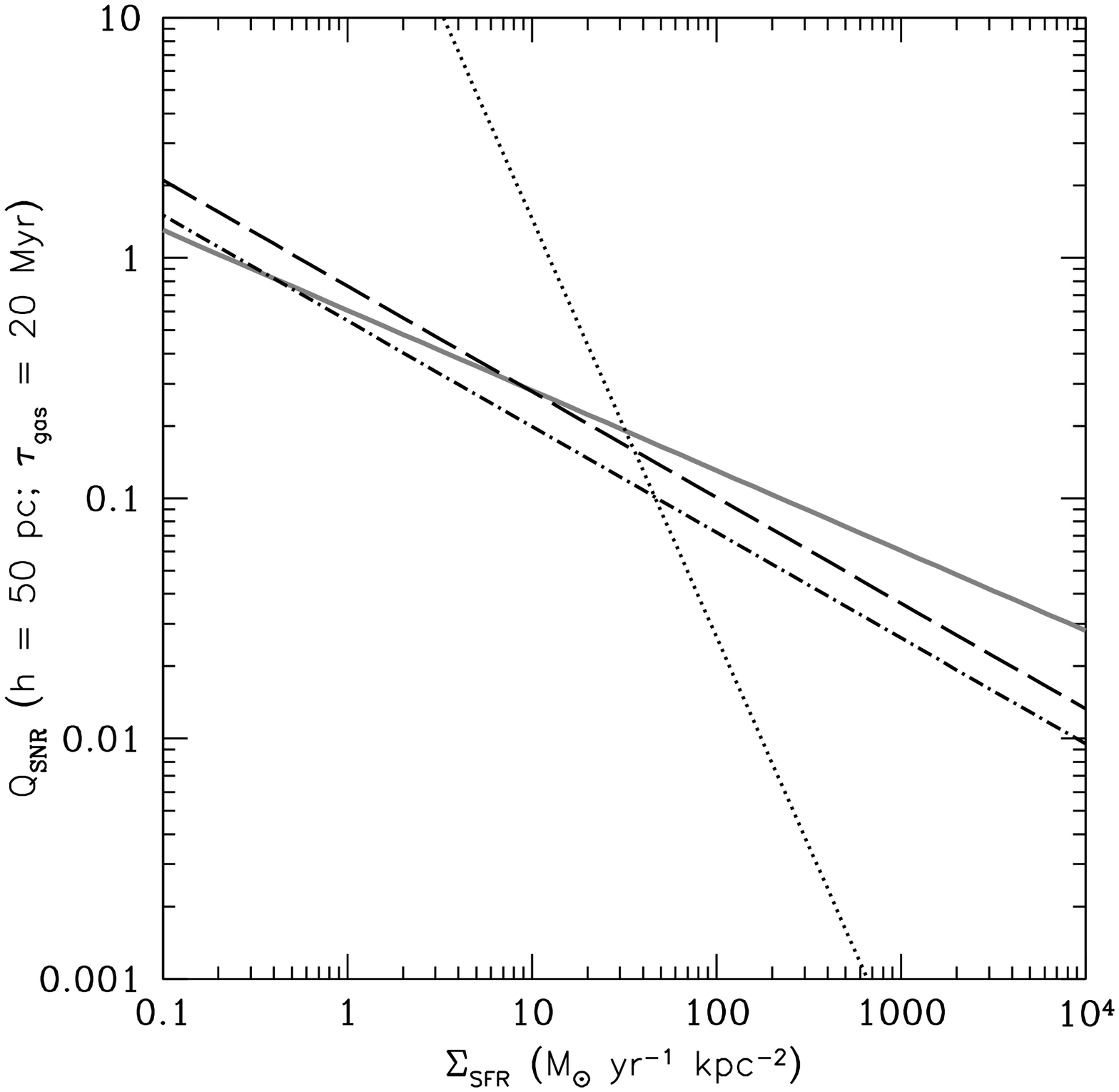}}
\figcaption{A plot of the overlap fraction $Q_{\rm SNR}$ of SNRs in the starburst ISM.  The black lines are for SNRs in the cold ISM.  The plotted lines assume different ISM pressures: the CC85 pressure (dashed), turbulent pressure with $\sigma = 50\ \kms$ (dash-dotted), and the hydrostatic pressure (dotted).  The estimates all drop below $\la 0.1$ for $\Sigma_{\rm SFR} \ga 50\ \Msun\ \yr^{-1}\ \kpc^{-2}$, indicating a transition to cold starburst ISMs.  The exact transition between hot and cold starbursts is unknown, though.  The grey line is the overlap fraction of SNRs in the hot ISM; values less than 1 indicates a truly diffuse hot phase. \label{fig:QSNR}}
\end{figure}

Because of its strong dependence on pressure, the overlap fractions differ greatly, but there is a universal trend towards smaller $Q_{\rm SNR}$ as $\Sigma_{\rm SFR}$ increases (Figure~\ref{fig:QSNR}).  In the weakest starbursts, $Q_{\rm SNR} \approx 1$, while $Q_{\rm SNR}$ drops to a few percent in powerful ULIRGs.  This trend is repeated in Table~\ref{table:BHotPredictions}, where I calculate the different $Q_{\rm SNR}$ for the GCCMZ, NGC 253, M82, and Arp 220.  The overlap factors are of order half in the NGC 253, and M82, and only a few percent in Arp 220.  There is a very wide disagreement in the $Q_{\rm SNR}$ for the GCCMZ, because the pressures in Table~\ref{table:Pressures} are themselves disparate.  If I use $P \approx 10^6\ \Kelv\ \cm^{-3}$ (the superwind, magnetic, radiation pressure), then $Q_{\rm SNR} \approx 1$.  Using the much greater molecular turbulence pressure gives values of a few percent, and the hydrostatic pressure implies that $Q_{\rm SNR} \approx 0$.

In practice, a few effects raise the overlap fractions from my fiducial parameters.  Star formation in starbursts occurs in clusters \citep{OConnell95,Kruijssen12}.  The combined effect of the stellar winds in the clusters digs out superbubbles.  In the low density superbubbles, the mechanical power of supernovae is not as easily lost to radiative losses.  In addition, some hot bubbles are heated by multiple supernovae.  Clustering increases the overlap fraction, since the overlap fraction depends more strongly on the number of supernovae in the bubble ($\propto E_{51}^{1.28}$) than on the number of bubbles ($\propto N$).  These effects may raise the overlap fraction a few times.  Evidently, $Q_{\rm SNR}$ \emph{is} enhanced in M82 enough for the hot wind to fill the volume, as demonstrated by the presence of hot X-ray gas \citep{Strickland07} and the rapid expansions of SNRs (see Section~\ref{sec:SNREvolution}).  In Arp 220, however, the overlap fraction would have to be raised by a factor $\ga 20$ for the hot phase to dominate.  \citet{Melioli04} found that the supernova heating efficiency is bistable in starbursts, tending either towards high radiative losses if dense molecular gas fills the starburst volume or efficient supernova heating once the hot phase grows enough.

High-redshift galaxies, both main sequence and true starburst, have greater scale heights.  At a given $\Sigma_{\rm SFR}$ and $\Sigma_g$, these galaxies should have smaller $\mean{\rho}$ and $\mean{\rho_{\rm SN}}$.  If the mean pressure in these galaxies is given by $\mean{\rho} \sigma^2 / 2$, then the decreased volumetric density implies that the overlap fraction is larger in these galaxies.  But both the CC85 wind pressure and the (homogeneous) hydrostatic pressure depend on column rather than volumetric density, so using these pressures implies that the overlap fraction decreases, as the larger scale height dilutes the SNRs.  

If the hot ionized medium is established, and the SNe go off in it, then what is the new overlap fraction?  The SNRs fade into the hot ISM before they suffer radiative losses \citep[compare with the cooling radii listed in Table 2 of][]{Draine91}.  In this case, the SNR ceases to be distinct from the background ISM when its internal energy density is comparable to the surrounding energy density:
\begin{equation}
R_{\rm max}^{\rm therm} \approx \left[\frac{E_{\rm SN}}{(4/3) \pi U_{\rm ISM}}\right]^{1/3} \approx \left[\frac{E_{\rm SN}}{2 \pi n_{\rm wind} k T_c}\right]^{1/3}
\end{equation}
Taking the CC85 wind pressure and density, I find 
\begin{align}
%Exact value: 34.5949 pc
\nonumber R_{\rm max}^{\rm therm} & \approx 35\ \pc\ \left(\frac{\Sigma_{\rm SFR}}{\Msun\ \yr^{-1}\ \kpc^{-2}}\right)^{-1/3}\\
& \times \left[E_{51}^{1/3} f_{\rm geom}^{2/3} \zeta^{-1/3} \left(\frac{\beta}{2}\right)^{-1/6} \left(\frac{\epsilon_{\rm therm}}{0.75}\right)^{-1/6} \right].
\end{align}
The overlap fraction is then
\begin{align}
%Exact value 0.522883
\nonumber Q_{\rm SNR}^{\rm hot} & \approx 0.52\  \left(\frac{\Sigma_{\rm SFR}}{\Msun\ \yr^{-1}\ \kpc^{-2}}\right)^{-1/3}\\
& \times \left[E_{51}^{4/3} f_{\rm geom}^{8/3} \zeta^{-4/3} \left(\frac{\beta}{2}\right)^{-1/6} \left(\frac{\epsilon_{\rm therm}}{0.75}\right)^{-7/6} \right].
\end{align}
with a SNR survival time
\begin{align}
\nonumber t_{\rm SNR}^{\rm hot} & \approx 47\ \kyr\ \left(\frac{\Sigma_{\rm SFR}}{\Msun\ \yr^{-1}\ \kpc^{-2}}\right)^{-1/3}\\
& \times \left[E_{51}^{1/3} f_{\rm geom}^{2/3} \zeta^{-1/3} \left(\frac{\beta}{2}\right)^{1/3} \left(\frac{\epsilon_{\rm therm}}{0.75}\right)^{-2/3} \right].
\end{align}

The overlap fraction is no bigger than in the cold phase, despite the lack of radiative losses to stop SNR growth, because the sound speed in the hot ISM is much larger, shortening the SNR survival time.  In all starbursts with $\Sigma_{\rm SFR} > 1\ \Msun\ \yr\ \kpc^{-2}$, $Q_{\rm hot} < 1$ (Figure~\ref{fig:QSNR}).  While $t_{\rm SNR}^{\rm hot}$ is short, the SNR is simply merging into a background of hot plasma, not being filled in by cold gas.  Instead, the low value of $Q_{\rm hot}$ means that while the hot phase fills some starbursts, the hot phase is truly diffuse as opposed to being a bunch of mostly-overlapping SNR bubbles.  Most of the hot plasma in weaker starbursts cannot be attributed to individual SNRs, but only to the collective action of supernovae over hundreds of kyr.

For convenience, I group starbursts into two varieties through this work.  I call the lower density starbursts, including the GCCMZ, NGC 253, and M82, ``hot'' starbursts because the hot phase fills most of their ISMs.  Higher density starbursts, including Arp 220 and submillimeter galaxies, are grouped as ``cold'' starbursts.  Both groups stand in contrast to the Milky Way and other $z \approx 0$ normal galaxies, where warm neutral and warm ionized gas fills most of the ISM, and thus might be called ``warm'' galaxies.  Figure~\ref{fig:ISMSketches} depicts the different ISM phase structure in warm, hot, and cold starbursts.

I address other objections to the existence of hot starbursts in Appendix~\ref{sec:HotObjections}.

\subsection{SNR evolution observed in starbursts}
\label{sec:SNREvolution}
I motivate the above picture by appealing to long-term evolution of SNRs in starbursts.  The properties of SNRs in nearby starbursts can be directly determined from the properties of compact radio and X-ray sources.  With Very Long Baseline Interferometry (VLBI), it is actually possible to directly determine both the size and expansion speeds of SNRs and radio SNe.  Since the late evolution of SNRs is strongly affected by which phase they are in, these observations are powerful tools for understanding the ISM phase structure of starbursts.

The basic idea, that supernovae going off in molecular clouds have highly radiative remnants whereas those that go off in the hot wind are in the Sedov phase when they fade away, was described by \citet{Chevalier01}.  Under the assumption that molecular clouds encase supernovae, and assuming an average density of $\mean{n_H} \approx 1000\ \cm^{-3}$, \citet{Chevalier01} predicted that most SNRs in M82 progress through the Sedov and enter the radiative phase rapidly.  By the time they reach the observed radii of most SNRs in M82, a few parsecs, they should be slowed down to a speed $\sim 500\ \kms$.  \citet{Chevalier01} argued that these SNRs would evolve slowly, explaining why the SNR's radio fluxes do not appear to vary with time.  

Yet VLBI observations repeatedly show this not to be the case for most SNRs.  Instead, the typical expansion speeds of the SNRs are roughly $10^4\ \kms$ \citep{Pedlar99,Beswick06,Fenech10}.  The speeds are, in fact, about the same as for new supernovae \citep{Brunthaler10}.  If the SNRs are embedded in the molecular medium, it would be difficult to explain these observations.  However, if the SNRs are in the hot wind, they only reach the Sedov phase when they have expanded to a radius $R_{\rm Sedov} = [3 M_{\rm ej} / (4 \pi \rho_{\rm ext})]^{1/3}$:
\begin{equation}
\label{eqn:RSedov}
%Exact value: 2.12933
R_{\rm Sedov} \approx 2.1\ \pc \left(\frac{M_{\rm ej}}{1\ \Msun}\right)^{1/3} \left(\frac{n_{\rm ext}}{1\ \cm^{-3}}\right)^{-1/3},
\end{equation}
where $M_{\rm ej}$ is the mass of the SN ejecta.\footnote{\citet{Beswick06} also criticized the ISM pressure $10^7\ \Kelv\ \cm^{-3}$ used by \citet{Chevalier01}, suggesting that it was much too high.  But as Table~\ref{table:Pressures} shows, that estimate is if anything \emph{too low} -- the energy densities of the molecular gas turbulence, superwind thermal gas, magnetic fields, radiation, hydrostatic pressure, and even the thermal pressure in some H II regions are all $\ga 10^7\ \Kelv\ \cm^{-3}$.  The first counterargument of \citet{Beswick06} is that the column of ionized gas inferred from the free-free absorption cutoff of supernova remnants is too small under such pressures, unless ionized gas only fills a small fraction.  But this is what I have argued in \citet{Lacki13-LowNu} -- the H II regions \emph{are} the WIM but fill only a few percent of the starburst volume.  Furthermore, the H II regions may be supported by turbulent pressure, so the densities estimated by \citet{Beswick06} might be too high.  Second, they quote the small thermal pressures inferred for M82's neutral gas.  As I discussed in \S~\ref{sec:UStarburst}, the pressure is dominated by turbulence, even in high density clumps.  Of course, none of this invalidates the basic conclusion that the SNRs are expanding quickly, in a low density medium.}

As for why the SNRs in M82 are not embedded within molecular clouds as originally supposed by \citet{Chevalier01}, note that the typical time after a stellar population forms until the first SNe go off is several Myr.  On the other hand, the turbulent gas within M82's molecular medium rearranges itself within one eddy time, $\ell_{\rm outer} / \sigma \approx 200\ \kyr\ (\ell_{\rm outer} / 10\ \pc) (\sigma / 50\ \kms)^{-1}$.  Thus, even ignoring the fact that stellar clusters may disrupt the molecular clouds they were born in \citep[e.g.,][]{Murray10} and the ``cruel cradle effect'' \citep[e.g.,][]{Kruijssen12}, the transient clumps of molecular gas that swathe a stellar nursery can be long gone by the time the first SNe go off.

What about Arp 220, where I argue that the molecular medium fills the entire starburst?  In Arp 220, VLBI observations paint a far different picture \citep[e.g.,][]{Smith98,Rovilos05,Lonsdale06,Parra07}.  Aside from an extremely large number of radio supernovae, the  radio sources observed in Arp 220 appear to be very small SNRs with radii $\sim 0.27 - 0.38\ \pc$ \citep{Batejat11}.  By noting the presence of the SNRs in observations from the prior decade, \citet{Batejat11} estimate that the expansion speed is at most $5000\ \kms$, implying that the SNe are well into the Sedov phase.  Comparing to equation~\ref{eqn:RSedov}, this is consistent with the SNe going off in a density $\sim 10^4\ \cm^{-3}$ \citep{Batejat11}.  By contrast, the Sedov radius for a CC85 wind with $n_e \approx 25\ \cm^{-3}$ is $0.7\ \pc\ (M_{\rm ej} / \Msun)^{1/3}$.  Assuming that SNe go off in random places in Arp 220's ISM, the small, slowly expanding SNRs observed by \citet{Batejat11} imply that the cold molecular gas is volume-filling.  

It would be interesting to see where the demarcation between hot wind and cold gas dominated starburst occurs by conducting VLBI observations of starbursts with properties between those of M82 and Arp 220.  One possible target is Arp 299; there have been several VLBI observations but no SNR expansion speeds reported yet \citep{Neff04,Ulvestad09,PerezTorres09}.

\subsection{What drives the turbulence in starbursts?}
\label{sec:TurbulenceSources}
Many drivers of turbulence in star-forming galaxies, tapping different energy sources, have been proposed.  These include gravitationally-powered disk-scale instabilities \citep{Fleck81,Wada02,Agertz09}, galactic accretion \citep{Klessen10}, the expansion of superbubbles and SNRs \citep{Norman96,Korpi99,Dib06,Joung06,Joung09}, H II region expansion \citep{Matzner02}, protostellar winds and jets \citep{McKee89,Li06}, and radiation pressure \citep{Murray10}.  A combination of any of these may operate in different classes of galaxy and in different ISM phases.  For example, the primary power source for Milky Way turbulence is probably superbubbles and supernovae, but with additional contributions from the other sources depending on the scale \citep{Miesch94,Norman96,MacLow04,Haverkorn08,Brunt09,Lee12}.

The hot superwind of starbursts is transparent to radiation and is not bound to the starburst.  The only possible drivers of turbulence are those with mechanical forcing: the expansion of superbubbles and SNRs, in particular.  Supernovae and stellar winds  are the power source for these; their mechanical luminosity is well known (see Appendix~\ref{sec:CC85Properties}).   

The situation is less clear in molecular clouds and in cold starbursts.  Supernovae go off in these regions, and might drive turbulence there as well \citep{Ostriker11}.  Supernova remnants may be highly radiative in these regions, with $\sim 10\%$ of their mechanical energy going into turbulent forcing \citep{Thornton98}.  On the other hand, radiation is potentially a powerful source of turbulence in the extreme infrared radiation pressure and optical depths in these environments (Table~\ref{table:Pressures}) (\citealt{Murray10,Ostriker11}; but see \citealt{Krumholz12}).  The Toomre ${\cal Q}$ is observed to be $\la 1$ in Arp 220 (see Section~\ref{sec:UStarburst}; \citealt{Kruijssen12}), a condition which promotes disk instabilities \citep{Ostriker11}.

The debate over the source of turbulence in nearby starbursts is related to the similar debate over turbulence driving in high-redshift galaxies.  High redshift disks are observed to host clumps and large velocity dispersions.  Although \citet{Green10} argued that a correlation between turbulence and star-formation rate proves star formation powers turbulence, other authors note that Toomre ${\cal Q}$ is near 1 in the clumps and propose that disk instabilities are the source of turbulence \citep{Genzel08,Burkert10}.

In any case, though, stellar mechanical luminosity sets a lower bound on the turbulent energy injection.  I therefore focus my discussion on turbulence powered by SNRs and other mechanical sources.  The addition of other turbulent drivers would lead to faster turbulent velocity dispersions, and stronger turbulent magnetic fields. 

\subsection{What about the cool and warm gas?}
\label{sec:WarmGas}
\begin{figure}
\centerline{\includegraphics[width=7cm]{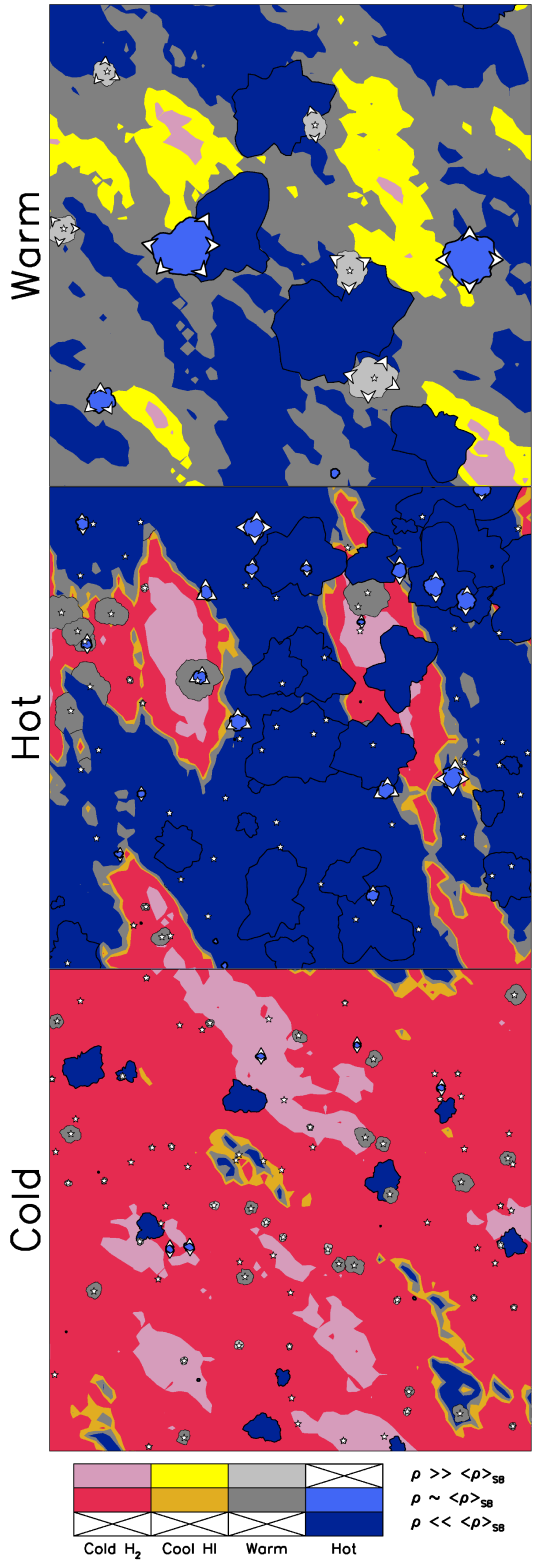}}
\figcaption{Schematic illustration of the ISM structure of galaxies and starbursts.  The Milky Way's ISM is ``warm'': its relatively quiescent conditions allow a rich phase structure, including warm gas with large filling factor.  H II regions and young SNRs are overpressured and expand (indicated by the outward arrows).  Molecular clouds are rare overdensities.  But in ``hot'' starbursts, the high density allows efficient cooling; most of the gas mass is cold H$_2$, with only trace amounts of cool and warm gas.  Frequent supernovae puncture out a volume-filling hot superwind.  H II regions are small and in equilibrium with the surrounding ISM.  Finally, in ``cold'' starbursts, the pressure is so high that even the SNRs fail to overlap.  The ISM is predominantly cold H$_2$. \label{fig:ISMSketches}}
\end{figure}

Throughout my discussion, I have largely ignored the diffuse warm phases, the Warm Neutral Medium (WNM) and the Warm Ionized Medium (WIM), as well as the Cool Neutral Medium (CNM).  HI gas makes up the great majority of the Milky Way's gas mass, and the WIM and WNM fill roughly half of its volume (see Figure~\ref{fig:ISMSketches}).  However, the molecular fraction of gas increases with increasing $\Sigma_g$ and $\Sigma_{\rm SFR}$; for starbursts, the majority of the gas mass is molecular \citep{Wong02,Ferriere07,Krumholz09}.

This still leaves a minority of the gas mass in neutral phases in weaker starbursts (e.g., \citealt{Ferriere07} and references therein for the GCCMZ; \citealt{Yun93} and \citealt{Wills98} for M82).  Yet its density should tend to be near the mean molecular gas density.  The neutral gas' pressure support is either thermal or neutral.  If the neutral gas were very underdense, it would need to have much higher turbulent speeds than the molecular gas ($\gg 50\ \kms$) to resist collapse.  Yet such speeds would eject the neutral gas.  On the other hand, the sound speed of warm gas, $\sim 16\ \kms$ is only a few times smaller than the turbulent speed of the molecular gas.  Thus, warm gas cannot remain in equilibrium if it is in very overdense, because the thermal pressure alone would exceed the surrounding turbulent pressure.  Cool neutral gas might be overdense if it had a small turbulent velocity, but as I show in Section~\ref{sec:ColdTurbulence}, the expected turbulent energy density depends weakly on the gas density. 

A WIM with a substantial filling fraction of starbursts is unlikely to exist in starburst regions \citep{Lacki13-LowNu}.  Briefly, the high densities of starburst gas leads to rapid recombination.  Therefore, the ionizing photons produced by the starburst can only ionize a few percent of the gas mass.  In addition, not all ionizing photons may be available, if some are destroyed by dust absorption or escape through the hot wind.  A low density WIM is inconsistent with the pressure constraints.  Instead, the only WIM that can survive is in H II regions that fill a small fraction of the volume and have densities comparable to the mean.  And indeed, such H II regions are known to exist through radio recombination line studies \citep{Zhao93,RodriguezRico05,RodriguezRico06,Law09}: the H II regions most likely \emph{are} the WIM in starbursts \citep{Lacki13-LowNu}.

In short, the WIM, WNM, and CNM, to the extent they exist in starbursts, have mean densities comparable to the molecular gas.  Since the calculation of turbulent speeds and magnetic fields does not depend on temperature, these quantities should be similar to those of the molecular gas as derived in Section~\ref{sec:ColdTurbulence}.  The one difference is the Mach number, which is much smaller in the WIM and WNM than in molecular gas because of the higher temperatures.  Smaller Mach numbers lead to smaller density contrasts in the turbulence \citep{Ostriker01}.  

It therefore appears that starburst ISMs are, in some sense, simpler than the Galactic ISM.  The famous three phases (cold, warm, and hot) of the Milky Way and other warm galaxies \citep{McKee77} are reduced to two in hot starbursts (cold and hot) and finally only one in cold starbursts, as depicted in Figure~\ref{fig:ISMSketches}.

\section{Turbulence and Magnetic Fields in Superwinds}
\label{sec:HotTurbulence}

\subsection{Motivation for magnetic fields in the superwind}
\label{sec:HotWindMotivation}
What evidence do we even have that there is turbulence and magnetic fields in the hot wind?  Although SNRs no doubt push into the wind, such motions do not necessarily fill the whole medium.  Nor is it obvious that a fluctuation dynamo operates in the hot wind.

Although we have little direct evidence regarding the magnetic fields and turbulence in superwinds, there is circumstantial evidence from the synchrotron emission.  VLBI observations resolve out most of a starburst's radio emission, indicating it is truly diffuse and not confined to small clumps \citep[e.g.,][]{Lonsdale06}.  In the GCCMZ, one of the few starburst regions we can resolve in low frequency radio, 74 MHz observations reveal synchrotron emission shining from beyond H II regions, implying the CRs are not confined to these regions \citep{Brogan03,Nord06}.  A MERLIN image of M82 also shows synchrotron emission at 400 MHz outside of a large H II region \citep{Wills97}.  I propose the simplest explanation for this diffuse radio emission is that CR $e^{\pm}$ actually do permeate the volume of the starburst and are radiating synchrotron throughout it.  In hot starbursts, this means that the CRs spend most of their time in the volume-filling phase, the superwind.  The diffuse synchrotron emission traces magnetic fields in the superwind.

In addition, the gamma-ray luminosities of M82 and NGC 253 imply that CR protons are being advected out by the starburst wind.  If they instead remained in the starburst, they would eventually be destroyed by pionic losses as they passed through dense gas clouds.  In this ``proton calorimeter'' limit \citep[c.f.,][]{Pohl94}, the energy that went into accelerating CR protons would then be converted into pions, about a third of it ending up as gamma-rays.  The actual gamma-ray luminosities of M82 and NGC 253 are only $\sim 40\%$ of the calorimetric limit, implying that CRs are escaping \citep{Lacki11-Obs,Abramowski12,Ackermann12}.  Furthermore, the hard GeV to TeV gamma-ray spectra of these starbursts imply that CRs escape through an energy-independent process, which is contrary to our normal understanding of CR diffusion \citep{Abramowski12}.  The GCCMZ also displays underluminous but hard gamma-ray emission and even synchrotron radio emission for its star-formation rate, which suggests rapid energy-independent CR escape \citep{Crocker11-Wind}.  Advection by the starburst wind naturally accounts for these observations.  But whereas the hot wind is consistently expelled out of the starburst, only $\sim 1\%$ of the cold and warm gas is ejected per advection time ($\sim 300\ \kyr$).  Otherwise, the gas mass would be depleted within a Myr, quenching star formation.  Unless the CRs ``know'' which warm and cold clouds are going to be expelled, advective escape implies they are primarily in the hot superwind.

The supernova observations I describe in \ref{sec:SNREvolution} support CR injection directly into the superwind too.  The rapidly expanding supernova remnants of M82 can only exist if they are in a low density phase, like the superwind.  Yet they are clearly glowing in synchrotron, implying CR $e^{\pm}$ acceleration.  Additional support comes from 400 MHz MERLIN observations of supernova remnants in M82.  These also emit synchrotron radiation, but a significant fraction display no free-free absorption \citep{Wills97}, again implying they are not in H II regions. 

Finally, radio haloes are observed around starbursts, indicating CR $e^{\pm}$ and magnetic fields are present in their winds \citep{Seaquist91}.  A similar kiloparsec-scale radio halo is observed around the GCCMZ \citep{Crocker11-Wind}.  Since there is warm and cold gas mixed with the wind, this by itself does not prove the CR $e^{\pm}$ and magnetic fields are in the hot phase per se.  Strong adiabatic losses may also cool the wind and the CR $e^{\pm}$ on kiloparsec scales from the starburst (CC85), so the physical conditions may not be equivalent to the hot ISM inside the starburst.  But the radio haloes are consistent with magnetic fields and CR $e^{\pm}$ being present in the hot wind itself.

The existence of turbulence (or at least volume-filling random motions), magnetic fields, and CRs throughout hot superwinds is therefore plausible.  Clearly this cannot be the whole picture: the (presumably pionic) gamma-ray emission from the GCCMZ, NGC 253, and M82 implies that CRs must at some point enter denser gas.  In fact, models of the nonthermal emission of M82 and NGC 253 are consistent with CRs experiencing the average density of gas in starbursts over their lifetimes (e.g., \citealt{Domingo05,Persic08,deCeaDelPozo09-M82,Rephaeli10,Lacki10-FRC1,Lacki13-XRay,YoastHull13}; higher than average densities are possible as shown by \citealt{Paglione12}), suggesting that they traverse all of the phases of the ISM.\footnote{But this is not the case in the GCCMZ, where the amount of gamma-ray emission is so low that CR protons are evidently not penetrating molecular clouds \citep{Crocker11-Wild}; this is also indicated by the lack of radio emission from secondary $e^{\pm}$ in GCCMZ molecular clouds \citep{Jones11}.}  The situation would then be analogous to the Milky Way, where CRs diffuse through the low density, warm/hot Galactic halo outside of the gas disk most of the time, but experience most of their gas interactions in the neutral and molecular Galactic disk.  As in the Milky Way, most of the synchrotron emission of starbursts comes from underdense hot plasma, but occasionally the CRs penetrate into molecular clouds and/or H II regions and experience energy losses from pion production, bremsstrahlung, and ionization.  This could happen if there are a sufficiently large number of molecular clouds, so that most magnetic field lines (and, thus, most CR paths) traverse a molecular cloud, as described in detail by \citet{YoastHull13}.

I note that this scenario, which motivates my calculation of the turbulence and magnetic fields in superwinds, bears some similarities to that presented by \citet{Becker09}.  \citet{Becker09} argued that the thermal, magnetic, CR, and turbulent energy densities in the hot ISM are in equipartition in actively star-forming galaxies.  They argued this meant that the timescales regulating each are equal, and derived magnetic field strengths by relating the CR diffusive escape time to the radiative cooling time.  My hypothesis is weaker, simply that turbulence and magnetic fields are in equipartition.  As it turns out, the four energy densities in the wind end up being roughly equal in weaker starbursts, but this is not explicitly required in this paper and the defining timescale is the flow crossing time (see Section~\ref{sec:Equipartition}).  In denser starbursts, equipartition between CRs and magnetic fields fails entirely.  

My discussion is guided by the CC85 theory of starburst winds, which gives approximate values for the wind temperature, central density, and pressure.

\subsection{Are superwinds collisional?}
\label{sec:CollisonlessWinds}

Many processes can shape the wind and alter the CC85 values, such as nonthermal pressure from magnetic fields and CRs, but one deserves special mentioning, the low collisionality of hot dilute plasmas.  The CC85 theory assumes that the wind is a hydrodynamic fluid, which has a locally defined pressure and temperature.  Yet, when the plasma has a small enough density, it becomes more collisionless, and its electrons and ions may have different temperatures ($T_e$ for electrons and $T_i$ for ions), or even no temperature at all.  Although \citet{Strickland07} briefly investigated the possibility that M82's wind plasma is collisionless, the issue bears greater examination when considering the very different conditions in wind fluids for starburst regions ranging from the GCCMZ to ULIRGs. 

Collisions are most effective at transferring energy when the two particles have the same mass.  While the effective cross sections for collisions between particles with the same charge are the same, lighter particles have a higher microscopic speed at a given energy, so their collision rate is faster.  Thus, the electron-electron collision frequency is highest, so electrons are the most collisional particles with the best-defined temperature.  The electron-electron collision frequency is $\nu_{ee} = 4 \pi n_e e^4 (\ln \Lambda)/ \sqrt{m_e (k_B T_e)^3}$.  The typical value of the Coulomb logarithm ($\ln \Lambda$) in a superwind plasma is $\sim 30$ \citep{Huba11}.  In a CC85 superwind, the electron-electron collision frequency is:
\begin{align}
%Exact value: 1.3176 kyr
\nonumber \nu_{ee}^{-1} & = 1.3\ \kyr\ \left(\frac{\Sigma_{\rm SFR}}{\Msun\ \yr^{-1}\ \kpc^{-2}}\right)^{-1}  \left(\frac{T_e}{T_c}\right)^{3/2}\\
              & \times \left[f_{\rm geom}^2 \zeta^{-1} \left(\frac{\ln \Lambda}{30}\right)^{-1} \left(\frac{\beta}{2}\right)^{-3} \left(\frac{\epsilon_{\rm therm}}{0.75}\right)^2 \right].
\end{align}
Note that the time for the wind to escape the starburst region is typically a few hundred kyr.  In all starburst superwinds, the electrons are collisional and have established a Maxwellian pressure and temperature distribution.

Since ions are heavier, they are slower and collide more rarely, with $\nu_{ii} = 4 \pi n_H e^4 / \sqrt{m_H (k_B T_e)^3}$ (for hydrogen ions):
\begin{align}
%Exact values: 65930.6 kyr
\nonumber \nu_{ii}^{-1} & = 66\ \kyr\ \left(\frac{\Sigma_{\rm SFR}}{\Msun\ \yr^{-1}\ \kpc^{-2}}\right)^{-1} \left(\frac{T_i}{T_c}\right)^{3/2}\\
              & \times \left[f_{\rm geom}^2 \zeta^{-1} \left(\frac{\ln \Lambda}{30}\right)^{-1} \left(\frac{\beta}{2}\right)^{-3} \left(\frac{\epsilon_{\rm therm}}{0.75}\right)^2 \right].
\end{align}
In the weakest starbursts, like the GCCMZ, the ion-ion collision approaches the advection time.  Thus, the ions in these regions will only approximately establish a Maxwellian distribution.  

Finally, electrons and ions are poorly coupled in wind plasmas because they have such different masses.  The light electrons are easily deflected by the much heavier ions with an electron-ion collision frequency that is just $\nu_{ei} = \nu_{ee} / (2 \sqrt{2})$.  Yet very little energy transfer occurs in these deflections, only about $m_e / m_H \sim 1/2000$ of the particle kinetic energy \citep{Kulsrud05}.  The timescale for energy transfer between the ions and electrons is set by the ion-electron collison frequency, which is much slower even than $\nu_{ii}$: $\nu_{ie} = \nu_{ee} m_e / (2\sqrt{2} m_i)$ or
\begin{align}
%Exact value: 6.85555 Myr 
\nonumber \nu_{ie}^{-1} & = 6.9\ \Myr\ \left(\frac{\Sigma_{\rm SFR}}{\Msun\ \yr^{-1}\ \kpc^{-2}}\right)^{-1} \left(\frac{T_{e,i}}{T_c}\right)^{3/2}\\
              & \times \left[f_{\rm geom}^2 \zeta^{-1} \left(\frac{\ln \Lambda}{30}\right)^{-1} \left(\frac{\beta}{2}\right)^{-3} \left(\frac{\epsilon_{\rm therm}}{0.75}\right)^2 \right].
\end{align}
Over a broad range of starburst conditions -- including the GCCMZ, NGC 253, and M82 -- the ion-electron collision time is comparable to or longer than the advection time.  The electrons and ions do not necessarily have the same temperature in these starbursts.  

Collisions only matter for the dynamics on the largest scales, those bigger than the mean free path.  The mean free path is given by $\lambda \approx v / \nu$, where $v$ is the thermal speed of the particle.  The mean free paths are similar for electrons and hydrogen ions.  For the specific case of electrons, with a thermal speed $v_e \approx \sqrt{k T_e / m_e}$, the mean free path is :
\begin{align}
%Exact value: 31.9472 pc
\nonumber \lambda_{ee} & = 32\ \pc\ \left(\frac{\Sigma_{\rm SFR}}{\Msun\ \yr^{-1}\ \kpc^{-2}}\right)^{-1}\\
             & \times \left[f_{\rm geom}^2 \zeta^{-1} \left(\frac{\ln \Lambda}{30}\right)^{-1} \left(\frac{\beta}{2}\right)^{-7/2} \left(\frac{\epsilon_{\rm therm}}{0.75}\right)^{5/2} \right].
\end{align}
The GCCMZ's scale height is actually not much bigger than the particle mean-free path -- collisionless dynamics play a role in all scales in that region.  In M82 and NGC 253, the superwind plasma is collisionless on scales smaller than a parsec.  

Therefore, unlike molecular clouds or the warm ionized medium of the Milky Way, a starburst superwind is \emph{not} a standard MHD fluid characterized by a single pressure and temperature.  Instead, it is governed by kinetic processes, with plasma waves and magnetic fields shaping the particle distributions on parsec and smaller scales.  Hydrodynamical models might leave out important physics, and quantities like the central temperature in the superwind are potentially inaccurate.  For the remainder of this paper, I assume that the CC85 temperature and density (equations~\ref{eqn:nC} and~\ref{eqn:SuperwindTc}) are correct, at least well enough to estimate the properties of turbulence in the superwind, but a kinetic treatment would be very useful.

Finally, what might be the signs of collisionless plasma?  First, the electron and ion temperatures can be different.  Fortunately, the different populations emit hard X-rays through distinct mechanisms.  X-ray line emission comes from ions, while bremsstrahlung continuum is mostly emitted by the easily-deflected electrons.  \citet{Strickland07} actually found that the diffuse hard continuum and iron line emission of M82 appeared to be tracing plasma of different properties.  Non-equilibrium between electrons and ions may account for this.  Second, as kinetic effects involving magnetic fields regulate the plasma, the magnetic field might be expected to introduce a preferred direction and \emph{anisotropy} to the wind plasma.  

\subsubsection{The Reynolds number: Can true turbulence exist in hot winds?}
\label{sec:Reynolds}
Given the low collisionality of wind plasma, can turbulence exist at all in the superwind?  In these conditions, the Braginskii viscosity is very high, preventing hydrodynamical turbulence from appearing even if collisions are effective.  The viscosity is dominated by ions moving along the magnetic field lines.  This characteristic Braginskii viscosity is given by 
\begin{equation}
\eta_0^i = 5.11 n_i k_B T_i \nu_{\rm ii}^{-1},
\end{equation} 
which is typically $\sim 600\ \gram\ \sec^{-1}\ \cm^{-1}$ for the superwind phase.  The Reynolds number of a flow in the superwind is then ${\cal R} = \rho v \ell / \eta_0^i$:
\begin{align}
%Exact number: 0.633845
\nonumber {\cal R} & = 0.63 \left(\frac{\Sigma_{\rm SFR}}{\Msun\ \yr^{-1}\ \kpc^{-2}}\right) \left(\frac{v}{1000\ \kms}\right) \left(\frac{\ell}{50\ \pc}\right)\\
         & \times \left[\zeta f_{\rm geom}^{-2} \left(\frac{\ln \Lambda}{30}\right) \left(\frac{\beta}{2}\right)^4 \left(\frac{\epsilon_{\rm therm}}{0.75}\right)^{-3}\right].
\end{align}
For standard hydrodynamical flows, turbulence appears when ${\cal R} \ga 10^3 - 10^4$, implying starburst superwinds are laminar flows: while random motions may be present, the energy in them does not cascade down to small scales and dissipate as in true turbulence \cite[see similar considerations for the Galactic coronal phase in][]{McIvor77}.  Motions driven on yet smaller scales will have smaller velocities, and still smaller Reynolds numbers. 

The problem with this argument is that collisionless processes can reduce the thermal ion mean free path and, with it, the viscosity.  First, as discussed in \S~\ref{sec:CollisonlessWinds}, starburst regions themselves can be smaller than the collisional mean free path, which means that collisionless processes must set the actual ion mean free path.  At most the effective kinematic viscosity is $c_s \ell_{\rm outer} / 3$, so that the effective Reynolds number is ${\cal R} \ga 3 v / c_S$ \citep{Subramanian06}.  The Reynolds number of the wind is therefore unlikely to be less than $\sim 1$.  While this lower bound is still not large enough to guarantee hydrodynamical turbulence, this is a conservative estimate that assumes that collisionless processes can deflect ions only on length-scales comparable to the starburst size itself. 

A very similar situation occurs in galaxy clusters, which are filled with hot ($\sim 10^8\ \Kelv$) and rarefied gas, just like starburst winds.  Even though the formal Reynolds number is $\la 1000$, turbulence appears to be present.  But the effective Reynolds number in galaxy clusters is thought to be raised by collisionless processes such as plasma instabilities \citep[e.g.,][]{Schekochihin05,Schekochihin06,Lazarian06,Subramanian06,Brunetti07}.  Similar processes may operate in starburst winds.  

\subsection{What is the outer scale of turbulence in the hot wind?}
\label{sec:HotOuterScale}
Turbulence dissipates more slowly on large scales than on small scales.  The final energy density of the turbulence, and its strength at each scale in the turbulent cascade, therefore depends on the scale most of the energy is injected.  In the Milky Way, this outer scale $\ell_{\rm outer}$ of the turbulence seems to depend on the location and ISM phase, with values ranging from a few parsecs for ionized gas in the Galactic spiral arms \citep{Haverkorn08} to about a hundred parsecs \citep{Haverkorn08,Chepurnov10}.  However, such large scales are not attainable in starbursts, simply because that is larger than the physical size of the starburst itself.  Instead, the largest possible scale of the turbulence is the scale height of the starburst, $h \approx 50\ \pc$.  In the GCCMZ, there is evidence that turbulence is injected on fairly large scales, $\ga 30\ \pc$ \citep{Shetty12}, which would be consistent with $\ell_{\rm outer} \approx h$.

For a lower bound, the outer scale of turbulence is plausibly the scale on which individual supernova remnants fade into the background ISM.  One specific possibility is that $\ell_{\rm outer}$ is the size of a supernova remnant whose energy density is equal to the pressure of the external medium: $\ell_1 \equiv [3 E_{\rm SN} / (4 \pi P_{\rm ext})]^{1/3}$.  This is consistent with the turbulent outer scale in the Milky Way: for an average supernova kinetic energy of $10^{51}\ \erg$ and Milky Way-like pressures, the outer scale is $\sim 180\ \pc\ [(P_{\rm ext}/ k_B) / (10^4\ \Kelv\ \cm^{-3})]^{1/3}$.  Assuming the external pressure is the thermal energy density in the hot superwind, the outer scale is then found as 
\begin{equation}
\ell_1 = \left[\frac{3 E_{\rm SN}}{4 \pi \times (3/2) n k T}\right]^{1/3} = \left[\frac{E_{\rm SN}}{2 \pi n K T}\right]^{1/3}.
\end{equation}

Another possible outer scale is that it is the radius $\ell_2$ of a supernova remnant whose shock speed $v$ is equal to the velocity dispersion of the surrounding ISM velocity field.  At this point, the supernova remnant ceases to be a distinct region in the ISM.  I calculate this scale by supposing that the supernova remnant is in the non-radiating Sedov phase, since the hot ISM cools inefficiently.  Then the kinetic energy of the supernova remnant is conserved with time, and assuming a uniform density, $v = \sqrt{3 E_{\rm SN} / (2 \pi r^3 \rho)}$.  The outer scale is then
\begin{equation}
\ell_2 = \left[\frac{3 E_{\rm SN}}{2 \pi \sigma^2 \rho}\right]^{1/3}.
\end{equation}
Note this scenario is the same as supposing that the energy density of the supernova is equal to the external turbulent pressure.  

Finally, most star-formation in starbursts occurs in large stellar clusters, some with masses far exceeding $10^6\ \Msun$ \citep{OConnell95,McCrady07,Kruijssen12}.  The combined energy input from the stars and supernovae in these clusters create large superbubbles.  The superbubbles also drive the ISM, but can do so on larger scales than individual supernovae \citep[e.g.,][]{Castor75,Silich07}.  If these superbubbles dominate energy injection, the outer scale of turbulence may plausibly be as large as $h$.  Turbulent driving can even occur over a wide range of scales; this is thought to be the case in the Milky Way \citep{Norman96,Joung06}.

\subsection{The energetics of superwind turbulence}
\label{sec:TurbDensity}
The turbulent energy injection rate into the superwind is $\epsilon_{\rm turb} \dot{E}_{\rm mech}$, where $\epsilon_{\rm turb}$ is the efficiency of conversion of mechanical power into turbulence.  If the turbulence survives for a timescale $t_{\rm turb}$, the turbulent energy density is
\begin{equation}
\label{eqn:UTurbBasic}
U_{\rm turb} = \frac{\epsilon_{\rm turb} \dot{E}_{\rm mech} t_{\rm turb}}{V} 
\end{equation}
where $V = 2 \pi R^2 h$ is the volume of the starburst.

If the turbulence has a characteristic speed $\sigma$, it is expected to dissipate into heat over one crossing time, $\ell_{\rm outer} / \sigma$, where $\ell_{\rm outer}$ is the outer scale \citep[e.g.,][]{Tennekes72}.  We can solve for $\sigma$ by using $U_{\rm turb} = \rho \sigma^2 / 2$ and equation~\ref{eqn:UTurbBasic}:
\begin{equation}
\label{eqn:sigmaAnalytic}
\sigma = \left[\frac{\epsilon_{\rm turb} \dot{E}_{\rm mech} \ell_{\rm outer}}{\rho_c \pi R^2 h}\right]^{1/3} = \left[\frac{\epsilon_{\rm turb} \dot{E}_{\rm mech}}{n_c \mu m_H (\pi R^2)} \left(\frac{\ell_{\rm outer}}{h}\right)\right]^{1/3}.
\end{equation}
Using the energy injection rate in eqn.~\ref{eqn:EDot} and superwind density in eqn.~\ref{eqn:nC}, I find a turbulent velocity dispersion of
\begin{align}
\label{eqn:sigmaTurb}
%Exact value: 1016.17 km/s
\nonumber \sigma & = 1020\ \kms\   \left(\frac{\ell_{\rm outer}}{h}\right)^{1/3}\\
                 & \times \left[\epsilon_{\rm turb} \zeta^{-1} f_{\rm geom}^{2} \left(\frac{\epsilon_{\rm therm}}{0.75}\right)^{1/2} \left(\frac{\beta}{2}\right)^{-3/2}\right]^{1/3}
\end{align}
Since both $\dot{E}_{\rm mech} / (\pi R^2)$ and $n_c$ increase linearly with $\Sigma_{\rm SFR}$, the star-formation rate cancels out.  If the $\ell_{\rm outer} \approx h$, the turbulence is weakly supersonic (${\cal M} \approx 1.1$).

The turbulent energy density is
\begin{equation}
\label{eqn:Uturb}
U_{\rm turb} = \frac{\rho_c^{1/3}}{2} \left[\frac{\epsilon_{\rm turb} \dot{E}_{\rm mech}}{\pi R^2} \left(\frac{\ell_{\rm outer}}{h}\right)\right]^{2/3}
\end{equation}
Note that since $\rho \propto \Sigma_{\rm SFR}$, we have $U_{\rm turb} \propto \Sigma_{\rm SFR}$.  The energy dissipated in the turbulence goes into heating the superwind.  Indeed, turbulent dissipation may be \emph{the} way that supernova mechanical energy is thermalized.

Smaller outer scales lead to quicker dissipation and smaller turbulent energy densities.  If the outer scale is set by the time it takes a SNR overpressure to fade into the ISM (Section~\ref{sec:HotOuterScale}), then:
\begin{align}
%Exact value: 34.5949 pc
\nonumber \ell_1 & = 35\ \pc \left(\frac{\Sigma_{\rm SFR}}{\Msun\ \yr^{-1}\ \kpc^{-2}}\right)^{-1/3}\\
                & \times \left[E_{51} f_{\rm geom}^{2} \zeta^{-1} \left(\frac{\beta}{2}\right)^{-1/2} \left(\frac{\epsilon_{\rm therm}}{0.75}\right)^{-1/2} \right]^{1/3}
\end{align}
The outer scale decreases from several tens of parsecs in the Galactic Center to only a few parsecs in extreme starbursts like Arp 220 (Table~\ref{table:BHotPredictions}).  If the outer scale is $\ell_1$, then the turbulent velocity dispersion is only
\begin{align}
%Exact value: 898.763 km/s
\nonumber \sigma_1 & = 900\ \kms\  \left(\frac{\Sigma_{\rm SFR}}{\Msun\ \yr^{-1}\ \kpc^{-2}}\right)^{-1/9}\\
                   & \times \left[h_{50}^{-3} E_{51} \epsilon_{\rm turb}^{3} f_{\rm geom}^{8} \zeta^{-4} \left(\frac{\epsilon_{\rm therm}}{0.75}\right) \left(\frac{\beta}{2}\right)^{-5} \right]^{1/9}.
\end{align}

If $\ell_{\rm outer}$ is set by the scale where SNRs merge with the ISM velocity field,
\begin{align}
%Exact value: 41.3018 pc
\nonumber \ell_2     & = 41\ \pc\ \left(\frac{\Sigma_{\rm SFR}}{\Msun\ \yr^{-1}\ \kpc^{-2}}\right)^{-3/11} \\
                     & \times \left[h_{50}^{2} E_{51}^{3} f_{\rm geom}^{2} \epsilon_{\rm turb}^{-2} \zeta^{-1}\ \left(\frac{\beta}{2}\right)^{-3/2} \left(\frac{\epsilon_{\rm therm}}{0.75}\right)^{1/2} \right]^{1/11}.
\end{align}
I then find
\begin{align}
%Exact value: 953.449 km/s
\nonumber \sigma_2 & = 950\ \kms\ \left(\frac{\Sigma_{\rm SFR}}{\Msun\ \yr^{-1}\ \kpc^{-2}}\right)^{-1/11}\\
& \times \left[h_{50}^{-3} E_{51} \epsilon_{\rm turb}^{3} f_{\rm geom}^{8} \zeta^{-4} \left(\frac{\epsilon_{\rm therm}}{0.75}\right)^{2} \left(\frac{\beta}{2}\right)^{-6}\right]^{1/11}.
\end{align}
Note that these turbulent velocities are slightly subsonic, and in fact decrease to only a few hundred kilometers per second for Arp 220 (Table~\ref{table:BHotPredictions}).  

\begin{deluxetable*}{llcccc}
\tablecaption{Predicted Turbulence and Magnetic Fields in Starburst Superwinds}
\tablehead{ & \colhead{Units} & \colhead{GCCMZ} & \colhead{NGC 253} & \colhead{M82} & \colhead{Arp 220 Nuclei}}
\startdata
\cutinhead{Basic superwind properties}
$f_{\rm geom}$     & \nodata                       & 1.41   & 1.29  & 1.15 & 1.41\\
$n_e$              & $\cm^{-3}$                    & 0.015  & 0.33  & 0.35 & 21\\
$P_{\rm wind} / k_B$ & $\Kelv\ \cm^{-3}$           & $1.1 \times 10^6$ & $2.4 \times 10^7$ & $2.5 \times 10^7$  & $1.5 \times 10^9$\\
$\nu_{ee}^{-1}$    & $\kyr$                        & 1.2    & 0.052 & 0.050 & 0.00083\\
$\nu_{ii}^{-1}$    & $\kyr$                        & 59     & 2.6   & 2.5  & 0.041\\
$\nu_{ie}^{-1}$    & $\kyr$                        & 6200   & 270   & 260  & 4.3\\
$\lambda_{ee}$     & $\pc$                         & 29     & 1.3   & 1.0  & 0.020\\
$Q_{\rm SNR}^{\rm turb}$ & \nodata                 & 0.028  & 0.29  & 0.11 & 0.013\\
$Q_{\rm SNR}^{\rm CC85}$ & \nodata                 & 0.86   & 0.31  & 0.23 & 0.060\\
$Q_{\rm SNR}^{\rm hydro}$ & \nodata                & $5.9 \times 10^{-5}$~\tablenotemark{a} & 0.097 & 0.020 & 0.00012\\
$Q_{\rm SNR}^{\rm hot}$ & \nodata                  & 1.2    & 0.34  & 0.27 & 0.10\\
\cutinhead{$\ell_{\rm outer} = h$}
$\sigma$           & $\kms$                        & 1300  & 1200  & 1100  & 1300\\
$U_{\rm turb} / k_B$ & $\Kelv\ \cm^{-3}$           & $1.7 \times 10^6$ & $3.4 \times 10^7$ & $3.0 \times 10^7$ & $2.4 \times 10^9$\\
$B_{\rm turb}$     & $\muGauss$   & 76    & 340   & 330   & 2900\\
$\ell_{\rm in}$    & $\km$                         & 1800  & 370   & 360   & 46\\
\cutinhead{$\ell_{\rm outer} = \ell_1$}
$\ell_{\rm outer}$ & pc                            & 33    & 12     & 12     & 3.0\\
$U_{\rm turb} / k_B$ & $\Kelv\ \cm^{-3}$           & $1.3 \times 10^6$ & $1.3 \times 10^7$ & $1.2 \times 10^7$ & $0.36 \times 10^9$\\
$\sigma$           & $\kms$                        & 1100  & 740    & 690    & 500\\
$B_{\rm turb}$     & $\muGauss$                    & 68    & 210    & 200    & 1100\\
$\ell_{\rm in}$    & $\km$                         & 1800  & 460    & 480    & 86\\
\cutinhead{$\ell_{\rm outer} = \ell_2$}
$\ell_{\rm outer}$ & pc                            & 35    & 16     & 16     & 4.9\\
$\sigma$           & $\kms$                        & 1100  & 820    & 770    & 590\\
$U_{\rm turb} / k_B$ & $\Kelv\ \cm^{-3}$           & $1.3 \times 10^6$ & $1.6 \times 10^7$ & $1.4 \times 10^7$ & $0.51 \times 10^9$\\
$B_{\rm turb}$     & $\muGauss$                    & 68    & 230    & 220    & 1300\\
$\ell_{\rm in}$    & $\km$                         & 1800  & 420    & 440    & 73
\enddata
\tablenotetext{}{I assume that the CC85 solution applies with $\beta = 2$, $\zeta = 1$, and $\epsilon_{\rm therm} = 0.75$; I also assume $\epsilon_{\rm turb}$ and $\epsilon_B$ are 1.}
\tablenotetext{a}{This value is for $f_{\rm gas} = 0.036$, as in Table~\ref{table:Pressures}.  If $f_{\rm gas} \approx 1$, then $Q_{\rm SNR}^{\rm hydro} \approx 0.0045$.}
\label{table:BHotPredictions}
\end{deluxetable*}

It is interesting that the predicted outer scales in the GCCMZ are of order the same length as the mysterious radio filaments, whose origin remain unknown. Turbulent may naturally concentrate magnetic fields into ribbons (see section~\ref{sec:Dynamoes}), perhaps providing an explanation for the filaments' existence.  This was also the conclusion of \citet{Boldyrev06}, who first suggested the filaments are turbulent structures; although they estimated $\ell_{\rm outer}$ as the radius of a SNR when it first runs into another SNR's shell.

\subsection{The Magnetic Field Strength}
\label{sec:BHot}
If a fluctuation dynamo operates in the starburst wind (Section~\ref{sec:Dynamoes}), then $B = \sqrt{8 \pi \epsilon_B U_{\rm turb}}$:
\begin{equation}
B_{\rm turb} = \sqrt{4 \pi \epsilon_B} \rho_c^{1/6} \left[\frac{\epsilon_{\rm turb} \dot{E}}{\pi R^2} \left(\frac{\ell_{\rm outer}}{h}\right) \right]^{1/3}.
\end{equation}

Substituting in the numerical values, I find
\begin{align}
\label{eqn:BAsSigmaSFR}
%Exact value: 57.3797 muGauss
\nonumber B_{\rm turb} & = 57\ \muGauss \left(\frac{\Sigma_{\rm SFR}}{\Msun\ \yr^{-1}\ \kpc^{-2}}\right)^{1/2} \left(\frac{\ell_{\rm outer}}{h}\right)^{1/3}\\
                       & \times \sqrt{\epsilon_B} \left[\epsilon_{\rm turb}^{2} \zeta f_{\rm geom}^{-2} \left(\frac{\epsilon_{\rm therm}}{0.75}\right)^{-1/2} \left(\frac{\beta}{2}\right)^{3/2} \right]^{1/6}.
\end{align}

A smaller outer scale leads to smaller turbulent energy densities, and thus smaller magnetic fields.  The magnetic field strength for an outer scale of $\ell_1$ is
\begin{align}
%Exact value: 50.7503 muGauss
\nonumber B_1 & = 51\ \muGauss\ \left(\frac{\Sigma_{\rm SFR}}{\Msun\ \yr^{-1}\ \kpc^{-2}}\right)^{7/18}\\
                         & \times \sqrt{\epsilon_B} \left[h_{50}^{-3} E_{51} \epsilon_{\rm turb}^{3} f_{\rm geom}^{-1} \zeta^{1/2} \left(\frac{\epsilon_{\rm therm}}{0.75}\right)^{-5/4} \left(\frac{\beta}{2}\right)^{7/4} \right]^{1/9}
\end{align}
and for an outer scale of $\ell_2$ is 
\begin{align}
%Exact value: 53.8383 muGauss
\nonumber B_2 & = 54\ \muGauss\ \left(\frac{\Sigma_{\rm SFR}}{\Msun\ \yr^{-1}\ \kpc^{-2}}\right)^{9/22}\\
                                & \times \sqrt{\epsilon_B} \left[h_{50}^{-3} E_{51} \epsilon_{\rm turb}^{3} \zeta^{3/2} f_{\rm geom}^{-3} \left(\frac{\epsilon_{\rm therm}}{0.75}\right)^{-3/4} \left(\frac{\beta}{2}\right)^{9/4} \right]^{1/11}.
\end{align}

\begin{figure}
\centerline{\includegraphics[width=8cm]{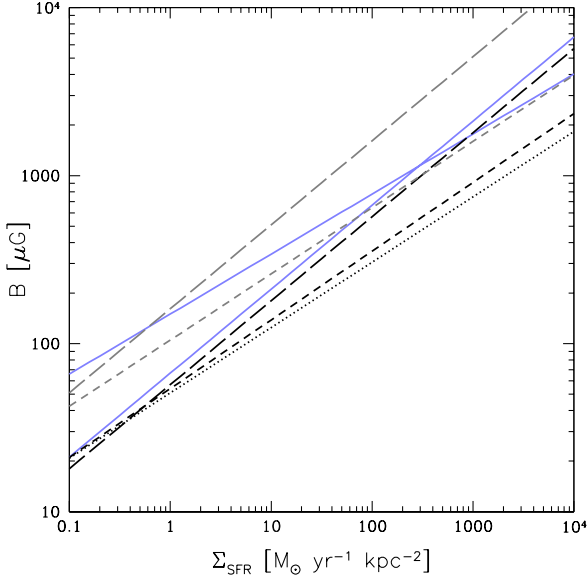}}
\figcaption{The estimated magnetic field strength in starburst superwinds (black) and cold/warm gas (grey).  The different line styles represent different outer scales: $\ell_0$ (long-dashed), $\ell_1$ (short-dashed), $\ell_2$ (dotted), and $\ell_3$ (short-dashed).  For comparison, the FIR-radio correlation implies $B$ strengths given by the blue solid lines \citep{Lacki10-FRC1}.  Note these lie within the range predicted by supernova-driven turbulence.\label{fig:B}}
\end{figure}

\subsubsection{Comparison with estimated magnetic field strengths}
\label{sec:ModelBComparison}
Using one-zone models of the multiwavelength (particularly radio and gamma-ray) emission of some nearby starburst regions, many papers have constrained the magnetic field strength of these individual starbursts under certain assumptions.  Examples of starbursts modelled this way include the Galactic Center Central Molecular Zone \citep{Crocker11-Wild,Crocker12,Lacki13-XRay}, NGC 253 \citep{Paglione96,Domingo05,Rephaeli10,Paglione12}, M82 \citep{Persic08,deCeaDelPozo09-M82,Paglione12,YoastHull13}, and the starburst nuclei of Arp 220 \citep{Torres04,Lacki13-XRay}.  These models often use a gas density that is equal to the average gas density and assume that CRs are accelerated with a similar efficiency as in the Milky Way.  Upper limits on the amount of Inverse Compton and bremsstrahlung emission constrain the density of synchrotron-emitting CR $e^{\pm}$, and therefore set a lower limit on $B$ \citep[e.g.,][]{Condon91,Thompson06,Crocker10}.  If the density is known, the amount of synchrotron emission sets a upper limit on $B$; if it is too high, too much power is radiated through the synchrotron mechanism, whereas with a lower $B$ most of the power goes into bremsstrahlung, ionization, and Inverse Compton losses.  But beware: it is possible that CRs experience a non-average density in the inhomogeneous ISM, leading to different allowed $B$ values \citep{Paglione12}.

It is worthwhile to compare my calculations of $B_{\rm turb}$ (Table~\ref{table:BHotPredictions}) to these model-constrained magnetic field strengths (Table~\ref{table:BasicProperties}).  The calculated magnetic field strengths are within the ranges found by modeling.  Thus, supernova-driven turbulence can plausibly generate pervasive magnetic fields in the starburst region that are responsible for the synchrotron radio emission.

\section{Turbulence and magnetic fields in cold and warm gas}
\label{sec:ColdTurbulence}
I showed in Section~\ref{sec:VolumeFillingPhase} that while the hot superwind should fill much of M82's volume, the high pressures in ULIRGs confines the hot gas too much for it to fill those starbursts.  Instead, most of the volume is in turbulent molecular gas.  Furthermore, even in hot starbursts, typical CRs probably enter the molecular medium at some point in their lives (see the discussion in section~\ref{sec:HotWindMotivation}).  

But what are the magnetic fields in the molecular medium?  If the molecular gas has some free charges, the turbulent dynamo pushes the magnetic energy density close to equipartition with the turbulence \citep{Stone98}, is implied by observations of Galactic molecular clouds that are near equipartition \citep{Crutcher99}.  A turbulent dynamo naturally explains this fact \citep{Balsara04}.  Typical turbulent velocity dispersions in starburst molecular gas are $25 - 150\ \kms$, with the lower end representative of motions in the GCCMZ \citep{Shetty12} and the high end found in $z \approx 2$ galaxies \citep{Green10}.  This implies a mean turbulent energy density of 
\begin{equation}
\frac{P}{k_B} = \frac{\mean{\rho} \sigma^2}{2 k_B} = 6.1 \times 10^8\ \Kelv\ \cm^{-3} \left(\frac{\mean{n_H}}{1000\ \cm^{-3}}\right) \left(\frac{\sigma}{100\ \kms}\right)^2, 
\end{equation}
and equipartition magnetic field strength:
\begin{equation}
%Exact value: 1.45038
B_{\rm turb} = 1.45\ \mGauss\ \epsilon_B \left(\frac{\mean{n_H}}{1000\ \cm^{-3}}\right)^{1/2} \left(\frac{\sigma}{100\ \kms}\right).
\end{equation}
Essentially, this is the hypothesis of \citet{Groves03} to explain the existence of the far-infrared--radio correlation, except that $\sigma$ is $\sim 100\ \kms$ instead of $10\ \kms$.  

As in Section~\ref{sec:HotTurbulence}, I assume that supernovae drive turbulence in the cold and warm gas (see section~\ref{sec:TurbulenceSources}).  What complicates the calculation is that molecular gas is neither homogeneous nor necessarily volume-filling.  There are at least three relevant densities for the problem.  The first is the density averaged over the volume of the entire starburst, $\mean{\rho}_{\rm SB} = M(H_2) / V_{\rm SB}$.  It is the density that is most closely related to the mass, which sets the \emph{total} kinetic energy of the turbulence and is what is actually measured.  The second relevant density is the density averaged only over the volume of the molecular clouds, $\mean{\rho}_{\rm MC} = M(H_2) / (V_{\rm SB} f_{\rm fill})$, where $f_{\rm fill}$ is the fraction of the starburst volume occupied by molecular clouds.  That is what sets the typical turbulent \emph{energy density} in the molecular clouds.  Finally, there is $\rho_{\rm ext}$, which is the density \emph{exterior} to a SNR expanding into a molecular cloud.  It sets the maximum size of the SNR, and presumably the outer scale of turbulence.  Roughly, $\rho_{\rm ext}$ is the median density in the volume of molecular clouds, and is somewhere in the range $\mean{\rho}_{\rm MC} / \sqrt{1 + b^2 {\cal M}^2}$ to $\mean{\rho}_{\rm MC}$ for isothermal turbulence \citep{Hopkins13-rhoDist}.

As in the superwind phase, the turbulent kinetic energy is roughly the power injected multiplied by the eddy-crossing time.  For the molecular gas, we have $M(H_2) \sigma^2/2 = \dot{E}_{\rm mol} \ell_{\rm outer} / \sigma$, where $\dot{E}_{\rm mol}$ is the power input into the molecular phase only.  We can solve for $\sigma$, getting
\begin{equation}
\sigma = \left[\frac{\dot{E}_{\rm mech} \epsilon_{\rm turb} \epsilon_{\rm mol}}{A \mean{\rho}_{\rm SB}} \left(\frac{\ell_{\rm outer}}{h}\right)\right]^{1/3}.
\end{equation}
The factor $\epsilon_{\rm mol} \equiv \dot{E}_{\rm mol} / \dot{E}_{\rm mech}$ represents the fraction of mechanical luminosity that is available for the cold gas because the sources are within that phase.  For example, if the molecular phase occupies a small fraction of the starburst volume, few of the supernovae go off in it; most do not stir up the molecular gas.  If sources are located randomly in the starburst, $\epsilon_{\rm mol} \approx f_{\rm fill}$; if they are instead clustered inside molecular clouds, $\epsilon_{\rm mol} \approx 1$.

If the outer scale of turbulence is some constant ratio of the starburst scale height, then the characteristic turbulent speed is
\begin{align}
%Exact value: 25.1481
\nonumber \sigma & = 25\ \kms\ \left(\frac{\Sigma_{\rm SFR}}{\Msun\ \yr^{-1}\ \kpc^{-2}}\right)^{1/3} \left(\frac{\mean{n}_{\rm SB}}{1000\ \cm^{-3}}\right)^{-1/3}\\
& \times \left[\epsilon_{\rm turb} \epsilon_{\rm mol} \left(\frac{\ell_{\rm outer}}{h}\right)\right]^{1/3}.
\end{align}
The equipartition magnetic field strength is 
\begin{align}
%Exact value: 364.744
\nonumber B_{\rm turb} & = 360\ \muGauss\ \left(\frac{\Sigma_{\rm SFR}}{\Msun\ \yr^{-1}\ \kpc^{-2}}\right)^{1/3} \left(\frac{\mean{n}_{\rm SB}}{1000\ \cm^{-3}}\right)^{1/6} \\
& \times \sqrt{\frac{\epsilon_B}{f_{\rm fill}}}\left[\epsilon_{\rm turb} \epsilon_{\rm mol} \left(\frac{\ell_{\rm outer}}{h}\right)\right]^{1/3}.
\end{align}
See Table~\ref{table:BColdPredictions} for these magnetic field strengths for the prototypical starbursts.

However, as with the superwind phase, turbulence can be much slower if $\ell_{\rm outer}$ is small.  The outer scales can easily be tiny in starbursts, where the high pressures confine supernova remnants to small radii.  Supernova remnants are highly radiative in dense molecular environments, unlike in the hot superwinds where the size evolution of the remnants could be understood by energy conservation.  At their maximum radius, the supernova remnants will then be in the momentum-conserving phase \citep{Chevalier74,McKee77}.  To do a more realistic calculation of the properties of the turbulence, I therefore set $\ell_{\rm outer} \approx \ell_3 \equiv R_{\rm max}$.  In addition, the surrounding ISM pressure is mostly from the turbulence itself, so I set $P_{\rm ext} = \mean{\rho}_{\rm MC} \sigma^2 / 2$.  Under these conditions, the outer scale is 
\begin{equation}
%Exact value 4.77819
\ell_3 = 4.8\ \pc\ \Psi_{\ell}\ \left(\frac{\Sigma_{\rm SFR}}{\Msun\ \yr^{-1}\ \kpc^{-2}}\right)^{-2/17} \left(\frac{\mean{n}_{\rm SB}}{1000\ \cm^{-3}}\right)^{-1/5},
\end{equation}
where the nuisance parameter $\Psi_{\ell}$ contains the information on the efficiency of the turbulence driving, geometry, filling factor, and density contrasts:
\begin{equation}
\Psi_{\ell} = \frac{E_{51}^{24/85} f_{\rm fill}^{27/85} h_{50}^{2/17}}{\epsilon_{\rm turb}^{2/17} \epsilon_{\rm mol}^{2/17} \Delta^{12/85}}
\end{equation}
with $\Delta = \rho_{\rm ext} / \mean{\rho}_{\rm MC}$.  The length scale is about an order of magnitude smaller than the scale height of the starburst, implying smaller turbulent speeds and magnetic fields by a factor of $\sim 2$.  The turbulent speed is
\begin{equation}
%Exact value: 11.4975
\sigma = 11\ \kms\ \Psi_{\sigma}\ \left(\frac{\Sigma_{\rm SFR}}{\Msun\ \yr^{-1}\ \kpc^{-2}}\right)^{5/17} \left(\frac{\mean{n}_{\rm SB}}{1000\ \cm^{-3}}\right)^{-2/5},
\end{equation}
with
\begin{equation}
\Psi_{\sigma} =  \frac{E_{51}^{8/85} \epsilon_{\rm turb}^{5/17} \epsilon_{\rm mol}^{5/17} f_{\rm fill}^{9/85}}{h_{50}^{5/17} \Delta^{4/85}}.
\end{equation}
Finally, the equipartition magnetic field strength is 
\begin{equation}
%Exact value: 166.758
B_{\rm turb} = 170\ \muGauss\ \Psi_B\ \left(\frac{\Sigma_{\rm SFR}}{\Msun\ \yr^{-1}\ \kpc^{-2}}\right)^{5/17} \left(\frac{\mean{n}_{\rm SB}}{1000\ \cm^{-3}}\right)^{1/10},
\end{equation}
with 
\begin{equation}
\Psi_B = \frac{E_{51}^{8/85} \epsilon_B^{1/2} \epsilon_{\rm turb}^{5/17} \epsilon_{\rm mol}^{5/17}}{f_{\rm fill}^{67/170} h_{50}^{5/17} \Delta^{4/85}}.
\end{equation}

The big question is, what are the values of $\Psi_{\ell}$, $\Psi_{\sigma}$, and $\Psi_B$, and the factors that comprise them?  It is generally thought that supernova remnants expanding in dense molecular media lose energy by radiation; according to the simulations of \citet{Thornton98}, only $\sim 10\%$ of the supernova mechanical energy is available for turbulence.  While $f_{\rm fill}$ is probably near 1 for cold starbursts, it may be much smaller in hot starbursts.  Presumably, $\epsilon_{\rm mol}$ is equal to $f_{\rm fill}$, but this is not necessarily the case if supernovae only go off in molecular clouds (although there is evidence against this for the weaker starbursts; Section~\ref{sec:SNREvolution}).  Finally, in the Mach 10 -- 100 turbulence of starburst molecular gas, $\Delta \approx 0.01$ -- $1$ \citep[e.g.,][]{Krumholz05,Hopkins13-rhoDist}.  

To answer these issues, I consider some ``natural'' values of $\Psi_{\ell}$, $\Psi_{\sigma}$, and $\Psi_B$.  In all cases, I set $\epsilon_{\rm turb} = 0.1$.  For the filling factor $f_{\rm fill}$, I use the somewhat arbitrary value of $25\%$ for the GCCMZ, NGC 253, and M82, but 100\% for Arp 220's nuclei.  Likewise, the $\epsilon_{\rm mol}$ geometrical factor is set equal to $f_{\rm fill}$.  Finally, I set $\Delta = 0.01$ for Arp 220's nuclei and $\Delta = 0.03$ for the other starbursts.  The resultant ``natural'' values for $\Psi_L$, $\Psi_{\sigma}$, and $\Psi_B$ are listed in Table~\ref{table:BColdPredictions}, as well as the ``corrected'' outer scales, velocity dispersions, and magnetic field strengths.  With these values, the turbulent driving is much weaker, and the available energy density must push more gas per unit volume (the smaller filling factors mean higher densities in those regions that \emph{do} have molecular gas).  The turbulent velocity dispersions are now smaller by a factor of $\sim 2$, or $\sim 4$ for Arp 220.  Likewise, the turbulent pressures are a factor 4--10 times smaller than those inferred from observations.  The resulting magnetic field strengths are lower, more in line with those inferred by modeling, and roughly equal to those in the turbulent superwind.

\begin{deluxetable*}{llcccc}
\tablecaption{Predicted Turbulence and Magnetic Fields in Starburst Molecular Gas}
\tablehead{ & \colhead{Units} & \colhead{GCCMZ} & \colhead{NGC 253} & \colhead{M82} & \colhead{Arp 220 Nuclei}}
\startdata
$\sigma_{\rm obs}$ & $\kms$                        & 25                & $\sim 50$         & $\sim 50$         & 80\\

\cutinhead{$U_{\rm turb} = \mean{\rho}_{\rm SB} \sigma_{\rm obs}^2 / 2$ with $\ell_{\rm outer} = 5\ \pc$ and $f_{\rm fill} = 1$}
$U_{\rm turb}$     & $\Kelv\ \cm^{-3}$             & $4.5 \times 10^6$ & $2.6 \times 10^7$ & $4.5 \times 10^7$ & $5.0 \times 10^9$\\
$B_{\rm turb}$     & $\muGauss$   & 130               & 300               & 400               & 4200\\

\cutinhead{Supernova driven turbulence with $\ell_{\rm outer} = h$ and $\Psi = 1$}
$\sigma$           & $\kms$                        & 45                & 160               & 120              & 160\\
$U_{\rm turb}$     & $\Kelv\ \cm^{-3}$             & $48 \times 10^6$  & $26 \times 10^7$  & $28 \times 10^7$ & $20 \times 10^9$\\
$B_{\rm turb}$     & $\muGauss$   		   & 410               & 950               & 980              & 8200\\

\cutinhead{$\ell_{\rm outer} = \ell_3$ with $\Psi$ = 1}
$\ell_{\rm outer}$ & pc                            & 5.2               & 4.4               & 4.0               & 1.1\\
$\sigma$           & $\kms$                        & 21                & 70                & 53                & 44\\
$U_{\rm turb}$     & $\Kelv\ \cm^{-3}$             & $1.1 \times 10^7$ & $5.1 \times 10^7$ & $5.1 \times 10^7$ & $1.6 \times 10^9$\\
$B_{\rm turb}$     & $\muGauss$  		   & 190               & 420               & 420               & 2300\\

\cutinhead{$\ell_{\rm outer} = \ell_3$ with ``natural'' $\Psi$ Values}
Natural $\Psi_{\ell}$ &                            & 1.6               & 1.6               & 1.6               & 2.5\\
Natural $\Psi_{\sigma}$ &                          & 0.34              & 0.34              & 0.34              & 0.63\\
Natural $\Psi_B$   &                               & 0.69              & 0.69              & 0.69              & 0.63\\
$\ell_{\rm outer}$ & pc                            & 8.6               & 7.1               & 6.5               & 2.8\\
$\sigma$           & $\kms$                        & 7.3               & 24                & 18                & 28\\
$U_{\rm turb}$     & $\Kelv\ \cm^{-3}$             & $5.0 \times 10^6$ & $2.4 \times 10^7$ & $2.4 \times 10^7$ & $0.61 \times 10^9$\\
$B_{\rm turb}$     & $\muGauss$   		   & 130               & 290               & 290               & 1500
\enddata
\label{table:BColdPredictions}
\tablenotetext{}{I assume $\epsilon_B = 1$ when calculating $B_{\rm turb}$.}
\end{deluxetable*}

\emph{Why aren't the observed $\sigma$ this low?} -- If the observed turbulent speeds are correct \citep[c.f.,][]{Ostriker11}, there must be more turbulent power than provided by highly radiative supernova remnants.  First, there could be other sources of turbulence within molecular clouds.  As discussed in \S~\ref{sec:TurbulenceSources}, protostellar outflows, radiation pressure, and starburst disk instabilities may all help drive turbulence in cold starbursts.

In addition, many of the young stars in the nuclei are likely in clusters \citep{Kruijssen12}.  By aggregating the energy input into a few, more powerful sources, the outer scale of the clusters' superbubbles will increase as should the turbulent energy density.  

On a broader level, turbulent power does not necessarily have to be directly injected into the molecular clouds.  Turbulent power on large scales can actually cascade down into molecular clouds on smaller scales.  Supernovae going off in the superwind near the molecular cloud could effectively stir the medium on 10 pc scales with minimal radiative losses.  The mechanical pushing and pulling and stirring of the superwind then drives turbulence on smaller scales in molecular clouds stuck within the eddies, bypassing the need for small-scale stirring by individual supernova remnants.  The driving of molecular cloud turbulence by external processes on larger scales likely happens in the Milky Way \citep{Brunt03,Brunt09}, as demonstrated by Larsons' relation where the fastest turbulent occurs on the largest scales \citep{Larson81,Miesch94,Heyer04}.  Indeed, Larsons' relation holds in the GCCMZ starburst too \citep{Shetty12}.

\emph{Why aren't the calculated $\sigma$ and $B$ much lower?} -- Yet what is just as striking is that the order of magnitude variation in $f_{\rm fill}$ and $\epsilon_{\rm turb}$ has a relatively small effect on $\ell_{\rm outer}$, $\sigma$, and $B$.  To take this even further, let $\epsilon_{\rm turb}$ vary over the entire range of 0.01 to 1.0, and independently vary $f_{\rm fill} = \epsilon_{\rm mol}$ over 0.1 to 1, and let $\Delta$ range from 0.01 to 1 too.  Even then, $\Phi_{\ell}$ varies at most by a factor of  5 (from 0.72 to 3.29), $\Phi_{\sigma}$ varies by at most a factor of 12 (from 0.10 to 1.24), and $\Phi_B$ varies at most a factor of 6 (from 0.26 to 1.6).  The turbulent energy density is only uncertain by 1.5 orders of magnitude.  

The weak dependencies of the nuisance parameters on these basic factors results from the physical factors having several counteracting effects.  For example, suppose we lowered $\epsilon_{\rm turb}$.  The main effect is to lessen the amount of turbulence, which overall pushes $\sigma$ down.  But with slower turbulent flows, the eddy-crossing time grows.  And, since the turbulent pressure is lower, SNRs are not confined to such small radii, the outer scale becomes longer, and the eddy-crossing time grows further.  Hence, $\sigma$ decreases, but quite slowly, as $\epsilon_{\rm turb}$ decreases.  Similar considerations apply for the filling factor.  

\emph{How does $B$ compare between the phases?} -- The great uncertainty in relevant quantities make it impossible to exactly predict $\ell_{\rm outer}$, $B$, or $\sigma$ from first principles.  But it is clear that the magnetic fields in the molecular material are at least as strong as those in the hot superwind.  If we push the turbulence to maximal values, with $\ell_{\rm outer} = h$ and $\epsilon_{\rm turb} = 1$, then $B$ may be as high as 3 to 7 times higher in the cold molecular gas than in the hot superwind.  However, those high levels of turbulence would manifest in turbulent speeds ($> 100\ \kms$; Table~\ref{table:BColdPredictions}) that are ruled out by observations.  In a more likely scenario with $\ell_{\rm outer} \approx R_{\rm max}$, the magnetic field strengths are roughly equal or somewhat greater than those in the hot superwind, but generally within a factor of $\sim 2$.  

As a result, the value of $B$ appears to be roughly constant as we move from one phase to the next in the starburst ISM.  This is good news for models of synchrotron emission from starburst galaxies.  Although the magnetic fields are probably stronger in molecular clouds, there is not orders of magnitude of variation between phases.  Probably the more important factor is the intermittent nature of turbulence, which in and of itself causes fluctuations of $B$ from one place to another.

\section{Equipartition in starbursts revisited}
\label{sec:Equipartition}
On some level, pressure balance between the phases must occur; otherwise, the overpressured phases will crush the others until they are in pressure balance.  In this sense, it's not a surprise that rough equipartition holds between the phases in starbursts (Table~\ref{table:Pressures}).  Of course, I also assume equipartition between turbulence and magnetic fields from the start. 

But why are most of the ``natural'' values of the various energy densities in starbursts so close to each other, \emph{even when considered independently of each other}?  The CC85 wind solution does not include any squeezing by an overpressured molecular medium; it is as if the molecular medium is not there at all -- yet the theoretical pressure is in rough equipartition with the molecular turbulence.  Likewise, my calculations of the turbulent molecular pressure did not depend on the pressure of the superwind pushing it into equipartition.  Furthermore, the pressure balance argument only applies to the \emph{total} pressure.  There is no a priori reason the thermal and turbulent pressures have to be anywhere near each other, and they are not in the molecular gas.  But they are in the superwind phase, and they are within an order of magnitude of each other in the H II regions.  Why is radiation in equipartition with the gas even in weak starbursts like the GCCMZ that are transparent to infrared photons?  And why is the CR energy density an order of magnitude below the turbulent pressure -- except for the ULIRGs, where the ratio plummets?  It's as if the phases \emph{started out} in pressure balance even before any mechanical coupling between them.

I explore these questions in this section.  Ultimately, the energy density is a combination of a power density and a timescale.  In several cases (turbulence, magnetic fields, CRs, superwind thermal pressure), the power source is the same -- supernovae mechanical energy, and the timescales -- roughly the superwind sound crossing time -- are near each other, explaining equipartition.  In other cases (radiation and H II thermal pressure), there is a fortuitous canceling between a large power and a short timescale.  

\subsection{Why are the turbulent pressures between the phases nearly equal?}
From equation~\ref{eqn:Uturb}, the turbulent energy density can be written as $U_{\rm turb} = [2 \rho \dot{\varepsilon}_{\rm turb}^2 \ell_{\rm outer}^2]^{1/3}$.  Therefore the ratio of the turbulent energy density in two phases is equal to
\begin{equation}
\frac{U_a}{U_b} = \left(\frac{\dot{\varepsilon}_a}{\dot{\varepsilon}_b}\right)^{2/3} \left(\frac{\rho_a}{\rho_b}\right)^{1/3}  \left(\frac{\ell_a}{\ell_b}\right)^{2/3}.
\end{equation}

It is clear that the turbulent energy density depends weakly on density, but relatively strongly on energy input.  Thus, even though the density in starburst molecular clouds is 100 to $10^4$ (depending on filling factor) times higher than that in the superwind, that alone would only boost the turbulent energy density by an order of magnitude.  Put another way, in the molecular clouds, the supernovae have to push more mass.  On the one hand, the turbulence is slower but on the other the eddy-crossing time is longer, and these effects partly cancel out.

But in molecular clouds and H II regions, the supernova remnants are highly radiative, injecting only 10\% of the input mechanical energy \citep{Thornton98}, reducing the molecular turbulent energy density by a factor $\sim 5$.  The outer scale lengths are fairly near each other, about twice as long in the superwind as in the molecular ISM and H II regions, if we consider the maximum radius of a SNR before fading into the ISM (see Tables~\ref{table:BHotPredictions} and~\ref{table:BColdPredictions}).  Therefore, we have $U_{\rm cold}/U_{\rm hot} \approx 10 \times (1/5) \times (1/2) \approx 1$.  

\subsection{Why are the thermal and turbulent pressures nearly equal in the superwind?}
The superwind is heated by supernovae in its starburst region.  The plasma in the superwind is hot enough that it is not affected by the starburst's gravity, and therefore explodes out of the starburst in roughly a dynamical (sound-crossing) time.  Thus, the superwind has a thermostat that regulates its internal thermal pressure, one directly related to the flow properties.  If the wind is too cool, it will stay in the starburst longer and heat up more; if it heats too much, it will expand quickly and prevent further heating.  

Note the similarity with the turbulent energy density: the thermal energy density is a power density times a flow-crossing time.  The only substantial difference is that here we want the time it takes the wind to cross the starburst rather than just the outer scale.  We can write
\begin{equation}
U_{\rm therm} = \frac{\dot{\varepsilon} h \psi}{c_S} = \frac{3}{2\gamma} \rho c_S^2,
\end{equation}
The $\psi$ factor corrects the thermal energy density to the CC85 solution and is equal to $3.08 \epsilon_{\rm therm} \mu \zeta / f_{\rm geom}^2 \approx 1$.  The solution for the sound speed is $c_S = [2\gamma \dot{\varepsilon}_{\rm mech} h \psi / (3\rho)]^{1/3}$.  Then we have
\begin{equation}
U_{\rm therm} = \left(\frac{3}{2\gamma}\right)^{1/3} (\dot{\varepsilon}_{\rm mech} h \psi)^{2/3} \rho^{1/3}.
\end{equation}
Once we compare to the turbulent energy density, it is clear that the thermal and turbulent energy densities must be in near equipartition in superwinds.  Their ratio is
\begin{equation}
\frac{U_{\rm therm}}{U_{\rm turb}} = \left(\frac{12}{\gamma}\right)^{1/3} \left(\frac{h}{\ell_{\rm outer}}\right)^{2/3} \left(\frac{\psi}{\epsilon_{\rm turb}}\right)^{2/3}.
\end{equation}
Since $\ell_{\rm outer}/h \approx 0.1 - 1$, we have $U_{\rm therm} / U_{\rm turb} \approx 0.3 - 2$.  

By contrast, the starburst's molecular medium has a much different thermostat mechanism than the superwind.  Molecular gas cools by radiating line emission and by collisions with dust grains.  Even in regions where it has the same power source of supernova-driven turbulence, the time it retains that heat is much smaller than the time it retains turbulent kinetic energy.  Therefore, the turbulence in the molecular gas is extremely supersonic.  But in the $10^7 - 10^8\ \Kelv$ superwind, radiative cooling is completely negligible so that the dynamical thermostat operates.\footnote{The superwind's ``thermostat'' is also a major difference with another home for extremely hot plasma, galaxy clusters, where the turbulent energy density is just a few percent of the thermal energy density \citep[e.g.,][]{Subramanian06}.  While a starburst's gravity is not strong enough to confine its superwind, a galaxy cluster's gravity does dominate the dynamics of its hot plasma.  Therefore, the turbulent energy density is a small fraction of the thermal energy of the virialized plasma.  Thermal energy from turbulent dissipation must accumulate in the plasma for over a Hubble time to start affecting the dynamics, a duration much longer than the Gyr eddy crossing time.}

\subsection{Why is the thermal pressure in H II regions within an order of magnitude of the turbulent pressure?}
The thermal pressures measured for H II regions within starburst regions (Table~\ref{table:Pressures}) tends to be a few times lower than the turbulent pressure I predict.  But why should this be true, when it is not true in the other dense phase, the molecular gas?  Neither the heating nor cooling of H II regions has anything to do with supernova-driven turbulence.  Put another way, why is the natural turbulent speed of a few tens $\kms$ for dense starburst gas just a bit higher than the sound speed of $\sim 15\ \kms$?

H II regions are heated by the photoelectric absorption of ionizing radiation of their OB stars.  We can estimate the heating rate per unit volume by dividing the ionizing luminosity by the Str\"omgren volume: $L_{\star} n_{\rm H\,II}^2 \alpha_B / Q_{\rm ion}^{\star}$, where $\alpha_B$ is the recombination coefficient, $L_{\star}$ is the ionizing power of the OB stars, and $Q_{\rm ion}^{\star}$ is the rate ionizing photons are emitted by those stars \citep[e.g.,][]{Draine11}.  Note the ratio $L_{\star} / Q_{\rm ion}^{\star}$ equals $\mean{h \nu_{\star}}$, the mean energy of an ionizing photon.  I find that the average volumetric heating rate in H II regions with a temperature $10^4\ \Kelv$ is
\begin{align}
\nonumber \dot{\varepsilon}_{\rm therm} & = 1.2 \times 10^{-17}\ \erg\ \cm^{-3}\ \sec^{-1} \left(\frac{n_{\rm H\,II}}{1000\ \cm^{-3}}\right)^2 \\
& \times \left(\frac{\mean{h \nu_{\star}}}{30\ \eV}\right)
\end{align}

The thermal energy within the H II regions is retained for one cooling time, $t_{\rm cool} = 3 n_{\rm H\,II} k T / \Lambda$, before being radiated away as line and free-free emission.  Given that the cooling coefficient has an approximate value $2 \times 10^{-24}\ \erg\ \cm^{-3}\ \sec^{-1} \times n_e n_H$ for Solar metallicity \citep{Osterbrock89}, the cooling time is
\begin{equation}
t_{\rm cool} = 33\ \yr\ \left(\frac{n_{\rm H\,II}}{1000\ \cm^{-3}}\right)^{-1}
\end{equation}
for a totally ionized plasma at $10^4\ \Kelv$.

By contrast, the turbulent volumetric power from supernova mechanical energy is much smaller:
\begin{align}
\nonumber \dot{\varepsilon}_{\rm turb} & = 1.1 \times 10^{-20} \erg\ \cm^{-3}\ \sec^{-1}\ \epsilon_{\rm turb} \left(\frac{\mean{n_H}_{\rm SB}}{1000\ \cm^{-3}}\right)\\
& \times \left(\frac{\tau_{\rm gas}}{20\ \Myr}\right)^{-1}.
\end{align}
At the same time, the flow crossing time is much longer than the cooling time:
\begin{equation}
t_{\rm eddy} = 200\ \kyr \left(\frac{\ell_{\rm outer}}{5\ \pc}\right) \left(\frac{\sigma}{25\ \kms}\right)^{-1}.
\end{equation}

We can now see why equipartition holds: although the heating rate is over ten thousand times the turbulent power, the heat is retained for only a ten-thousandth of the time.  The ratio of thermal to turbulent energy density in H II regions is
\begin{align}
\nonumber \frac{U_{\rm therm}}{U_{\rm turb}} & = \frac{\dot{\varepsilon}_{\rm therm}}{\dot{\varepsilon}_{\rm turb}} \frac{t_{\rm cool}}{t_{\rm eddy}}\\
\nonumber                                    & \approx 0.18 \epsilon_{\rm turb} \left(\frac{n_{\rm H\,II}}{\mean{n_H}_{\rm SB}}\right) \left(\frac{\tau_{\rm gas}}{20\ \Myr}\right) \left(\frac{\mean{h\nu_{\star}}}{30\ \eV}\right)\\
                                             & \times \left(\frac{\ell_{\rm outer}}{5\ \pc}\right)^{-1} \left(\frac{\sigma}{25\ \kms}\right).
\label{eqn:HIITurbThermRatio}
\end{align}
Thus, thermal H II region pressure is naturally $\sim 5$ times smaller in H II regions of average density, at least in nuclear starburst regions where $\tau_{\rm gas}$ and $\ell_{\rm outer}$ are small.

This coincidence may not apply to high redshift main sequence galaxies or other environments with long $\tau_{\rm gas}$.  On the one hand, $\tau_{\rm gas}$ is much longer in these regions, $\sim 0.5$ Gyr.  That means the galaxies have lower SFR for their gas masses, and hence, less turbulent power per unit mass.  Turbulence might naturally be weaker.  There might be a sign of this in Table~\ref{table:Pressures}: in the GCCMZ where $\tau_{\rm gas} \approx 400\ \Myr$, the H II region thermal energy densities are higher than the other energy densities (equation~\ref{eqn:HIITurbThermRatio} implies a ratio of 3.6).  If the thermal energy density is much greater than the surrounding ISM pressure, the H II regions will expand until they are in equilibrium: thus $n_H / \mean{n_H}_{\rm SB}$ would be lower.  On the other hand, in the lower gas densities of main sequence galaxies, the turbulent outer scale might be larger, increasing $t_{\rm eddy}$.

\subsection{Why is the radiation pressure near equipartition with turbulence?}
The near equipartition between radiation and turbulent energy densities in starburst regions is another coincidence.  The amount of power released as light by a starburst, $L_{\rm bol} \approx 2.15 \times 10^{43} \erg\ \sec^{-1} \times ({\rm SFR} / \Msun\ \yr^{-1})$ \citep{Kennicutt98}, is much greater than the mechanical power.  On the other hand, the light escapes very quickly, in a light crossing time if the starburst is not optically thick: $t_{\rm rad} \approx \tau_{\rm IR} h / c$, where the IR optical depth $\tau_{\rm IR}$ is typically a few or less \citep{Krumholz12}.  These two factors cancel each other out.  

From the timescale argument, we again have
\begin{align}
\nonumber \frac{U_{\rm turb}}{U_{\rm rad}} & \approx \frac{\dot{E}_{\rm mech} \epsilon_{\rm turb} \ell_{\rm outer} / \sigma}{L_{\rm bol} \tau_{\rm IR} h / c}\\
& \approx 3.5 \frac{\epsilon_{\rm turb}}{\tau_{\rm IR}} \frac{\ell_{\rm outer}}{h} \left(\frac{\sigma}{1000\ \kms}\right)^{-1}.
\end{align}
Thus, by coincidence, the turbulent pressure in hot superwinds nearly equals the radiation pressure.  In the densest cold starbursts, $\epsilon_{\rm turb}$ is only $\sim 0.1$, $\ell_{\rm outer} \approx 0.1 h$ and $\tau_{\rm IR} \ga 1$.  On the other hand, the turbulent speeds in the cold ISM are $\la 100\ \kms$, so the turbulence resides for longer.  In these regions, $U_{\rm rad}$ goes from being a few times smaller than $U_{\rm turb}$ to a few times larger than $U_{\rm turb}$.  

\subsection{Why is $U_{\rm CR}$ near $U_B$ only in weaker starbursts?}
\label{sec:CREquipartition}
Whether equipartition between CRs and magnetic fields holds in starburst regions is a matter of debate \citep[e.g.,][]{Thompson06}.  Generally, the equipartition assumption gives reasonable results for hot starbursts but too small $B$ estimates for cold starbursts.  For example, for M82 and NGC 253, the equipartition formula gives results of $\sim 190\ \muGauss$ \citep{Beck12}, even after being corrected for secondaries and strong losses \citep{Lacki13-Equip}.  This is in the range derived from detailed modeling and this paper's estimate for $B$ (see section~\ref{sec:BHot}), so equipartition appears plausible.  However, the magnetic field strengths derived for ULIRGs like Arp 220 are typically only $\sim 500\ \muGauss$ \citep{Thompson06,Lacki13-Equip}.  This is inconsistent with modeling (section~\ref{sec:ModelBComparison}) and my own estimate for $B$, which indicate magnetic field strengths of several mG.  

The CR energy density is
\begin{equation}
\label{eqn:UCRBasic}
U_{\rm CR} = \frac{\epsilon_{\rm CR} \dot{E}_{\rm mech} t_{\rm CR}}{V} 
\end{equation}
where $t_{\rm CR}$ is the residence time for CRs in the superwind and $\epsilon_{\rm CR} \approx 0.1$ is the CR acceleration efficiency.  The bulk of CR energy is in protons.  Protons in starbursts either escape through advection if they traverse low density ISM, or lose energy to pionic losses if they experience a high enough density.  The gamma-ray observations of M82 and NGC 253 indicate that $F_{\rm cal} \approx 20 - 40\% < 1$ in these galaxies.  Thus, it appears that advection sets the lifetimes of CRs in these starbursts.

We therefore have
\begin{equation}
\label{eqn:UCRAdvection}
U_{\rm CR}^{\rm adv} = \frac{\epsilon_{\rm CR} \dot{E}_{\rm mech}}{2 \pi R^2 v_{\rm wind}} 
\end{equation}
where $v_{\rm wind}$ is the average speed that CRs are transported out of the starburst.  Note that the outflow speed rapidly accelerates at the sonic point \citep{Chevalier85}; within the starburst, CRs flow outwards at a slower speed, $v_{\rm wind} \approx 300$ -- $600 \kms$, until they reach the starburst edge.  Comparing with $U_B = U_{\rm turb}$ in equation \ref{eqn:UTurbBasic}, I find a ratio of
\begin{equation}
\frac{U_{\rm CR}^{\rm adv}}{U_B} = \left(\frac{\epsilon_{\rm CR}}{\epsilon_{\rm turb}}\right) \left(\frac{\ell_{\rm outer}}{h}\right) \left(\frac{\sigma}{v_{\rm wind}}\right)
\end{equation}
Because only $\sim 10\%$ of the mechanical energy is converted into CRs, $\epsilon_{\rm CR} / \epsilon_{\rm turb} \approx 1/10$.  On the other hand, the turbulent outer scale in the hot wind may be a few times smaller than the scale height, and $v_{\rm wind}$ is probably a few times smaller than 1.  Overall, we then find that $U_{\rm CR}^{\rm adv}/U_B \sim 1$.  This arises not primarily because the CRs and magnetic field drive each other.  Instead, equipartition happens because both CRs and magnetic fields draw from the same power source -- supernovae -- and have similar characteristic times, the sound/turbulent crossing time.\footnote{But if diffusive reacceleration operates in superwinds, it plausibly could drive turbulence and the CR energy density much closer to equipartition (see Section~\ref{sec:DCascCalc}).}  

Likewise, the ratio of CR and thermal pressure in the superwind remains constant as long as advection dominates CR transport: it is roughly $U_{\rm CR} / U_{\rm therm} \approx \psi (\epsilon_{\rm CR} / \epsilon_{\rm therm}) (c_S / v_{\rm wind}) \approx 1/3$.  CRs do not provide the majority of the pressure, but they do track the thermal pressure because supernovae power both the superwind and CRs.

Even if the ratio of CR and magnetic energy densities vary by an order of magnitude, the estimated equipartition magnetic field strength is still reasonable.  As \citet{Lacki13-Equip} showed, for a given observation of a starburst, the dependence of the estimated magnetic field strength depends only as $\xi^{2/7.2}$ where $\xi = U_B / U_{\rm CR}$ if CR $e^{\pm}$ are cooled by bremmstrahlung.  Thus, an order of magnitude variation in $\xi$ leads to only a factor 2 change in $B$.  Note the estimated $U_{\rm CR}$ has a stronger dependence $\sim \xi^{-3.2/7.2}$, so its estimate is less trustworthy.

However, in dense starbursts, proton calorimetry is unavoidable.  Then the CR energy density is \emph{not} set by the sound-crossing time, but by the pionic time, which can be arbitrarily shorter:
\begin{equation}
t_{\pi} \approx 5000\ \yr \left(\frac{n_H}{10^4\ \cm^{-3}}\right)^{-1}
\end{equation}
from \citet{Mannheim94}. Thus, once proton calorimetry is achieved, equipartition between magnetic fields and CRs fails.  In Arp 220, equipartition fails by nearly two orders of magnitude (assuming the CRs are evenly distributed throughout the starburst; see section~\ref{sec:CRStoch}).  As a result, $\xi$ approaches 50 and the equipartition estimate for $B$ falls short by a factor of $\sim 3$.

\section{The turbulent cascade and CR diffusion}
\label{sec:Cascade}
Most of the observational probes of starburst turbulence (see Section~\ref{sec:Implications}) only probe the fluctuation spectrum on large scales, where most of the turbulent power is.  But we often want to know what the full spectrum of turbulence is; for example, the microphysics is relevant for the dissipation into heat.  In the Milky Way, there are several observational constraints on turbulence, such as radio scintillation \citep{Scalo04}.  Yet most of these probes are impractical for starburst regions other than the GCCMZ, due to their great distances and high extinctions.  

The diffusion of CRs is intimately related to the magnetic fluctuation spectrum on small scales, especially on the gyroradius of individual CRs.  As a general rule, strong fluctuations slow down CRs.  The CR diffusion rate is therefore a powerful probe of the small-scale turbulence, although in practice, interpreting it is difficult.  The synchrotron emission of nearby starbursts is resolved, so these studies are in principle possible, and the rate of CR diffusion is a key quantity in predicting this resolved emission \citep[c.f.,][]{Torres12}.  Extremely strong turbulence can also energize extant CRs through second-order Fermi acceleration, pushing them towards equipartition.  

CRs have an important role in the neutral ISM of starbursts: they provide a low level of ionization, allowing the gas to couple to magnetic fields.  Without some form of ionization, the turbulent dynamo described in section~\ref{sec:Dynamoes} could not operate in the mass-dominant neutral ISM.  Ultraviolet photons are unlikely to penetrate through the thick columns of starburst molecular gas.  Because of their penetrating power, CRs are a good candidate for ionizing \citep{Papadopoulos10-CRDRs}.  But if the CR diffusion is too slow, the CRs are confined to small bubbles, and we must look to some other source of radiation \citep{Lacki12-GRDRs,Lacki12-SLRs}.

\subsection{What is the spectrum of turbulent fluctuations?}
\label{sec:TurbSpectrum}
Although power is injected at large scales, the turbulent cascade spans orders of magnitude in galaxies.  At each scale,  eddies decay into smaller eddies in one flow-crossing time.  The cascade is finally terminated by microscopic processes like viscosity, converting the power into heat.  The differential spectrum of power in modes with wave number between $k$ and $k + dk$ is generally described as a power law with spectral index $-q$: $E(k) \propto k^{-q}$.

A simple but powerful guess for the spectrum of hydrodynamic turbulence is Kolmogorov turbulence.  In this theory, turbulent energy cascades through each length scale without significant dissipative losses.  That means that the energy at each scale $\ell$ is essentially the amount of power injected at the outer scale multiplied by a residence time, which is taken to be the flow-crossing time at that scale, $\sigma(\ell) / \ell$.  Then, as seen from eqn.~\ref{eqn:Uturb}, the turbulent velocity at that scale is just $\sigma \propto \ell^{1/3}$, with a turbulent energy density of $U_{\rm turb} (\ell) \propto \ell^{2/3}$.  The power spectrum is the well-known Kolmogorov spectrum $E(k) \propto k^{-5/3}$.  While Kolmogorov turbulence was derived under the assumption of nonmagnetic, hydrodynamic turbulence, the same power law is a common feature in simulated MHD turbulence \citep{Cho02}.  \citet{Goldreich95} derived a spectral index of $q = 5/3$ for Alfven waves in MHD turbulence, although these waves are anisotropic.  Furthermore, a Kolmogorov-like power law is observed to hold in the Milky Way WIM from parsec scales all the way down to megameter scales (the so-called ``Big Power Law in the Sky''; \citealt{Armstrong95}).  Kolmogorov-like turbulence is also observed in Galactic HI gas \citep[e.g.,][]{Roy08}, where the presence of a turbulent cascade is supported by the existence of Larson's laws \citep{Larson81,Miesch94,Shetty12}.

\begin{figure}
\centerline{\includegraphics[width=8cm]{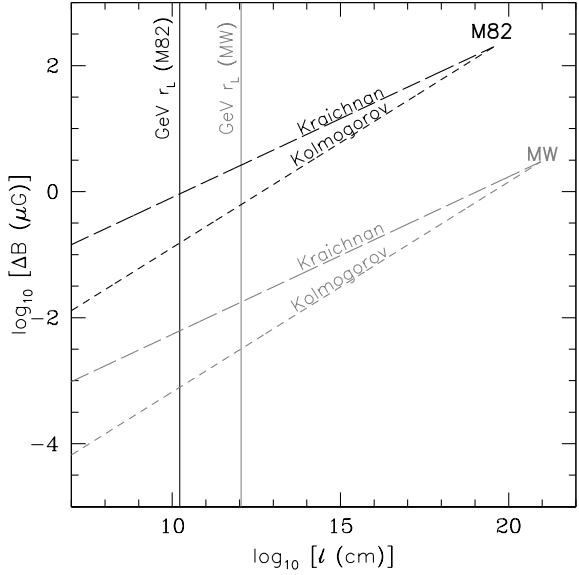}}
\figcaption{A comparison of the strength of magnetic fluctuations at different scales in the Milky Way (grey) and M82's superwind (black).  There is more turbulence to begin with in M82, because of the energetic environment, and it starts from a smaller scale.  The larger fluctuations at the Larmor radii of CRs means the diffusion constants are smaller.\label{fig:BSpectra}}
\end{figure}

Besides the Kolmogorov spectrum, the ``Kraichnan'' spectrum with $q = 3/2$ is a common hypothesis.  There is more power in small-scale magnetic fluctuations in the Kraichnan spectrum than in the Kolmogorov spectrum (Figure~\ref{fig:BSpectra}).  

\subsection{The inner scale of magnetic fluctuations}
\label{sec:lInner}
Aside from the outer scales and energy injection rates calculated in the previous sections, the inner scale is the final ingredient needed to calculate the amount of turbulence at each scale.  The physics thought to terminate the turbulent cascades is quite different for ionized and neutral matter.  In the Milky Way warm ionized medium, the inner scale is only of order $\sim 10^7\ \cm$ \citep{Spangler90}.  \citet{Spangler90} suggested that the inner scale was the larger of two lengths, both of which are related to the ion cyclotron frequency $\Omega_i = e B / (m_p c)$.  The first is ion inertial length, $v_A / \Omega_i$, at which a cyclotron resonance could damp waves; the second is the ion Larmor radius $v_{\rm therm} / \Omega_i$ \citep{Spangler90}.  In starburst superwinds with $U_B = U_{\rm turb} \approx U_{\rm therm}$, both scales are nearly equal.  Despite the low densities and high temperatures of the superwind that increase $v_A$ and $v_{\rm therm}$, the strong magnetic fields of starbursts imply that the inner scales of their turbulence is quite small, of order tens to a few thousand kilometers.  These values are listed in Table~\ref{table:BHotPredictions}.  Note that, by construction, these inner scales are all smaller than the gyroradius for a GeV ion.

The situation is much different for starburst molecular gas.  In this case, the magnetic fields directly act on the free ions and electrons created by occasional ionizing radiation in the otherwise neutral gas.  But it is the neutral gas that contains virtually all of the mass and the kinetic energy.  The first barrier to the turbulent cascade is ambipolar diffusion, in which the ions drift relative to the neutral gas.  On small length scales, ions diffuse out of eddies faster than they are transported by the gas, resulting in friction \citep{Zweibel83}.   The effective Reynolds number for ambipolar diffusion drops below 1 on scales:
\begin{equation}
\label{eqn:lADCold}
%Exact number: 0.02593068 pc = 8.001429e16 cm
\ell_{AD} \approx 8.0 \times 10^{16}\ \cm\ {\cal M}_A^{-1} \left(\frac{B}{2\ \mGauss}\right) \left(\frac{x_e}{10^{-5}}\right)^{-1} \left(\frac{n ({\rm H_2})}{10^4\ \cm^{-3}}\right)^{-3/2}
\end{equation}
which could serve as an inner scale for magnetic fluctuations \citep{Klessen00}.  The Alfven Mach number $\sigma/v_A$ is denoted by ${\cal M}_A$ in this equation.  This role of ambipolar diffusion, particularly concerning whether it actually generates small-scale fluctuations itself, is disputed, though \citep{Tagger95,Balsara01,Elmegreen04}.

The more fundamental barrier is that the magnetic field couples to the neutrals only through neutral-ion collisions.  However, these are relatively infrequent, with the typical mean free path of an ion being $\ell_N \approx v_{\rm therm} / (\mean{\sigma v} n_{H2})$ \citep{Lithwick01}.  The MHD cascade therefore must terminate at $\ell_N$, because the ions have no way of responding to the motions of neutrals without colliding with them.  According to \citet{Osterbrock61}, the cross section for ion-H$_2$ collisions\footnote{Strictly speaking, this $\mean{\sigma v}$ is for heavy ions colliding with H$_2$ molecules.  However, at the temperatures of $100\ \Kelv$ typical of starburst molecular gas, this cross section should still roughly work \citep{Osterbrock61}.} is given by $\mean{\sigma v} \approx 1.9 \times 10^{-9}\ \cm^{-3}\ \sec^{-1}$.  Then the inner scale of MHD turbulence in starburst molecular gas is
\begin{equation}
\label{eqn:lInCold}
\ell_N \approx 1.4 \times 10^{12}\ \cm \left(\frac{T}{100\ \Kelv}\right)^{1/2} \left(\frac{n_(H_2)}{100\ \cm^{-3}}\right)^{-1}.
\end{equation}
Unlike in the superwinds, the inner scale is much larger than the gyroradius of a GeV ion, which has important implications for CR diffusion.  However, the inner scale gets progressively smaller in denser regions.

\subsection{Calculation of CR diffusion constants}
\label{sec:DCascCalc}

\begin{deluxetable*}{lllccccccccc}
\tablecaption{Predicted CR diffusion constants}
\tablehead{\colhead{Starburst} & \colhead{Phase} & \colhead{Assumptions} & \colhead{$r_L (1\ \GeV)$} & \multicolumn{2}{c}{$D_{\rm casc}(1\ \GeV)$} & \colhead{$D_{\rm fast} (1\ \GeV)$} & \colhead{$D_{\rm eddy}$} & \colhead{$D_{\rm FLRW}$} & \multicolumn{3}{c}{$t_{\rm reacc} (1\ \GeV)$} \\ & & & \colhead{(cm)} & \colhead{$q = 5/3$} & \colhead{$q = 3/2$} & & & & \colhead{$q = 5/3$} & \colhead{$q = 3/2$} & \colhead{Fast} \\ & & & & & & & & & \colhead{(kyr)} & \colhead{(kyr)} & \colhead{(kyr)}}
\startdata
GCCMZ               & Hot  & $\ell_{\rm outer} = h$      & $4.4 \times 10^{10}$ & 1.0  & 0.026 & 0.15 & 20 & 1500 & 18 & 0.45 & 2.6\\
                    &      & $\ell_{\rm outer} = \ell_1$ & $5.0 \times 10^{10}$ & 0.81 & 0.023 & 0.14 & 12 & 1000 & 18 & 0.52 & 3.2\\
                    &      & $\ell_{\rm outer} = \ell_2$ & $4.9 \times 10^{10}$ & 0.84 & 0.023 & 0.14 & 12 & 1100 & 18 & 0.51 & 3.1\\
                    & Cold & $\ell_{\rm outer} = h$; $\Psi = 1$               & $8.2 \times 10^9$    & 0.57 & 0.011  & 0.35 & 0.69  & 1500 & 8200  & 160 & 4900\\
                    &      & $\ell_{\rm outer} = R_{\rm max}$; $\Psi = 1$     & $1.7 \times 10^{10}$ & 0.17 & 0.0053 & 0.24 & 0.034 & 160  & 11000 & 340 & 15000\\
                    &      & $\ell_{\rm outer} = R_{\rm max}$; ``Natural'' $\Psi$ & $2.5 \times 10^{10}$ & 0.26 & 0.0082 & 0.63 & 0.019 & 260  & $1.4 \times 10^5$ & 4400 & $3.4 \times 10^5$\\
\hline
NGC 253             & Hot  & $\ell_{\rm outer} = h$      & $9.7 \times 10^9$ & 0.61 & 0.012  & 0.073 & 19  & 1500   & 12 & 0.24 & 1.4\\
                    &      & $\ell_{\rm outer} = \ell_1$ & $1.6 \times 10^{10}$ & 0.27 & 0.0076 & 0.057 & 2.7 & 360 & 14 & 0.39 & 3.0\\
                    &      & $\ell_{\rm outer} = \ell_2$ & $1.4 \times 10^{10}$ & 0.32 & 0.0083 & 0.060 & 3.9 & 480 & 14 & 0.35 & 2.6\\
                    & Cold & $\ell_{\rm outer} = h$; $\Psi = 1$               & $3.5 \times 10^9$ & 0.44 & 0.0074 & 0.12  & 2.4   & 1500   & 500  & 8.4 & 140\\
                    &      & $\ell_{\rm outer} = R_{\rm max}$; $\Psi = 1$     & $7.9 \times 10^9$ & 0.11 & 0.0033 & 0.081 & 0.095 & 140    & 650  & 19  & 470\\
                    &      & $\ell_{\rm outer} = R_{\rm max}$; ``Natural'' $\Psi$ & $1.2 \times 10^{10}$ & 0.18 & 0.0050 & 0.21  & 0.053 & 220 & 8600 & 250 & 10000\\
\hline
M82                 & Hot  & $\ell_{\rm outer} = h$      & $1.0 \times 10^{10}$ & 0.62 & 0.013  & 0.078 & 17  & 1500 & 14 & 0.29 & 1.8\\
                    &      & $\ell_{\rm outer} = \ell_1$ & $1.7 \times 10^{10}$ & 0.28 & 0.0077 & 0.061 & 2.5 & 360  & 17 & 0.47 & 3.7\\
                    &      & $\ell_{\rm outer} = \ell_2$ & $1.5 \times 10^{10}$ & 0.33 & 0.0086 & 0.064 & 3.8 & 490  & 16 & 0.42 & 3.1\\
                    & Cold & $\ell_{\rm outer} = h$; $\Psi = 1$               & $3.4 \times 10^9$ & 0.43 & 0.0072 & 0.14  & 1.9   & 1500   & 810   & 14 & 250\\
                    &      & $\ell_{\rm outer} = R_{\rm max}$; $\Psi = 1$     & $7.9 \times 10^9$ & 0.11 & 0.0031 & 0.089 & 0.065 & 120    & 1100  & 32 & 900\\
                    &      & $\ell_{\rm outer} = R_{\rm max}$; ``Natural'' $\Psi$ & $1.1 \times 10^{10}$ & 0.17 & 0.0048 & 0.23  & 0.037 & 200 & 14000 & 410 & 20000\\
\hline
Arp 220 Nuclei      & Hot  & $\ell_{\rm outer} = h$      & $1.2 \times 10^9$ & 0.30  & 0.0042 & 0.024 & 20   & 1500 & 5.3 & 0.073 & 0.43\\
                    &      & $\ell_{\rm outer} = \ell_1$ & $3.0 \times 10^9$ & 0.063 & 0.0016 & 0.015 & 0.46 & 91   & 7.2 & 0.19  & 1.7\\
                    &      & $\ell_{\rm outer} = \ell_2$ & $2.5 \times 10^9$ & 0.083 & 0.0019 & 0.017 & 0.89 & 150  & 6.8 & 0.16  & 0.14\\
                    & Cold & $\ell_{\rm outer} = h$; $\Psi = 1$               & $4.1 \times 10^8$ & 0.21  & 0.0025  & 0.041 & 2.4   & 1500 & 250   & 2.9 & 48\\
                    &      & $\ell_{\rm outer} = R_{\rm max}$; $\Psi = 1$     & $1.4 \times 10^9$ & 0.026 & 0.00070 & 0.022 & 0.015 & 34   & 370   & 10  & 320\\
                    &      & $\ell_{\rm outer} = R_{\rm max}$; ``Natural'' $\Psi$ & $2.3 \times 10^9$ & 0.055 & 0.0014  & 0.055 & 0.024 & 86   & 2000  & 51  & 2000
\enddata
\tablenotetext{}{All spatial diffusion constants are in units of $\DiffTableUnits$ and are for $\beta_{\rm CR} = 1$.}
\label{table:DPredictions}
\end{deluxetable*}

A CR is largely unaffected by magnetic inhomogeneities smaller than its Larmor radius $r_L$, and it basically follows a magnetic field line if it curves on scales  larger than its Larmor radius.  Magnetic fluctuations of the same size as the Larmor radius can interact with CRs resonantly and deflect it.  This is the gyroresonance.  Since the inhomogenieties $\Delta B (r_L)$ at the Larmor radius are much typically smaller than the large-scale $B$ setting the size of the Larmor radius, the CRs are deflected by only a small angle $\sim \Delta B (r_L) / B$ with each Larmor orbit.  If the fluctuations are random, it takes $\sim (B / \Delta B (r_L) )^2$ orbits to be isotropized; from this the mean free path of a CR can be calculated \citep{Kulsrud05}. 

\emph{Kolomogorov spectrum} -- If isotropic Kolmogorov spectrum turbulence extends down to the gyroradius of the CR, it provides a minimum level of magnetic fluctuations that deflect the CR.  The mean free path of a CR is no more than
\begin{equation}
\lambda_{\rm CR} \approx r_L^{1/3} \ell_{\rm outer}^{2/3} 
\end{equation}
assuming the magnetic field is entirely turbulent \citep{Schlickeiser02}.  The spatial diffusion constant is $D_{\rm casc} \approx \lambda v_{\rm CR} / 3$, or
\begin{equation}
D_{\rm casc} \approx r_L^{1/3} \ell_{\rm outer}^{2/3} \beta_{\rm CR} c / 3,
\end{equation}
where $\beta_{\rm CR} c$ is the CR's speed.  

Using the outer scales and magnetic fields I found in sections~\ref{sec:HotTurbulence} and~\ref{sec:ColdTurbulence}, I calculate CR diffusion constants and list them in Table~\ref{table:DPredictions}.  These are listed for CR protons with energies of 1 GeV, the peak of the CR energy density spectrum.  The diffusion constants are very small compared to the values of $3 \times 10^{28}\ \cm^2\ \sec^{-1}$ in the Solar neighborhood: the high volumetric rate of supernovae drives vast amounts of turbulent power through each scale of the cascade, and the high magnetic field strengths lead to small Larmor radii and smaller mean free paths.  The time for CRs to escape the starburst through diffusion is then $t_{\rm diff} \approx h^2 / D$, about $3\ \Myr$.  Then CR diffusion is much slower than advection, as supported by the hard gamma-ray spectra of M82 and NGC 253, which indicate that the dominant transport process is energy-independent \citep{Abramowski12}.

\emph{Fast modes in MHD turbulence and the Kraichnan spectrum} -- In the modern theory of MHD turbulence developed by \citet{Goldreich95}, most of the turbulent power is in the form of Alfven waves.  The waves are extremely anisotropic at small scales, with the wavelength parallel to the magnetic field much longer than the wavelength perpendicular to the magnetic field ($\lambda_{\|} / \lambda_{\bot}) \propto (\ell_{\rm outer} / \lambda_{\bot})^{1/3}$.  The power spectrum of these waves is in fact Kolmogorov, $k_{\bot}^{-5/3}$ \citep{Goldreich95}.  The problem for CR confinement is that CRs are thought to scatter on fluctuations moving parallel to the magnetic field.  Thus, by the point that $\lambda_{\|}$ equals $r_L$, $\lambda_{\bot}$ is tiny, and the fluctuations are too weak to confine CRs.

Fortunately, Alfven waves are not the only MHD modes present in magnetic turbulence.  Simulations of MHD turbulence reveal that a fraction of the turbulent power goes into producing fast magnetosonic waves \citep{Cho02,Cho03}, which are isotropic and thus can scatter CRs \citep{Yan02}.  The normalization of the cascade spectrum is set by this fraction.  Although this fraction is small when the magnetic field is ordered or when $v_A \ll c_S$ \citep{Cho02,Cho03}, it may be of order unity for starbursts where the magnetic field is essentially all turbulent and is in equipartition with or far exceeds the thermal pressure.  

Even better for the confinement of CRs, the fast magnetosonic wave spectrum decays has a Kraichnan spectrum at small scales \citep{Cho02} (see Figure~\ref{fig:BSpectra}).  

If the magnetic fluctuation strength due to fast modes is $B_{\rm fast}$, and if a fraction $\epsilon_{\rm fast}$ of turbulent power is in these modes, then $(\Delta B_{\rm fast} / B) \approx (r_L / \ell_{\rm outer})^{1/2} \epsilon_{\rm fast}$.  We then have, by a similar argument to the Kolomogorov spectrum, $\lambda_{\rm CR} \approx \epsilon_{\rm fast}^{-1} \sqrt{r_L \ell_{\rm outer}}$.  The diffusion constant for the Kraichnan spectrum is 
\begin{align}
\label{eqn:DCascKraich}
%Exact value: 7.17052e24
D_{\rm casc} & \approx \frac{\beta_{\rm CR} c}{3 \epsilon_{\rm fast}} \sqrt{r_L \ell_{\rm outer}}.
\end{align}

\citet{Schlickeiser98} calculated the diffusion constant of CRs scattering on fast modes using quasilinear theory (see also \citealt{Schlickeiser02}) in which they included both the resonance and a ``transit time damping'' contribution (described in section~\ref{sec:OtherD}).  They found that the diffusion constant is limited by CRs with small pitch angles, resulting in a diffusion constant that is roughly $\sim \beta_{\rm CR} c / v_A$ times larger than in equation~\ref{eqn:DCascKraich}.  The exact value of the diffusion constant for fast modes with $q = 3/2$ in \citet{Schlickeiser98} is
\begin{align}
D_{\rm fast} & \approx \frac{4}{3\pi {\cal Z}(5/2)} \epsilon_{\rm fast}^{-1} \sqrt{\frac{1}{2\pi} \frac{\beta_{\rm CR}^3 c^3}{v_A} r_L \ell_{\rm outer}},
\end{align}
where ${\cal Z}$ is the zeta function.

I list $D_{\rm casc} (q = 3/2)$ and $D_{\rm fast}$ for supernova-driven turbulence within starbursts in Table~\ref{table:DPredictions}.  We see that the diffusion constants are very small, and tiny compared to those inferred for the Milky Way ($\sim 10^{28}$ -- $10^{29}\ \cm^2\ \sec^{-1}$).  If these diffusion constants apply, then standard CR diffusion plays no role in the large scale motion of CRs through and out of starbursts.  

\emph{Diffusive reacceleration} -- The scattering of CRs off magnetic turbulence also results in momentum diffusion, with CRs gaining roughly $\sim (v_A/c)^2$ in energy per mean free path through second-order Fermi acceleration.  While this process is slow in the Milky Way, it could be much faster in starbursts because the mean free paths are shorter and the turbulent speeds are faster.  The effective momentum diffusion constant is $D_{pp} \approx p^2 v_A^2 / D_{\rm casc}$, and the associated timescale for momentum gain is $t_{\rm reacc} \approx 9 D_{\rm casc} / v_A^2$ \citep{Strong07}.\footnote{For scattering off fast modes, \citet{Schlickeiser02} finds that $D_{pp}$ is $\ln(c/v_A)$ times larger and $t_{\rm reacc}$ is $\ln(c/v_A)$ times shorter.}  

If the diffusion constants calculated above are correct, diffusive reacceleration is extremely fast in starburst winds.  The reacceleration timescale is only $\sim 10\ \kyr$ for Kolmogorov turbulence and just a few centuries for Kraichnan turbulence (Table~\ref{table:DPredictions}).  As this is much shorter than the escape time, at face value, this implies the CR energy density grows exponentially.  Of course, the turbulent energy would be damped by this process, so what must happen is that CRs, turbulence, and magnetic fields approach equipartition.  If this is true, CR acceleration in starburst winds is not primarily a process happening at discrete supernova remnants, but is a collective process throughout the hot ISM \citep[c.f.,][]{Amano11,Melia11}.  The efficiency for converting supernova mechanical power into CR kinetic energy could then be higher than the canonical value of 10\%.  Note also that some of the energy would go into MeV suprathermal particles, not just the GeV-TeV particles observed by gamma-ray telescopes.

Diffusive reacceleration is much slower in molecular regions, where the turbulent speeds are $\la 100\ \kms$.  The reacceleration times are $\ga \Myr$ for Kolmogorov timescales and $\sim 10\ \kyr$ for Kraichnan turbulence (see Table~\ref{table:DPredictions}).  Note that the radiative lifetime of CRs in ULIRGs is just a few kyr; therefore, reacceleration probably does not affect the overall energetics of CRs in these environments.  

\subsection{Turbulent mixing}
\label{sec:TurbulentMixing}
The transport of CRs involves not just the flow of CRs relative to matter but the flow of the matter containing the magnetic fluctuations.  For small enough scattering diffusion constants, like those found in Section~\ref{sec:DCascCalc}, the CRs are effectively frozen into the fluid.  Thus the motion of the ISM itself becomes increasingly important.  While the large-scale bulk advection is well known to carry away CRs, the small-scale chaotic advection has not been explored as much.

Turbulence is well known for its ability to mix fluids.  The chaotic flows knead regions with different properties into one another, smoothing out the coarse-grained distribution.  While it does not directly even out the fine-grained distribution, the inhomogeneities are on such tiny scales that even very low levels of microphysical diffusion smooth out the fine-grained distribution \citep[see the review by][]{Shraiman00}.  Turbulence is suspected to be a prime mover behind the transport of metals in the ISM of star-forming regions \citep[e.g.,][]{Roy95,deAvillez02,Scalo04,Pan10}.  

CRs are another ``pollutant'' injected by star-forming regions that turbulence will spread.  We can define an effective diffusion constant for the turbulent mixing as
\begin{equation}
D_{\rm mix} \approx \ell_{\rm outer} \sigma
\end{equation}
\citep{Tennekes72}.  Turbulent mixing is most effective in regions with high $\sigma$, particularly the hot superwind phase.  Another noteworthy feature of turbulent mixing is that it is \emph{energy independent}.  Thus, observations of the gamma-ray spectral shape do not constrain turbulent mixing.

But, while we can define a ``diffusion constant'' for the process, there are several important differences with standard diffusion.  Turbulence creates inhomogeneities in the flow; for transonic or supersonic turbulence, the fluctuations are extremely strong.  If CRs are frozen into the flow, turbulent mixing squeezes and stretches them too.  The resulting strong fluctuations in the density of CRs are purely random and not directly related to the locations of CR sources.  On top of that, adiabatic expansion and compression of the CRs enhances density gradients, as $P \propto V^{-4/3}$ for relativistic particles.  Furthermore, turbulent mixing is generally intermittent on large scales, so that even if the underlying fluid were homogeneous, there would be fluctuations in the density of CRs simply because the mixing cannot even out the CR abundance perfectly \citep{Shraiman00}.  By contrast, standard diffusion reduces inhomogeneities, simply because there are more CRs in a high density fluctuation to flow out than CRs outside the fluctuation to flow in.

The essential difference is that standard diffusion works to even out the \emph{density} of CRs, the number of CRs per volume, whereas turbulent mixing works to even out the \emph{abundance} of CRs, the number of CRs per mass \citep[c.f.,][]{Ensslin11}.  The former can be thought of as a ``Eulerian'' process, where the CRs follow the volume.  This is reflected by the Eulerian methods used to treat diffusive propagation in models of galactic CR populations like GalProp \citep[as described in][]{Strong98}.  But turbulent mixing is a ``Lagrangian'' process, where the CRs follow the mass, suggesting a possible need for Lagrangian methods when modeling starburst CR populations.  An example of these methods are provided by MHD simulations of CR populations in galaxy clusters, where Lagrangian CR transport is included (often as the only form of CR transport) \citep[e.g.,][]{Hanasz03,Pfrommer08}.

The ${\cal M} \approx 100$ turbulence in the molecular medium has especially extreme density contrasts and extreme intermittency.  Current CR models assume that CRs experience essentially steady losses in a uniform medium.  But actual losses in a supersonic turbulent medium may be highly irregular.  Suppose CRs are injected into such a medium at random locations.  At any given time, the median gas density in the volume is $\sim \mean{\rho} / {\cal M}$ to $\mean{\rho}$, but most of the mass is in clumps with density $\sim {\cal M} \mean{\rho}$.  Thus, most CRs could be injected into underdense regions, where they are presumably frozen into the flow.  In these regions, losses are much slower than supposed by one-zone models.  Yet the flow rearranges itself on an eddy crossing time, $\sim \ell_{\rm outer} / \sigma \approx 10\ \kyr$, and most of the fluid containing the CRs ends up in high density clumps.  As the flow is squeezed into these clumps, the radiative losses become orders of magnitude stronger -- up to a factor ${\cal M}^2 \approx 10^4$ faster.  At some point, the radiative losses become faster than the advection times, and the CRs are destroyed.  Thus, CR losses could occur in short bursts.  

Although proton calorimetry implies the total gamma-ray luminosity does not depend on the proton cooling time, adiabatic compression of the CRs converts mechanical energy into CR kinetic energy.  When the CRs are destroyed, that energy is mostly radiated away as gamma rays and neutrinos.  Turbulent mixing in supersonic turbulence can therefore decouple the power injected in CRs from the power radiated as gamma rays.  The intermittent destruction of CRs and the implications for proton calorimetry need to be studied. 

\subsection{Comparison with Observations}
\emph{The CR diffusion constant around 30 Doradus} -- By comparing the resolved synchrotron and Inverse Compton emission of the super star cluster 30 Doradus in the Large Magellanic Cloud, \citet{Murphy12} derived the CR diffusion constant and its energy dependence at GeV energies in that region.  The star-formation rate, $0.15\ \Msun\ \yr^{-1}$, and size of the region the CRs diffused through, $\sim 100\ \pc$, correspond to a star-formation surface density of $5\ \Msun\ \kpc^{-2}\ \yr^{-1}$, of the same order as in the GCCMZ.  

The measured CR diffusion constant is $10^{27}\ \cm^2\ \sec^{-1} (E/\GeV)^{0.7}$ for relativistic particles \citep{Murphy12}.  They conclude that the smallness of the diffusion constant is due to the high level of turbulence surrounding 30 Doradus.   At 1 GeV, the measured value is indeed comparable to the expected diffusion constant in the GCCMZ for a Kolmogorov spectrum of turbulence.  The steep energy dependence is, however, incompatible with the standard Kolmogorov cascade.  But according to \citet{Blasi12}, diffusion constants with $D \propto E^{0.7}$ arise when CRs amplify waves from a previously existing Kolmogorov spectrum as they self-confine.  This process decreases $D$.

Yet the 100 pc surrounding 30 Doradus may not be typical of starburst regions as a whole.  There is only one super star cluster, whereas starburst regions contain many (dozens of similar mass in M82, for example; \citealt{OConnell95}).  Furthermore, most of the star formation responsible for accelerating the CRs occurs in R136, with a diameter of a few pc, at the center of 30 Doradus \citep[e.g.,][]{Hunter95}, so the estimates of turbulence I derived above do not apply to the CRs 100 pc away.  For example, perhaps the CRs are amplifying relatively weak turbulence in these distant regions to produce a small but steeply increasing diffusion constant.  In contrast, star formation occurs on 100 pc scales throughout starbursts.  Observations of the GCCMZ, M82, and NGC 253 exclude such steep energy dependencies within their starbursts, suggesting different physics.

\emph{The TeV emission of the GCCMZ and NGC 253} -- If the TeV emission from the GCCMZ region is in fact powered by star formation, it indicates that CR diffusion is quite slow at TeV energies \citep{Crocker11-Wild}.  The spectrum observed by HESS remains relatively hard, with a power law index of $\sim 2.3$, until the highest observed energies of 10 TeV \citep{Aharonian06}.  This implies energy-independent escape for CRs with energies of up to 100 TeV, or $D \la 4 \times 10^{27}\ \cm^2\ \sec^{-1}$ 

Likewise, the hard GeV-TeV gamma-ray spectra of M82 and NGC 253 set stringent constraints on the rate of energy-dependent diffusion in these starbursts.  According to \citet{Abramowski12}, the lack of a spectral break in the gamma-ray spectrum of NGC 253 as observed with \emph{Fermi} and HESS indicates that $D \la 3 \times 10^{27}\ \cm^2\ \sec^{-1}$ for protons with energies up to 30 TeV.  We can compare this number with the predicted cascade diffusion constants $D_{\rm casc}$ listed in Table~\ref{table:DPredictions}.  I find that the 30 TeV diffusion constants are $\sim 10^{28}\ \cm^2\ \sec^{-1}$ for the Kolomogorov spectrum and $\sim 10^{27}\ \cm^2\ \sec^{-1}$ for the Kraichnan spectrum.  

While there is mild tension between the $D_{\rm casc}$ and the observed TeV spectrum for the Kolmogorov case, the Kraichnan case is in line with observations.  It is interesting that above 100 TeV, the predicted $D_{\rm casc}$ for $k = 3/2$ finally exceeds the limits set by HESS, and a break in the proton spectrum should appear.  Unfortunately, at such high energies (corresponding to $\ga 10\ \TeV$ gamma rays), NGC 253 becomes opaque as photons are destroyed by the $\gamma\gamma$ absorption process \citep{Inoue11,Lacki13-XRay}.  The GCCMZ, however, is transparent at these energies \citep{Lacki13-XRay}, and observable up to a few hundred TeV \citep{Moskalenko06}.

\subsection{What truly confines CRs in starbursts?}

\subsubsection{Problems with confinement by turbulent fluctuations}
\emph{Does the turbulence cascade in hot winds?} -- As noted in Section~\ref{sec:lInner}, the estimates of $D_{\rm casc}$ in hot winds only apply if the turbulence can cascade to small scales.  Many authors have argued that the turbulence is halted at large scales in the hot medium of the Milky Way (and by extension, in hot starbursts).  A common objection is that the small Reynolds number prevents any true cascade of magnetic energy to small scales \citep{Hall80}.  But, as I have argued, the collisionless nature of the wind implies the hydrodynamical Reynolds number is not necessarily the relevant quantity (Section~\ref{sec:Reynolds}).

A more subtle problem raised by detailed theories of MHD turbulence is the possibility that the fast modes most likely to confine CRs are damped in hot ISM.  According to \citet{Yan04}, fast modes are collisionally damped on scales comparable to the mean free path of particles in the medium.  Since this length is just a few parsecs, both in the Milky Way hot ISM and in starburst hot winds, this could prevent the cascade from reaching scales relevant for CRs.  In the GCCMZ, though, $\ell_{\rm outer}$ is nearly equal to the mean free path, so it is possible the cascade is not damped collisionally.

\emph{Neutrals in cold gas} -- The long ambipolar diffusion (equation~\ref{eqn:lADCold}) and neutral-ion drag scales (equation~\ref{eqn:lInCold}) in molecular gas pose severe problems for CR confinement in molecular gas in starbursts, especially for ULIRGs where virtually the entire volume may be low density molecular gas.  Only CRs with $\ga \TeV$ energies have $r_L \ga \ell_{\rm N}$.  This problem is analogous to the situation for MeV positrons propagating in neutral gas within the Milky Way, but at least in the Galaxy, most of the volume is ionized \citep{Higdon09,Jean09,Prantzos11}.  

\subsubsection{Other confinement processes}
\label{sec:OtherD}
\emph{Self-confinement} -- Streaming CRs, with an anisotropic momentum distribution, excite Alfven waves that effectively confine the CRs \citep{Kulsrud69}.  In cold fully-ionized plasmas, these waves force CRs to stream at speeds smaller than $v_A$.  

In the hot wind of starbursts, the Alfvenic speed is actually faster than the bulk speed of the wind in the core of the starburst ($\sim 300\ \kms$), and is approximately equal to the asymptotic wind speed ($v_{\infty} \approx 1600\ \kms$).  Thus, even if self-confinement did operate, diffusive escape could actually still dominate over advection if no other process intervened.  In addition, \citet{Holman79} argued that self-confinement fails to limit CRs to speeds below the sound speed $c_S$, although $c_S \approx v_A$ in hot ISM.  

The Alfven speed is much slower in the cold gas, but self-excited Alfven waves may be strongly damped by neutral atoms in these phases.  \citet{Higdon09} noted similar problems for confining Galactic MeV positrons in neutral gas.  On the other hand, the CR energy densities are much higher than in the Galaxy, providing much stronger driving for the build up of such waves.  A calculation of whether the streaming instabilities are enough to overcome drag is needed.

In the warm ionized medium of H II regions, self-confinement should be effective.

\emph{Field line random walk} -- Another possibility is that CRs do follow magnetic field lines, but the field lines themselves are so twisted that they confine CRs within the starbursts.  This is the Field Line Random Walk (FLRW) mechanism, which is important in the ``compound diffusion'' discussed by \citet{Lingenfelter71}.  The field lines are randomized on scales $\ell_{\rm peak}$ where the magnetic fluctuation spectrum peaks.  The worst possible case for confinement is when most of the turbulent power is near the outer scale.  The effective diffusion constant for FLRW
\begin{equation}
D_{\rm FLRW} \approx \ell_{\rm peak} c / 3,
\end{equation}
as long as $r_L < \ell_{\rm peak}$.  These values when $\ell_{\rm peak} = \ell_{\rm outer}$ are listed in Table~\ref{table:DPredictions}.

While this mechanism does not work if the outer scale is nearly the size of the starburst, confinement is effective if the outer scale is just a few parsecs, because the starburst would have a scattering optical depth to CRs of order $\sim 10$ and the typical number of scatterings before escape would be $\sim 10^2$.  This effective diffusion constant, $D_{\rm FLRW} \approx 1.5 \times 10^{29}\ \cm^2\ \sec^{-1}\ (\ell_{\rm peak} / 5\ \pc)$, is comparable to $D$ in the Milky Way: in the Milky Way, the mean free path is also of order a few parsecs.

\emph{Firehose and mirror plasma instabilities} -- The need for some kind of CR confinement in the Milky Way's hot ISM prompted suggestions that magnetic fluctuations could be generated directly on small scales.  These fluctuations could be generated by the firehose and mirror instabilities, which operate when the pressure is anisotropic enough in the plasma \citep{Hall80}.  It's not clear these could work in starburst wind, though.  A key point in \citet{Hall80}'s argument is that the magnetic pressure is much smaller than the thermal pressure; this allows the instabilities grow.  But in the starburst wind, the magnetic pressure approaches the thermal pressure, inhibiting the growth of these instabilities.  \citet{Schekochihin05} came to this conclusion for galaxy cluster plasma too: the plasma instabilities driven by anisotropic pressure work when magnetic pressure is much smaller than the thermal pressure, but die away as magnetic fields become strong.  

\emph{Transit time damping} -- Although the gyroresonance is the most obvious way to scatter CRs, compressible turbulent MHD waves can also deflect CRs through transit time damping (TTD) \citep{Schlickeiser98,Yan04}.  TTD results from the ``mirror'' force as magnetic field strengths fluctuate.  TTD naturally occurs in the fast modes, and in fact is accounted for in the $D_{\rm fast}$ given by \citet{Schlickeiser98}.  Unlike the gyroresonance, the cascade of fast modes does not have to reach $r_L$ to scatter CRs.  According to \citet{Yan04}, TTD is what confines CRs in Galactic neutral clouds and the hot phase, so its role should also be considered for starbursts.

\emph{Confinement by H II regions and SNRs} -- CRs that attempt to escape may be trapped by pockets where confinement is effective.  Candidate CR traps include H II regions or supernova remnants.  While the bubbles have a filling factor much less than 1, their covering factor can be $\ga 1$, so that the typical CR is deflected back inwards as it tries to escape.

\subsubsection{Summary of confinement processes in starbursts}
\begin{figure*}
\centerline{\includegraphics[width=16cm]{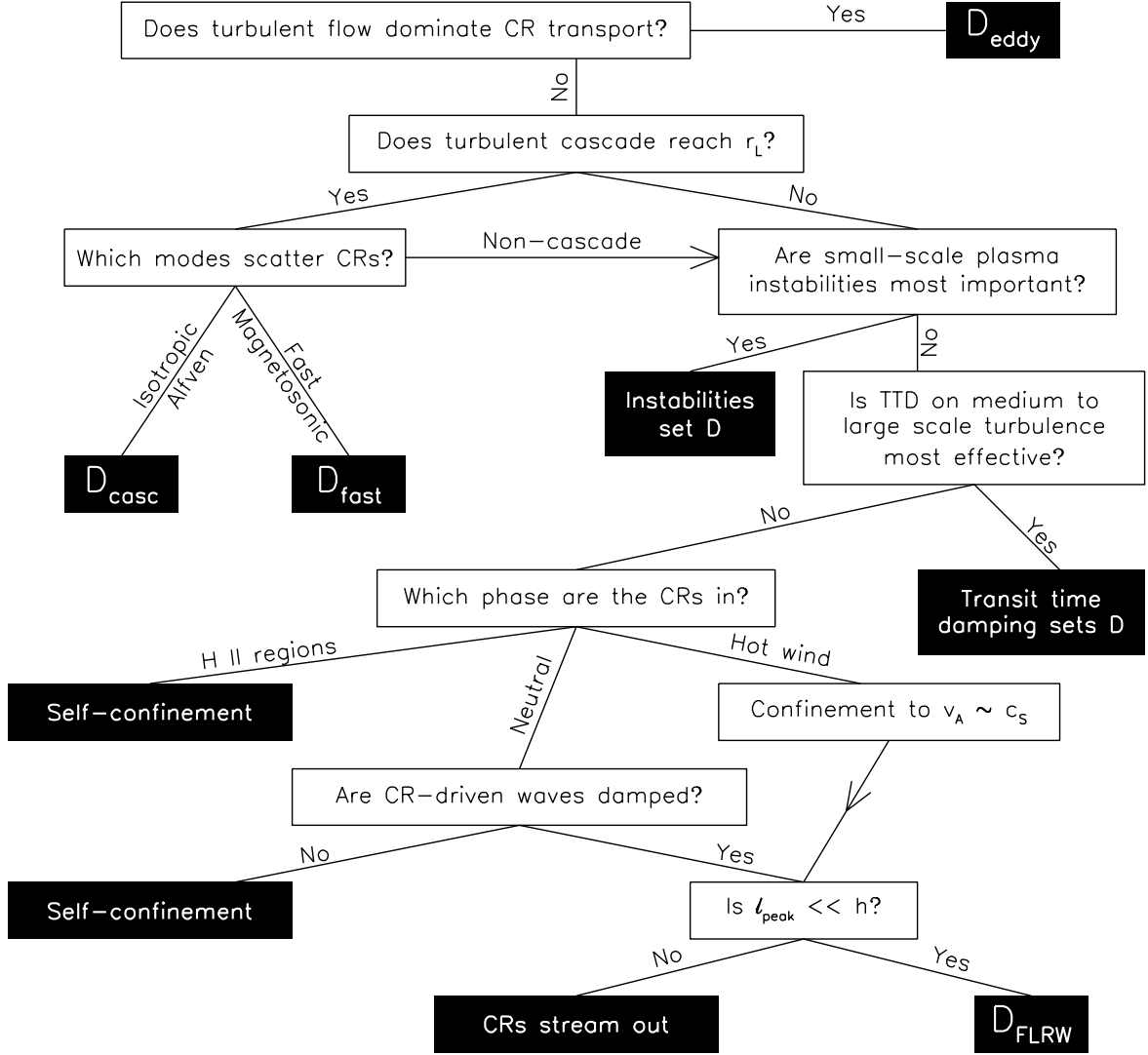}}
\figcaption{Flowchart summarizing the processes that could set the rate of CR diffusion in starbursts.  Although a cascade of the strong starburst turbluence implies the small $D$ given by $D_{\rm casc}$ and $D_{\rm fast}$, clearly there are many assumptions. \label{fig:CRDiffFlowchart}}
\end{figure*}

Many processes are conjectured to confine CRs in the Milky Way, and with our far lesser knowledge of starbursts, it is difficult to say which one is most important in those regions.  Figure~\ref{fig:CRDiffFlowchart} lays out the possibilities and the conditions they hold in.  It is impossible to even answer basic questions of CR diffusion in starbursts with certainty.  The diffusion may be Lagrangian (turbulent mixing) or Eulerian (other kinds of confinement), energy independent (turbulent mixing and FLRW) or dependent (scattering by turbulence), and may be faster (FLRW, self-confinement in the hot phase) or slower (turbulent mixing, scattering by turbulence) than advection.

What we can probably say with high likelihood is that the diffusion is no slower than that given by turbulent mixing ($D_{\rm eddy}$) and no faster than that given by field line random walk ($D_{\rm FLRW}$).  The few observations we do have suggest that diffusion is either slow and/or energy-independent, consistent with scattering by a turbulent cascade reaching the gyroscale, turbulent mixing, or a field line random walk.

\subsection{Do CRs fill all of starbursts?}
\label{sec:CRStoch}
Do CRs in fact reach all of the gas, as is commonly supposed?  Suppose that CR accelerators in starbursts inject an instantaneous pulse of CRs.  The CRs diffuse away from the accelerator, roughly filling a sphere of radius $R_{\rm diff} = \sqrt{D t}$, where $D$ is a constant diffusion constant that applies throughout the starburst.  The CRs reside for a time $t_{\rm CR}$, after which they are lost.  Therefore, I consider only CR bubbles that are younger than $t_{\rm CR}$.   The maximum size the CR bubbles reach is $R_{\rm diff}^{\rm max} = \sqrt{D t_{\rm CR}}$.  If the rate at which CR accelerators appear and inject a burst of CRs is $dN_{\rm acc}/dt$, then the volume filled by CRs is
\begin{equation}
V_{\rm CR} = \int_0^{t_{\rm CR}} \frac{dN_{\rm acc}}{dt} \times \frac{4}{3} \pi R_{\rm diff}(t)^3 dt.
\end{equation}
Now, supposing that SNRs are (short-lived) CR accelerators, we have $dN_{\rm acc}/dt = \Gamma_{\rm SN}$.  The estimated volume filled by CRs is
\begin{equation}
V_{\rm CR} = \frac{8 \pi}{15} \Gamma_{\rm SN} D^{3/2} t_{\rm CR}^{5/2}.
\end{equation}
The overlap fraction of these CRs is just $Q_{\rm CR} = V_{\rm CR} / V_{\rm SB}$; if $Q_{\rm CR} \ga 1$, then the CRs are volume-filling.

Although the true CR diffusion constants in starbursts are unknown, the $D_{\rm eddy}$ from turbulent mixing sets a lower limit to the distance CRs can traverse.  I compute the $R_{\rm diff}^{\rm max}$ and $Q_{\rm CR}$ for the prototypical starbursts, and list them in Table~\ref{table:CRStoch}.

\begin{deluxetable*}{lllllcc}
\tablecaption{Minimum CR filling factors}
\tablehead{\colhead{Starburst} & \colhead{$V_{\rm SB}$} & \colhead{$t_{\rm CR}$} & \colhead{Phase} & \colhead{Assumptions} & \colhead{$R_{\rm diff}^{\rm max} (D_{\rm eddy})$} & \colhead{$Q_{\rm CR} (D_{\rm eddy})$} \\ & \colhead{($\pc^3$)} & \colhead{($\kyr$)} & & & ($\pc$) & }
\startdata
GCCMZ               & $3.1 \times 10^6$ & 160 & Hot  & $\ell_{\rm outer} = h$      & 100 & 48 \\
                    &                   &     &      & $\ell_{\rm outer} = \ell_1$ & 78  & 21 \\
                    &                   &     &      & $\ell_{\rm outer} = \ell_2$ & 81  & 24 \\
                    &                   &     & Cold & $\ell_{\rm outer} = h$; $\Psi = 1$               & 19  & 0.31\\
                    &                   &     &      & $\ell_{\rm outer} = R_{\rm max}$; $\Psi = 1$     & 4.3 & 0.0034\\
                    &                   &     &      & $\ell_{\rm outer} = R_{\rm max}$; ``Natural'' $\Psi$ & 3.2 & 0.0015\\
\hline
NGC 253             & $7.1 \times 10^6$ & 160 & Hot  & $\ell_{\rm outer} = h$      & 99 & 830\\
                    &                   &     &      & $\ell_{\rm outer} = \ell_1$ & 38  & 46\\
                    &                   &     &      & $\ell_{\rm outer} = \ell_2$ & 46  & 80\\
                    &                   &     & Cold & $\ell_{\rm outer} = h$; $\Psi = 1$               & 36 & 39\\
                    &                   &     &      & $\ell_{\rm outer} = R_{\rm max}$; $\Psi = 1$     & 7.1 & 0.30\\
                    &                   &     &      & $\ell_{\rm outer} = R_{\rm max}$; ``Natural'' $\Psi$ & 5.3 & 0.13\\
\hline
M82                 & $2.8 \times 10^7$ & 160 & Hot  & $\ell_{\rm outer} = h$      & 96 & 620\\
                    &                   &     &      & $\ell_{\rm outer} = \ell_1$ & 36 & 33\\
                    &                   &     &      & $\ell_{\rm outer} = \ell_2$ & 45 & 63\\
                    &                   &     & Cold & $\ell_{\rm outer} = h$; $\Psi = 1$               & 32 & 23\\
                    &                   &     &      & $\ell_{\rm outer} = R_{\rm max}$; $\Psi = 1$     & 5.9 & 0.14\\
                    &                   &     &      & $\ell_{\rm outer} = R_{\rm max}$; ``Natural'' $\Psi$ & 4.4 & 0.061\\
\hline
Arp 220 Nuclei      & $3.1 \times 10^6$ & 5   & Hot  & $\ell_{\rm outer} = h$      & 18  & 12\\
                    &                   &     &      & $\ell_{\rm outer} = \ell_1$ & 2.8 & 0.041\\
                    &                   &     &      & $\ell_{\rm outer} = \ell_2$ & 3.8 & 0.11\\
                    &                   &     & Cold & $\ell_{\rm outer} = h$; $\Psi = 1$               & 6.3  & 0.50\\
                    &                   &     &      & $\ell_{\rm outer} = R_{\rm max}$; $\Psi = 1$     & 0.50 & $2.4 \times 10^{-4}$\\
                    &                   &     &      & $\ell_{\rm outer} = R_{\rm max}$; ``Natural'' $\Psi$ & 0.63 & $4.9 \times 10^{-4}$
\enddata
\tablenotetext{}{These filling factors are calculated using the turbulent mixing diffusion constants listed in Table~\ref{table:DPredictions}.}
\label{table:CRStoch}
\end{deluxetable*}

The first thing to note is that $R_{\rm diff}^{\rm max}$ and $Q_{\rm CR}$ are much larger in the hot winds, because the turbulent speeds are themselves much larger.  For the GCCMZ, NGC 253, and M82, CRs should easily fill the hot wind.

Second, the predicted $Q_{\rm CR}$ is typically smaller than 1 if I use the $D_{\rm eddy}$ listed for the cold ISM in Table~\ref{table:DPredictions}.  Of course, in the hot starbursts, only a small fraction of the volume is filled by the molecular clouds and other neutral material, so the use of the cold $D_{\rm eddy}$ is inappropriate.  Instead, the relevant question is whether the CRs can diffuse all of the way through a molecular cloud before they are destroyed.  From Table~\ref{table:CRStoch}, it appears that CRs that survive for 160 kyr can traverse at least $R_{\rm diff}^{\rm max} \ga 5$ parsecs through starburst molecular clouds, roughly the size of the molecular clouds.  On the other hand, in overdense molecular clouds, CRs are not carried by the wind.  Rather, they face strong pionic losses that shorten their lifetimes, so it's still an open question whether CRs make it all the way through the clouds before being destroyed.

Finally, in the Arp 220 nuclei, it appears that CRs may not fill the entire starburst region if they diffuse at only $D_{\rm eddy}$.  The turbulent speeds are slow in the molecular gas of these cold starbursts, and the outer scale of turbulence is probably quite small.  Instead, CRs are predicted to only fill $0.01 - 0.1\%$ of the volume, residing in about $10^4$ bubbles that are each $\la 1\ \pc$ wide.  

If that were true, the CR energy density in these bubbles would be much greater than even the ISM turbulent pressures listed in Table~\ref{table:Pressures}.  Thus, the bubbles would expand until they did reach equipartition, partly because of a greater volume and partly due to adiabatic losses.  Indeed, the expansion of such bubbles might themselves be a source of such turbulence in starbursts, on scales smaller than 10 pc.\footnote{This mechanism could be a way around extreme radiative losses by SNRs in cold starbursts, setting a floor to the amount of mechanical energy driven into the ISM by supernovae, but only if the CR acceleration efficiency is $\sim 10\%$ as in the Milky Way and if CR diffusion is very slow.  The fact that ULIRGs lie on the far-infrared--radio correlation  \citep{Condon91} supports the idea that the CR acceleration efficiency is the same.  Thus, CRs could play an important dynamical role in cold starbursts.  Note that, unlike in other scenarios where CRs are dynamically important \citep[e.g.,][]{Socrates08}, the CRs do not directly drive a wind, and do not fill the starburst.  Their dynamical role arises only because they are very inhomogeneous, concentrating their energy density into a few small bubbles.}  This just further underscores how little we know about CR diffusion in these environments.  Outside of the CR-filled bubbles, some alternate source of ionization is necessary if diffusion is this slow, such as gamma rays \citep{Lacki12-GRDRs} or short-lived radioactive isotopes \citep{Lacki12-SLRs}.

These lower bounds can be improved by using exact analytic \citep[e.g.,][]{Aharonian96} or numerical solutions for CR diffusion, as done for proton calorimetric homogeneous starbursts in \citet{Torres12}.

\section{Observational implications}
\label{sec:Implications}
\subsection{Doppler shifts in X-ray lines from superwind turbulence}
A key prediction of this work is that the volume-filling hot superwind is turbulent, with random motions that are about as fast as the large-scale outflow itself.  Ions embedded in the wind are Doppler shifted when they radiate, not just from the bulk flow but from the transonic turbulence.  In fact, Fe K line emission is directly observed from M82's superwind, so a finely-resolved X-ray spectrum should be able to trace these motions \citep{Strickland07,Strickland09}.  

The Doppler shift of the superwind ions from turbulence is $\Delta E \approx E \sigma/c$:
\begin{equation}
\Delta E \approx 22\ \eV\ \left(\frac{\sigma}{1000\ \kms}\right) \left(\frac{E}{6.7\ \keV}\right).
\end{equation}
\emph{Suzaku}'s spectrometer has an energy resolution of 130 eV at 6 keV \citep{Koyama07}.  The soft X-ray spectrometer on Astro-H, scheduled for a 2014 launch, will have an energy resolution of 7 eV in the range 0.3 to 12 keV \citep{Takahashi12}.  Thus, it can easily detect the broadening due to turbulence as well as the bulk outflow.

Unfortunately, while the line widths can indicate the presence of the random motions that are likely present due to supernovae, they cannot so easily indicate whether there is a true turbulent cascade.  The small-scale motions in a turbulent cascade necessarily have smaller amplitudes, and are harder to distinguish.  Furthermore, the angular resolution of X-ray spectrometers is poor, so the Doppler shifts of many eddies will all be blurred together.

\subsection{The far-infrared--radio correlation in starbursts}

Some of the strongest constraints on magnetic and CR densities in starbursts comes from the existence of the far-infrared--radio correlation (FRC).  Star-forming galaxies have a nearly constant ratio between their far-infrared luminosities and their 1.4 GHz synchrotron radio luminosity; this is the FRC \citep{Helou85,Condon92,Yun01}.  In denser galaxies, the far-infrared luminosity simply traces the star-formation rate.  The reason why the synchrotron emission traces the star-formation rate so well is unclear.  In starburst regions, $e^{\pm}$ traced by radio emission generally cool before escaping \citep{Volk89-Calor}.  The radio luminosity apparently is set by a conspiracy between the ratio of pionic secondary $e^{\pm}$ to primary $e^-$ fraction \citep{Rengarajan05,Thompson07}, and the ratio of synchrotron cooling timescales to $e^{\pm}$ lifetime \citep{Condon91,Thompson06,Lacki10-FRC1}.  In normal galaxies, by contrast, primary $e^-$ are overwhelmingly responsible for radio emission, $e^-$ mainly escape, and losses like bremsstrahlung are unimportant \citep{Chi90,Lisenfeld96-Normal,Bell03,Strong10}. 

\citet{Lacki10-FRC1} and \citet{Lacki10-FRC2} showed that the effects of bremsstrahlung and ionization cooling could balance the presence of secondary $e^{\pm}$, if the CRs experience average density gas.  However, for this balancing to take place, the magnetic field strength must go as
\begin{equation}
\label{eqn:BFRC-SigmaG}
B_{\rm FRC1} \approx 400\ \muGauss \left(\frac{\Sigma_g}{\gcm2}\right)^{0.7}
\end{equation}
or
\begin{equation}
%n_MW = 0.241992 cm^-3 
\label{eqn:BFRC-rho}
%Exact value 386
B_{\rm FRC2} \approx 390\ \muGauss \left(\frac{\mean{n}}{1000\ \cm^{-3}}\right)^{0.5},
\end{equation}
if the \citet{Kennicutt98} Schmidt law held (Figure~\ref{fig:B}).  

A turbulent dynamo powered by supernovae mechanical power can explain why equation~\ref{eqn:BFRC-SigmaG} holds.  The normalization and density dependence of the predicted $B$ for both the hot wind and cold gas are roughly in line with the FRC constraint (sections~\ref{sec:HotTurbulence} and~\ref{sec:ColdTurbulence}), as shown in Figure~\ref{sec:ModelBComparison}.  In detail, however, there are discrepancies that might shed light on whether supernovae-driven turbulence sets $B$ in starbursts.  First, the magnetic field strength is stronger at low density than equation~\ref{eqn:BFRC-SigmaG} predicts.  Second, the magnetic field strength grows more weakly with $\Sigma_{\rm SFR}$ than in equation~\ref{eqn:BFRC-SigmaG} if $\ell_{\rm outer}$ also shrinks.  

To examine these effects, I run one-zone models of the CR population in starbursts, similar to those described in \citet{Lacki10-FRC1}, but using the $B_{\rm turb}$ I derived in sections~\ref{sec:HotTurbulence} and \ref{sec:ColdTurbulence}.  I consider both a constant $\tau_{\rm gas} = 20\ \Myr$ and the \citet{Kennicutt98} Schmidt Law.  From these, I calculate the 1.4 GHz synchrotron emission and spectral index, as well as the 1-100 GeV gamma-ray emission and spectral index.  I ignore free-free emission and free-absorption for the radio, and leptonic emission for the gamma rays.  I assume an advection speed of $v = 300\ \kms$.  Most importantly, I assume that the CRs experience average density gas ($\mean{\rho}_{\rm SB}$), as most other one-zone models do.  

\begin{figure*}
\centerline{\includegraphics[width=8cm]{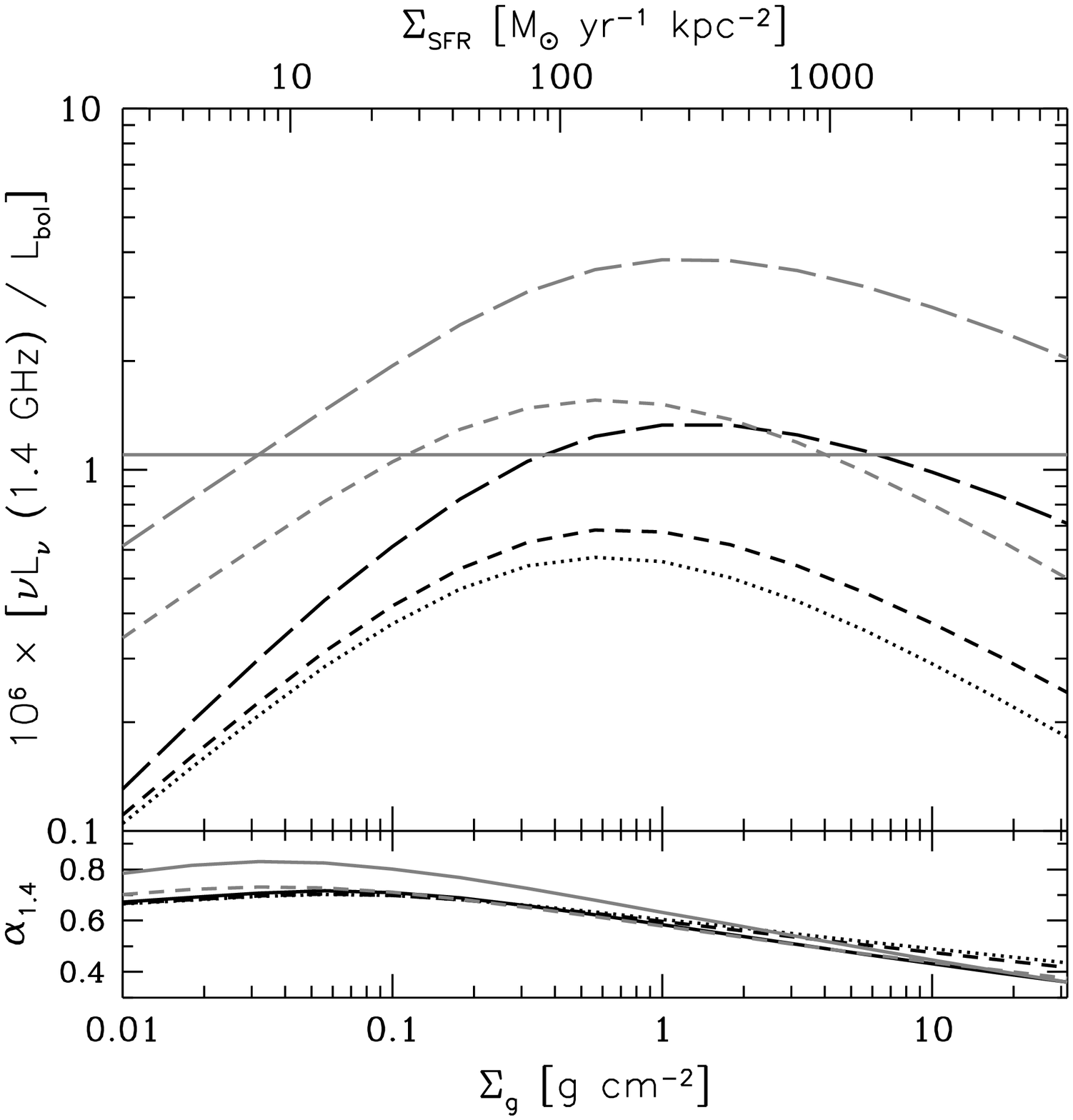}\includegraphics[width=8cm]{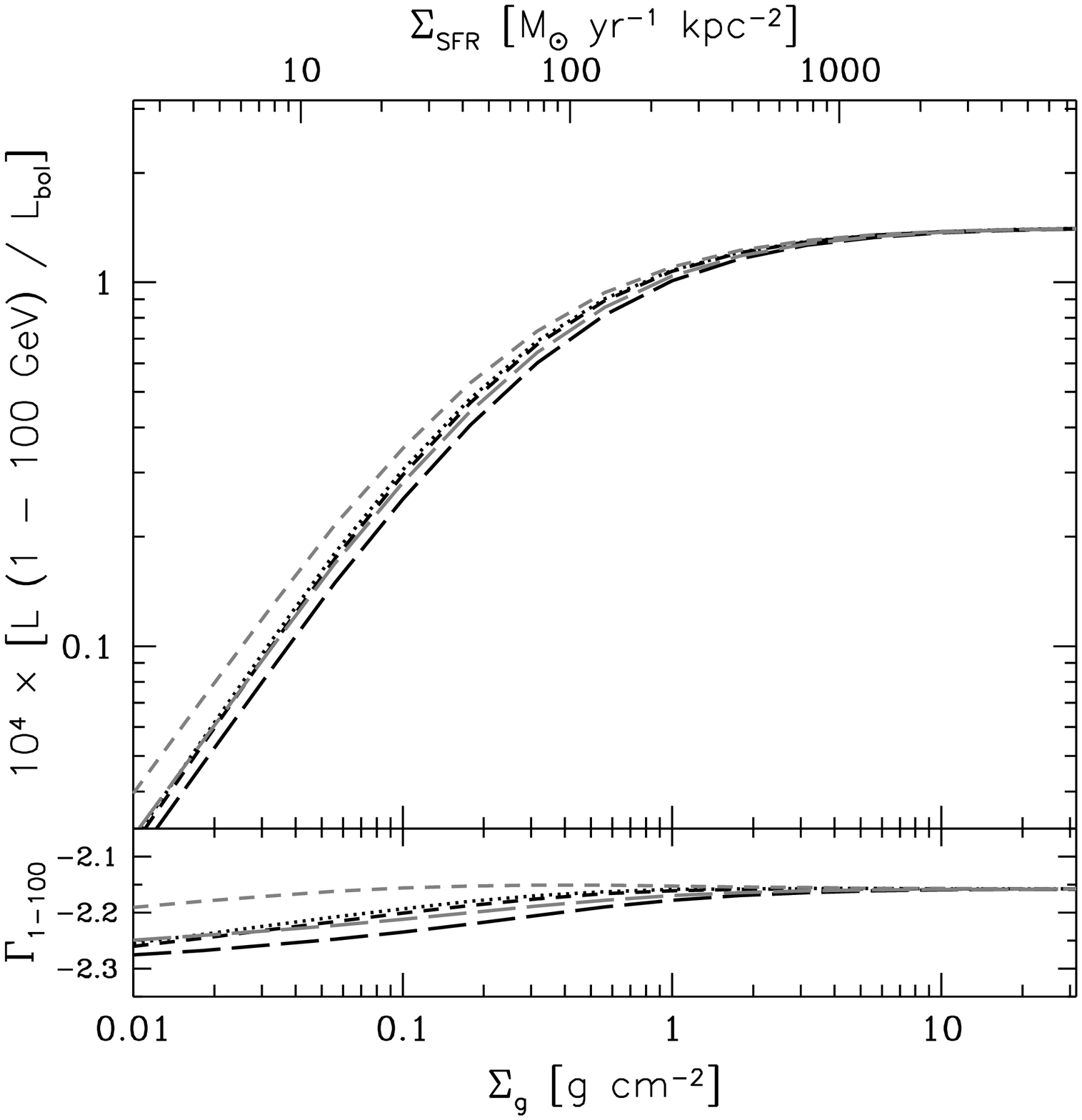}}
\figcaption{Properties of the synchrotron radio (1.4 GHz) emission (left) and gamma-ray emission (right) in my models.  These models assume that the gas consumption time is 20 Myr in all starbursts.  Spectral indices are plotted at the bottom.\label{fig:FRCt20}\label{fig:GeVt20}}
\end{figure*}

Figure~\ref{fig:FRCt20} is a plot of the predicted FRC (left) and gamma-ray--SFR correlation (right) in starbursts with $\tau_{\rm gas} = 20\ \Myr$.  There is considerable variation in $\nu L_{\nu} (1.4\ \GHz) / L_{\rm bol}$ for different assumptions about the ISM phase and the outer scale of turbulence.  At $\Sigma_{\rm SFR} \ga 25\ \Msun\ \yr^{-1}\ \kpc^{-2}$ both the hot $B$ with $\ell_{\rm outer} = h$ and the cold $B$ with $\ell_{\rm outer} = R_{\rm max}$ predict the FRC within a factor of 2.  In starbursts with smaller $\Sigma_{\rm SFR}$, advection carries out $e^{\pm}$ before they radiate much radio emission, so the FRC is predicted to be broken in all models.  The gamma-ray emission likewise starts off weak at small $\Sigma_{\rm SFR}$ because of advection but grows to a constant fraction of the bolometric luminosity as proton calorimetry sets in.  Another universal feature of the models is that the radio/SFR ratio peaks at $\Sigma_{\rm SFR} \approx 150$ -- $500\ \Msun\ \yr^{-1}\ \kpc^{-2}$.  This is because the ratio of density to $\Sigma_{\rm SFR}$ remains constant for a constant $\tau_{\rm gas}$, instead of decreasing for the K98 Schmidt law in which $\rho \propto \Sigma_{\rm SFR}^{0.7}$.  In addition, when $\ell_{\rm outer}$ shrinks with $\Sigma_{\rm SFR}$, $B$ grows more weakly than $\sqrt{\Sigma_{\rm SFR}}$.  Because of these effects, bremsstrahlung and ionization losses grow ever more important with $\Sigma_{\rm SFR}$.  Finally, the radio spectra grow flatter with increasing $\Sigma_{\rm SFR}$ \citep{Thompson06}.  This is observed \citep{Condon91,Clemens08,Williams10}; whether it is caused by these cooling processes or free-free absorption is debatable \citep{Clemens10,Leroy11,Lacki13-LowNu}.

\begin{figure}
\centerline{\includegraphics[width=8cm]{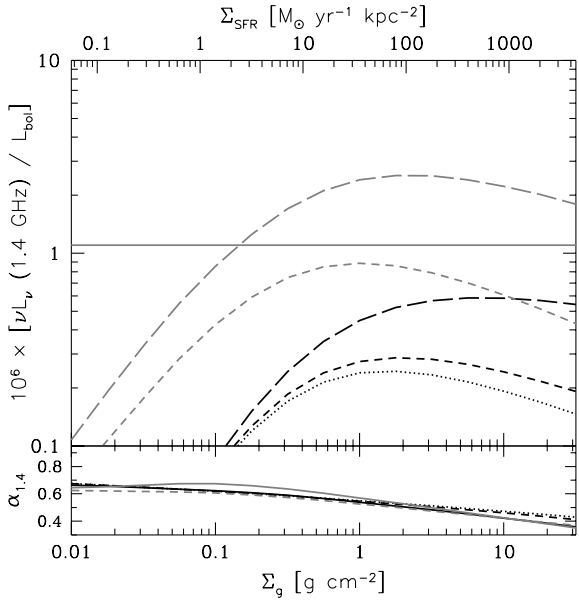}}
\figcaption{Properties of the synchrotron radio (1.4 GHz) emission if the K98 Schmidt law applies, as assumed in \citet{Lacki10-FRC1}.  Spectral indices are plotted at the bottom.\label{fig:FRCK98}}
\end{figure}

In contrast, when I assume the K98 Schmidt law applies, it is harder to reproduce the observed FRC with supernova-driven turbulence (see Figure~\ref{fig:FRCK98}).  This is because in the K98 Schmidt law, $\Sigma_{\rm SFR}$ is much lower than found in starbursts like M82 and NGC 253 given their $\Sigma_g$.  As a result, there is less turbulent energy density, and very small $B$ for low $\Sigma_{\rm SFR}$.  Thus the FRC is broken by advection for the hot wind $B$ until $\Sigma_{\rm SFR} \ga 300\ \Msun\ \yr^{-1}\ \kpc^{-2}$.  Indeed, \citet{Lacki10-FRC1} noted this difficulty in reproducing the FRC with the K98 Schmidt Law in starbursts with winds.  At higher surface densities, the FRC stabilizes.  As in the $\tau_{\rm gas} = 20\ \Myr$ models, $\alpha_{1.4}$ decreases and the GeV emission per star-formation increases with $\Sigma_{\rm SFR}$.  

\subsection{How polarized is the synchrotron emission?}
Synchrotron emission is notable for being highly polarized in an ordered magnetic field.  But if the emission comes from isotropic turbulence, the polarization of radiation emitted varies on a sightline.  When a region is not even resolved, the fluctuating polarization on different sightlines cancel even further.  Thus, synchrotron radiation from highly turbulent starbursts appears depolarized.  Polarization can still be observed if the turbulence has a preferred direction throughout the starburst, perhaps through shearing \citep{Sokoloff98}. 

The simplest model of radio polarization is a ``cell'' model, where magnetic fields are coherent over cells of size $\ell_{\rm cell}$.  Different cells have uncorrelated polarizations.  The polarization along a sightline can be viewed as a random walk away from zero, with each step being the polarization from one cell.  On a sightline of length $s$, each cell contributes $\ell_{\rm cell}/s$ of the emission, and there are $s/\ell_{\rm cell}$ cells, so the degree of polarization is roughly $\sqrt{\ell_{\rm cell}/s}$.  Assuming the sightline length is the starburst radius and the cell size is the outer scale of turbulence, the polarization towards any resolved sightline through a starburst is of order $\sim 10$ -- $50\%$.

The polarization signal nearly vanishes if the starburst is unresolved, though.  In this case, each cell contributes only $\ell_{\rm cell}^3/V_{\rm SB}$ of the total emission and there are $V_{\rm SB}^3/\ell$ cells, leading to a polarization of $\sim \sqrt{\ell_{\rm cell}^3/V_{\rm SB}}$.  In M82 and NGC 253, the integrated polarization should be $\la 1\%$, and it essentially vanishes in Arp 220.

As discussed in the next section on Faraday rotation, the observed polarization is higher in the GCCMZ.

\subsection{Faraday rotation and the magnetic field structure}
Faraday rotation is the twisting of a radio wave's polarization through an ionized, magnetized medium.  It is a frequency-dependent effect, but can be parameterized by the rotation measure (RM) that depends on the electron density, sightline length, and magnetic field structure.  The necessary free electrons are present in the superwind and H II region, as well as the molecular medium through the process of CR ionization.  The magnetic field structure is not known well (see section~\ref{sec:Dynamoes}). \citet{Bhat13} finds, from simulations of MHD turbulence, that the rotation measures average to zero but have a dispersion in a turbulent magnetic field.  The dispersion is
\begin{equation}
\sigma_{\rm RM} = 0.81\ {\rm rad}\ {\rm m}^{-2} \bar{\sigma_{\rm RM}} \left(\frac{\mean{n_e}}{\cm^{-3}}\right) \left(\frac{B_{\rm eq}}{2\sqrt{3}\ \muGauss}\right) \frac{\sqrt{\ell_{\rm outer} s}}{\pc}.
\end{equation}
Here, $B_{\rm eq}$ is the magnetic field strength if $\epsilon_B = 1$, but note that $\epsilon_B < 1$ in \citet{Bhat13}.  The factor $\bar{\sigma_{\rm RM}}$ equals 0.4, and $s$ is the sightline length.

I take the starburst radius as the sightline length for the hot and cold ISM.  For the molecular medium, I assume that the ionization fraction is $x_e = 10^{-5}$, as expected from cosmic ray ionization \citep{Lacki12-GRDRs}.  Then the free electron density is $n_e = 10^{-5} n_H$.  In the Galactic Center, the CR ionization rate may be relatively low, however \citep{Crocker11-Wild}.  I list the expected dispersions in Faraday rotation measures in Table~\ref{table:HII-RMs}.

The $\sigma_{\rm RM}$ grow with $\Sigma_{\rm SFR}$, as both $n_e$ and $B$ increase.  The superwind has more free electrons, so its Faraday rotation measure is larger.  According to these Faraday measures, the typical sightline should become Faraday thick at frequencies $\nu_F = c \sqrt{\rm RM}$ of $\sim 700\ \MHz$ in the GCCMZ, and $\sim 5\ \GHz$ in NGC 253, M82, and Arp 220's nuclei.  At lower frequencies, the frequency dependence of the polarization angle becomes very complicated for the synchrotron emission occurring within the turbulent magnetic fields.  The rotation measure itself then fluctuates with wavelength \citep[e.g.,][]{Chi97,Sokoloff98}.\footnote{In these cases, rotation measure is defined as the derivative of polarization angle with wavelength squared \citep{Chi97,Sokoloff98}.}

Observations of the Faraday rotation measure in starburst are complicated by the large number of H II regions present in these areas.  Although the magnetic field structures are probably not much different than in the molecular medium, H II regions are both dense and almost completely ionized.  I estimate the rotation measure signals toward typical starburst H II regions in Table~\ref{table:HII-RMs}, using a sightline of 5 pc.  The regions become Faraday thick at $\nu \gg 10\ \GHz$.  Any weaker Faraday signal from the superwind or molecular medium behind a H II region is completely scrambled.  

Worse, the H II regions may have fairly large covering fractions, at least in weaker starbursts, despite filling a very small fraction of the volume.  H II regions are directly visible in free-free absorption towards the GCCMZ, and appear to cover roughly $\sim 1/2$ the region \citep{Brogan03,Nord06}.  The covering fraction of H II regions in M82 and NGC 253 is unknown.  According to a very preliminary fit I did in \citet{Lacki13-LowNu}, a free-free absorption dip in the low frequency spectrum of the M82 starburst measured by \citet{Adebahr13} is consistent with a covering fraction near 1.  However, in Arp 220, radio recombination lines indicate very dense H II regions with very small covering fraction \citep{RodriguezRico05}. The Faraday signal in Arp 220 may therefore be relatively pure.  

\begin{deluxetable*}{lllcccc}
\tablecaption{Predicted Faraday rotation measure dispersions}
\tablehead{\colhead{Starburst} & \colhead{Phase} & \colhead{Assumptions} & \colhead{$x_e$} & \colhead{s} & \colhead{$\sigma_{\rm RM}$} & \colhead{$\nu_F$} \\ & & & & \colhead{($\pc$)} & \colhead(${\rm rad}\ \meter^{-2}$) & \colhead{$\GHz$}}
\startdata
GCCMZ               & Hot  & $\ell_{\rm outer} = h$      & 1 & 100 & 7.5 & 0.82\\
                    &      & $\ell_{\rm outer} = \ell_1$ & 1 & 100 & 5.3 & 0.69\\
                    &      & $\ell_{\rm outer} = \ell_2$ & 1 & 100 & 5.6 & 0.71\\
                    & Cold & $\ell_{\rm outer} = h$; $\Psi = 1$               & $10^{-5}$ & 100 & 10.5 & 0.97\\
                    &      & $\ell_{\rm outer} = R_{\rm max}$; $\Psi = 1$     & $10^{-5}$ & 100 & 1.6  & 0.38\\
                    &      & $\ell_{\rm outer} = R_{\rm max}$; ``Natural'' $\Psi$ & $10^{-5}$ & 100 & 1.4  & 0.36\\
                    & H II & $\ell_{\rm outer} = h$; $\Psi = 1$               & 1         & 5 & $2.3 \times 10^5$ & 150\\
                    &      & $\ell_{\rm outer} = R_{\rm max}$; $\Psi = 1$     & 1 & 5 & $3.6 \times 10^4$     & 57\\
                    &      & $\ell_{\rm outer} = R_{\rm max}$; ``Natural'' $\Psi$ & 1 & 5 & $3.2 \times 10^4$     & 53\\
\hline
NGC 253             & Hot  & $\ell_{\rm outer} = h$      & 1 & 150 & 940  & 9.2\\
                    &      & $\ell_{\rm outer} = \ell_1$ & 1 & 150 & 280  & 5.0\\
                    &      & $\ell_{\rm outer} = \ell_2$ & 1 & 150 & 360  & 5.6\\
                    & Cold & $\ell_{\rm outer} = h$; $\Psi = 1$               & $10^{-5}$ & 150 & 13  & 1.1\\
                    &      & $\ell_{\rm outer} = R_{\rm max}$; $\Psi = 1$     & $10^{-5}$ & 150 & 1.7 & 0.39\\
                    &      & $\ell_{\rm outer} = R_{\rm max}$; ``Natural'' $\Psi$ & $10^{-5}$ & 150 & 1.5 & 0.37\\
                    & H II & $\ell_{\rm outer} = h$; $\Psi = 1$               & 1 & 5 & $2.4 \times 10^5$ & 150\\
                    &      & $\ell_{\rm outer} = R_{\rm max}$; $\Psi = 1$     & 1 & 5 & $3.1 \times 10^4$ & 53\\
                    &      & $\ell_{\rm outer} = R_{\rm max}$; ``Natural'' $\Psi$ & 1 & 5 & $2.8 \times 10^4$ & 50\\
\hline
M82                 & Hot  & $\ell_{\rm outer} = h$      & 1 & 300 & 1300 & 11\\
                    &      & $\ell_{\rm outer} = \ell_1$ & 1 & 300 & 390  & 5.9\\
                    &      & $\ell_{\rm outer} = \ell_2$ & 1 & 300 & 510  & 6.8\\
                    & Cold & $\ell_{\rm outer} = h$; $\Psi = 1$               & $10^{-5}$ & 300 & 34  & 1.7\\
                    &      & $\ell_{\rm outer} = R_{\rm max}$; $\Psi = 1$     & $10^{-5}$ & 300 & 4.1 & 0.61\\
                    &      & $\ell_{\rm outer} = R_{\rm max}$; ``Natural'' $\Psi$ & $10^{-5}$ & 300 & 3.6 & 0.57\\
                    & H II & $\ell_{\rm outer} = h$; $\Psi = 1$               & 1 & 5 & $4.4 \times 10^5$ & 200\\
                    &      & $\ell_{\rm outer} = R_{\rm max}$; $\Psi = 1$     & 1 & 5 & $5.3 \times 10^4$ & 69\\
                    &      & $\ell_{\rm outer} = R_{\rm max}$; ``Natural'' $\Psi$ & 1 & 5 & $4.7 \times 10^4$ & 65\\

\hline
Arp 220 Nuclei      & Hot  & $\ell_{\rm outer} = h$      & 1 & 100 & $4.0 \times 10^5$ & 190\\
                    &      & $\ell_{\rm outer} = \ell_1$ & 1 & 100 & $3.8 \times 10^4$ & 59\\
                    &      & $\ell_{\rm outer} = \ell_2$ & 1 & 100 & $5.8 \times 10^4$ & 72\\
                    & Cold & $\ell_{\rm outer} = h$; $\Psi = 1$               & $10^{-5}$ & 100 & 7100 & 25\\
                    &      & $\ell_{\rm outer} = R_{\rm max}$; $\Psi = 1$     & $10^{-5}$ & 100 & 300  & 5.2\\
                    &      & $\ell_{\rm outer} = R_{\rm max}$; ``Natural'' $\Psi$ & $10^{-5}$ & 100 & 300  & 5.2\\
                    & H II & $\ell_{\rm outer} = h$; $\Psi = 1$               & 1 & 5 & $1.6 \times 10^8$ & 3800\\
                    &      & $\ell_{\rm outer} = R_{\rm max}$; $\Psi = 1$     & 1 & 5 & $6.6 \times 10^6$ & 770\\
                    &      & $\ell_{\rm outer} = R_{\rm max}$; ``Natural'' $\Psi$ & 1 & 5 & $6.6 \times 10^6$ & 770
\enddata
\label{table:HII-RMs}
\tablenotetext{}{The H II region calculations assumes that $\mean{\rho}$, $\sigma$, $\ell_{\rm outer}$, and $B$ are the same as in the cold case.  The only difference is in the ionization fraction and sightline length.}
\end{deluxetable*}

We can compare these Faraday rotation dispersions to a few observations, particularly in the well-studied GCCMZ.  Sightlines through the GCCMZ have very large RMs, reaching $\sim 3000\ \radm2$ on some sightlines \citep[e.g.,][]{Roy05,Law11}.  These values are completely inconsistent with my predictions for the superwind and molecular medium, but they are smaller than the predicted H II region RMs.  However, the observed rotation measures evince a large-scale regular field on scales $\ga 150\ \pc$, with a twisted poloidal structure \citep{Law11}.  The synchrotron emission from the mysterious nonthermal filaments within the GCCMZ is polarized, with both the filaments themselves and their polarization aligned perpendicular to the Galactic plane, again supporting a poloidal magnetic field \citep{YusefZadeh84,Tsuboi86}.  

Other starbursts display evidence for large-scale fields too.  Using Faraday rotation measurements, \citet{Heesen11} discovered a helical magnetic field within the wall of NGC 253's starburst outflow (although not in the starburst itself) using RMs.  In M82 itself, polarized submillimeter emission indicates the alignment of dust with a large scale magnetic field \citep{Greaves00,Jones00}.  Regular magnetic fields are completely unexpected in a generic turbulence model, although the wind introduces a preferred direction.  A preferred direction in the density distribution of gas was invoked by \citet{Boldyrev06}, who propose the Galactic Center filaments are structures made by supernova-driven turbulence, to explain the alignment of the filaments.

\citet{Robishaw08} observed Faraday rotation towards OH megamasers within Arp 220 and other ULIRGs.  While the significance is very weak in Arp 220, the mean RM $\sim 1250\ \radm2$ is roughly expected from the molecular medium.  On the other hand, \citet{Robishaw08} found much larger RMs $\gg 10^4\ \radm2$ in III Zw 35; these are hard to explain.

\subsection{The magnetic fields within starburst SNRs}
High resolution radio observations have revealed populations of SNRs in starbursts (Section~~\ref{sec:SNREvolution}).  By assuming that equipartition between CRs and magnetic fields holds, one can estimate the magnetic fields within the SNRs from their radio flux.  These estimates imply that magnetic field strengths are greater than the ambient ISM \citep[e.g.,][]{Batejat11}.

Indeed, we expect for theoretical reasons that the post-shock magnetic field in a SNR is greater than the ambient ISM magnetic field.  First, the shock itself compresses the incoming gas and its frozen-in magnetic fields, increasing $B$ by a factor $f_{\rm comp} \sim 3$--7 \citep{vanDerLaan62,Reynolds81,Volk02}:
\begin{equation}
B_{\rm SNR}^{\rm comp} = f_{\rm comp} B_{\rm ISM}.
\end{equation}
Second, plasma instabilities lead to magnetic field amplification in the post-shock region.  Through this process, the shock converts some fraction $\epsilon_{\rm amp} \approx 0.01$ of the upstream material's kinetic energy density (in the shock frame) into magnetic fields \citep[e.g.,][]{Lucek00,Berezhko04}:
\begin{equation}
B_{\rm SNR}^{\rm amp} = \sqrt{8 \pi \epsilon_{\rm amp} \rho_{\rm ext} v_s^2}
\end{equation}
where $v_s$ is the SNR shock speed and $\rho_{\rm ext}$ is the density of the material the shock is expanding into.  Note that $\rho_{\rm ext}$ is not necessarily the mean density of the entire starburst ($\mean{\rho}_{\rm SB}$) -- the SNR sits in some ISM phase that can be overdense or underdense, and there can be fluctuations in the phase's density (c.f. Section~\ref{sec:ColdTurbulence}).  Which of these processes sets the SNR magnetic field strength in starbursts is a matter of debate \citep{Thompson09,Chomiuk09,Batejat11}.

I address this question using the $B$ estimates developed in this paper.  Suppose the SNR goes off in a phase with average density $\mean{\rho}_{\rm phase}$.  Then, the ambient magnetic field energy density is $(\epsilon_B/2) \mean{\rho}_{\rm phase} \sigma^2$.  By analogy with Section~\ref{sec:ColdTurbulence}, we can define the ratio of the ISM density locally surrounding the SNR with the average density of that ISM phase: $\Delta \equiv \rho_{\rm ext} / \mean{\rho}_{\rm phase}$.  For the hot wind plasma, which has subsonic or transonic turbulence, $\Delta \approx 1$.  In the cold molecular gas, where the turbulence has high Mach number, the median $\Delta$ could be in the range $0.01$--$1$, depending on the unknown density distribution of gas \citep{Padoan97,Ostriker01,Hopkins13-rhoDist}.  Then the ratio of the energy densities in the compressed magnetic field and in the amplified magnetic field are:
\begin{align}
\nonumber \frac{U_B^{\rm comp}}{U_B^{\rm amp}} & = \frac{f_{\rm comp}^2 \epsilon_B}{2 \epsilon_{\rm amp} \Delta} \left(\frac{\sigma}{v_S}\right)^2\\
& = 800 \left(\frac{f_{\rm comp}}{4}\right)^2 \left(\frac{\epsilon_{\rm amp}}{0.01}\right)^{-1} \left(\frac{\epsilon_B}{\Delta}\right) \left(\frac{\sigma}{v_S}\right)^2.
\end{align}

The most important factor controlling whether the SNR magnetic field is dominated by compressed or amplified magnetic fields is the ratio of $\sigma$ and $v_S$.  For young SNRs, the shock speed is very large and typically of order $10^4\ \kms$.  But the ISM turbulent speed varies by a factor $>10$ between the hot superwind ($\sigma \sim 1000\ \kms$) and the cold molecular gas ($\sigma \sim 50\ \kms$).  Thus, which mechanism is most important depends on SNR location within a starburst.  In the hot superwind, compression of magnetic fields dominates: $U_B^{\rm comp} / U_B^{\rm amp} \approx 8$.  In cold molecular gas, post-shock amplification wins: $U_B^{\rm comp} / U_B^{\rm amp} \approx 0.02$.

In a hot starburst, the typical supernova goes off in the volume-filling superwind.  The majority of SNRs therefore have radio fluxes set by the compression of hot wind magnetic fields.  However, the \emph{brightest} SNRs with the strongest magnetic fields probably are located within the overdense molecular clouds, where post-shock amplification operates \citep[as suggested by][]{Thompson09}.  The implication is that the radio luminosity function of SNRs in starbursts arises from a complicated distribution of environments; why the observed radio luminosity function seems to be the same between galaxies and starbursts is unclear \citep{Chomiuk09}.  In cold starbursts, amplification alone determines SNR magnetic fields until the SNR shocks have slowed down below $5000\ \kms$, as long as $\Delta \ga 1$.

\begin{deluxetable*}{lllcccccc}
\tablecaption{Predicted magnetic fields in SNRs}
\tablehead{\colhead{Starburst} & \colhead{Phase} & \colhead{Assumptions} & \colhead{\multirow{2}{*}{$\displaystyle \frac{\rho_{\rm ext}}{\mean{\rho}_{\rm SB}}$}} & \colhead{$\Delta$} & \colhead{$B_{\rm SNR}^{\rm comp}$~\tablenotemark{a}} & \multicolumn{3}{c}{$B_{\rm SNR}^{\rm amp}$~\tablenotemark{b}} \\ & & & & & & \colhead{$v = 2500\ \kms$} & \colhead{$v = 5000\ \kms$} & \colhead{$v = 10000\ \kms$}}
\startdata
GCCMZ               & Hot  & $\ell_{\rm outer} = h$      & $4 \times 10^{-5}$ & 1 & 310$\ \muGauss$ & 21$\ \muGauss$  & 42$\ \muGauss$  & 84$\ \muGauss$ \\
                    &      & $\ell_{\rm outer} = \ell_1$ & $4 \times 10^{-5}$ & 1 & 270$\ \muGauss$ & 21$\ \muGauss$  & 42$\ \muGauss$  & 84$\ \muGauss$ \\
                    &      & $\ell_{\rm outer} = \ell_2$ & $4 \times 10^{-5}$ & 1 & 270$\ \muGauss$ & 21$\ \muGauss$  & 42$\ \muGauss$  & 84$\ \muGauss$ \\
                    & Cold & $\ell_{\rm outer} = h$; $\Psi = 1$               & 1    & 1 & 1.6$\ \mGauss$  & 3.2$\ \mGauss$  & 6.4$\ \mGauss$ & 13$\ \mGauss$\\
                    &      & $\ell_{\rm outer} = R_{\rm max}$; $\Psi = 1$     & 1    & 1 & 770$\ \muGauss$ & 3.2$\ \mGauss$  & 6.4$\ \mGauss$ & 13$\ \mGauss$\\
                    &      & $\ell_{\rm outer} = R_{\rm max}$; ``Natural'' $\Psi$ & 0.12 & 0.03 & 530$\ \muGauss$ & 1.1$\ \muGauss$ & 2.2$\ \mGauss$ & 4.4$\ \mGauss$\\
\hline
NGC 253             & Hot  & $\ell_{\rm outer} = h$      & 0.002 & 1 & 1.4$\ \mGauss$   & 100$\ \muGauss$ & 200$\ \muGauss$ & 400$\ \muGauss$\\
                    &      & $\ell_{\rm outer} = \ell_1$ & 0.002 & 1 & 850$\ \muGauss$  & 100$\ \muGauss$ & 200$\ \muGauss$ & 400$\ \muGauss$\\
                    &      & $\ell_{\rm outer} = \ell_2$ & 0.002 & 1 & 930$\ \muGauss$  & 100$\ \muGauss$ & 200$\ \muGauss$ & 400$\ \muGauss$\\
                    & Cold & $\ell_{\rm outer} = h$; $\Psi = 1$               & 4    & 1 & 3.8$\ \mGauss$ & 2.1$\ \mGauss$  & 4.2$\ \mGauss$ & 8.5$\ \mGauss$\\
                    &      & $\ell_{\rm outer} = R_{\rm max}$; $\Psi = 1$     & 4    & 1 & 1.7$\ \mGauss$ & 2.1$\ \mGauss$  & 4.2$\ \mGauss$ & 8.5$\ \mGauss$\\
                    &      & $\ell_{\rm outer} = R_{\rm max}$; ``Natural'' $\Psi$ & 0.12 & 0.03 & 1.2$\ \mGauss$ & 730$\ \muGauss$ & 1.5$\ \mGauss$ & 2.9$\ \mGauss$\\
\hline
M82                 & Hot  & $\ell_{\rm outer} = h$      & 0.001 & 1 & 1.3$\ \mGauss$   & 100$\ \muGauss$ & 210$\ \muGauss$ & 410$\ \muGauss$\\
                    &      & $\ell_{\rm outer} = \ell_1$ & 0.001 & 1 & 800$\ \muGauss$  & 100$\ \muGauss$ & 210$\ \muGauss$ & 410$\ \muGauss$\\
                    &      & $\ell_{\rm outer} = \ell_2$ & 0.001 & 1 & 890$\ \muGauss$  & 100$\ \muGauss$ & 210$\ \muGauss$ & 410$\ \muGauss$\\
                    & Cold & $\ell_{\rm outer} = h$; $\Psi = 1$               & 4    & 1 & 3.9$\ \mGauss$ & 2.8$\ \mGauss$  & 5.6$\ \mGauss$ & 11$\ \mGauss$\\
                    &      & $\ell_{\rm outer} = R_{\rm max}$; $\Psi = 1$     & 4    & 1 & 1.7$\ \mGauss$ & 2.8$\ \mGauss$  & 5.6$\ \mGauss$ & 11$\ \mGauss$\\
                    &      & $\ell_{\rm outer} = R_{\rm max}$; ``Natural'' $\Psi$ & 0.12 & 0.03 & 1.2$\ \mGauss$ & 970$\ \muGauss$ & 1.9$\ \mGauss$ & 3.9$\ \mGauss$\\
\hline
Arp 220 Nuclei      & Hot  & $\ell_{\rm outer} = h$      & 0.002 & 1 & 12$\ \mGauss$  & 800$\ \muGauss$ & 1.6$\ \mGauss$ & 3.2$\ \mGauss$\\
                    &      & $\ell_{\rm outer} = \ell_1$ & 0.002 & 1 & 4.5$\ \mGauss$ & 800$\ \muGauss$ & 1.6$\ \mGauss$ & 3.2$\ \mGauss$\\
                    &      & $\ell_{\rm outer} = \ell_2$ & 0.002 & 1 & 5.3$\ \mGauss$ & 800$\ \muGauss$ & 1.6$\ \mGauss$ & 3.2$\ \mGauss$\\
                    & Cold & $\ell_{\rm outer} = h$; $\Psi = 1$               & 1    & 1 & 33$\ \mGauss$  & 18$\ \mGauss$  & 37$\ \mGauss$  & 74$\ \mGauss$\\
                    &      & $\ell_{\rm outer} = R_{\rm max}$; $\Psi = 1$     & 1    & 1 & 9.2$\ \mGauss$ & 18$\ \mGauss$  & 37$\ \mGauss$  & 74$\ \mGauss$\\
                    &      & $\ell_{\rm outer} = R_{\rm max}$; ``Natural'' $\Psi$ & 0.01 & 0.01 & 5.8$\ \mGauss$ & 1.8$\ \mGauss$ & 3.7$\ \mGauss$ & 7.4$\ \mGauss$
\enddata
\tablenotetext{a}{Assumes that $f_{\rm comp} = 4$.  I use the $B_{\rm turb}$ values listed in Tables~\ref{table:BHotPredictions} and~\ref{table:BColdPredictions}.}
\tablenotetext{b}{Assumes that $\epsilon_{\rm amp} = 0.01$.  I also assume the SNR shock is in the ambient ISM and not in a stellar wind bubble.}
\label{table:BSNRPredictions}
\end{deluxetable*}

I have listed the predicted $B_{\rm SNR}$ for both magnetic field compression and amplification in Table~\ref{table:BSNRPredictions}.  The SNR magnetic fields are expected to be intense, ranging from a few hundred $\muGauss$ in the GCCMZ to nearly 0.1 G in the youngest SNRs in Arp 220.  The magnetic fields are typically stronger in the cold molecular gas by an order of magnitude.  The magnetic field strengths in Arp 220's SNRs, which are observed to have expansion speeds of $\sim 5000\ \kms$ \citep{Batejat11}, are roughly $40\ \mGauss$.  This is essentially the equipartition estimate of \citet{Batejat11} from the SNRs' radio fluxes.  

\section{Conclusion}
\label{sec:Conclusion}
The power of supernovae and stellar winds can drive the chaos of the interstellar medium in starbursts.  The mechanical luminosity couples efficiently to the gas, stirring it and driving turbulence.  The high speed random motions amplify magnetic fields in all phases of the ISM.  Supernovae shocks accelerate CRs, which both provide a source of ionization and are observational tracers of the magnetic fields that result from the stirring.

In low density, low pressure (``hot'') starbursts, the expanding supernova remnants successfully sweep up most of the ISM and leave behind rarefied, hot plasma that erupts as a wind.  Further supernova explosions both provide needed mass to sustain the wind and, I argue, turbulent driving in the hot plasma.  The winds are not thermalized gases as usually pictured, but are typically collisionless on parsec scales.  The conditions in the wind are analogous to galaxy clusters, where turbulence and magnetic fields are well studied.  Although the hydrodynamical Reynolds number is very low, plasma processes could decrease the viscosity and permit turbulence.  I show that the hot wind's turbulence reaches speeds of $\sim 1000\ \kms$ (${\cal M} \approx 1$).  The magnetic field strength grows through the fluctuation dynamo, reaching $\sim 80\ \muGauss$ in the GCCMZ and $\sim 300\ \muGauss$ in M82 and NGC 253.  The dissipation of the turbulence plausibly provides the efficient heating that sustains the wind.  

High pressure (``cold'') starbursts confine SNRs and are filled with dense molecular gas (Figure~\ref{fig:ISMSketches}).  Although expanding supernova remnants lose much of their mechanical energy through radiation in molecular clouds and in cold starbursts, they still are able to drive turbulence of speeds $\sim 20\ \kms$.  The equipartition magnetic fields are a few times higher than those predicted in the superwind.  In the extreme conditions of Arp 220's nuclei, I predict turbulent magnetic fields with strengths of $\sim 2\ \mGauss$.  Similar levels of turbulence and magnetic field strengths should hold for warm neutral gas and H II regions.

I argue that CRs in hot starbursts spend most of their time in the hot wind.  The observed synchrotron emission largely reflects the magnetic fields in these winds.  From time to time, they dive into dense molecular gas where they emit gamma rays \citep[c.f.,][]{YoastHull13}. 

The importance of supernova mechanical power in starbursts is reflected in the near equipartition of energy densities in starbursts.  Turbulence, magnetic fields, CRs, and the thermal energy of the superwind all are powered by supernovae.  Since their residence time is typically of order the hot wind sound crossing time of the starburst, they naturally tend to equipartition.  In the special case of CRs, equipartition should roughly hold in hot starbursts, but it likely fails in dense starbursts when their residence time is instead set by the pion loss time. 

The strong levels of turbulence and the small outer scales possible in starbursts should confine CRs very effectively if the turbulent cascade reaches down to small scales.  In fact, the motion of the ISM may be more important than the CR diffusion within the ISM.  Thus, CR propagation may be Lagrangian in starbursts instead of Eulerian as in the Milky Way.  Even if there is no way of confining CRs, the CRs are forced to follow the tortuous paths of the turbulent magnetic field lines themselves, giving a FLRW diffusion constant.  Thus, turbulence rules how CRs propagates in starbursts.

In Section~\ref{sec:Implications}, I considered some observational implications of this scenario.  First, X-ray lines from starburst wind should display turbulent broadening roughly comparable to their thermal broadening.  Second, I demonstrated that the predicted turbulent magnetic fields are roughly consistent with the FIR-radio correlation.  Third, if the magnetic field is indeed mostly turbulent, the synchrotron emission should be depolarized, especially when integrated over the entire starburst.  Fourth, I calculate Faraday rotation measures for each of the phases of the starburst ISM.  Fifth, there should be milliGauss magnetic fields in starburst SNRs; magnetic field compression sets $B$ in the hot superwind, whereas post-shock amplification sets it in cold molecular gas.  Finally, as I noted in section~\ref{sec:TurbulentMixing}, if turbulent advection is the main mode of CR transport, then it may result in fluctuations in the synchrotron emission with no obvious correlate.

Many questions remain about how turbulence, CRs, and magnetic fields interplay in starburst environments.
\begin{itemize}
\item First, what is the true ISM structure of starbursts?  Does hot superwind plasma fill weaker starbursts like M82 and the GCCMZ?  Does cold molecular gas really fill ULIRGs?  I argued that CRs fill most of the volume of the starbursts, and their radio synchrotron emission indicates the magnetic field strength in the volume-filling phase.  Is that true?  

\item What is the nature of turbulence in the hot superwind plasma, if it is present?  Are there really energy cascades down in spatial scales?  Does the kinetic energy end up as heat?  If so, the microphysics of turbulence may leave subtle observational signs, such as heating ions and electrons at different rates.  For example, maybe turbulently heated electrons explains the anomalous hard X-ray continuum emission of M82 \citep{Strickland07}.  

\item I evaluated the turbulence and its equipartition magnetic fields in superwinds with properties from CC85 solution.  \emph{Chandra} X-ray observations of M82 suggest the solution is adequate for real starbursts \citep{Strickland09}.   But formally speaking, the CC85 solution treats the superwind as an ideal gas and assumes no magnetic fields, cosmic rays, or turbulence.  There is need to simulate the superwind plasma self-consistently.  Do collisionless effects or strong magnetic fields alter its properties?

\item How does the turbulent dynamo operate in starbursts?  While there have been numerous MHD simulations of turbulent dynamoes, starbursts differ in that they have large-scale advective winds.  Winds twist motions, so even the purely compressive motions from supernovae explosions may be partly transformed into solenoidal motions as they are blasted by the wind.  A simulation of that effect should be illuminating.

\item What is the role of intermittence in starburst turbulence?  What sort of magnetic field structures does it produce?  Is the assumption of a single magnetic field strength in each phase adequate for models?  Does it lead to unsteady CR destruction and time variability in nonthermal emission?

\item What is the spectrum of magnetic fluctuations?  Is most of the energy concentrated at large scales (as in the Kolmogorov spectrum)?  If so, then Faraday rotation measurements may prove useful.  Does the spectrum fall slowly enough to be very good at trapping CRs (as in the Kolmogorov and Kraichnan cases)?  

\item What is the role of turbulent mixing of gas/plasma in transporting CRs?  It would be interesting to construct ``Lagrangian'' models of CR transport in starbursts, as opposed to the ``Eulerian'' models we have now like GalProp.  Are there order unity fluctuations in the CR energy density that arise simply due to turbulent eddies?  These would be entirely uncorrelated with star-formation sites.  In supersonic cold gas, if the CRs are frozen into the gas, they essentially live for one eddy-crossing time, before being destroyed when advected into a turbulent clump.  How do such transient phenomena square with current steady-state models of CR populations?

\item What happens to the turbulent cascade at small scales in cold molecular gas?  Are there enough magnetic fluctuations to freeze CRs into the gas?  If not, what traps the CRs in ULIRGs -- or do they stream away entirely?  The situation is analogous to that for MeV positrons in the Milky Way \citep{Higdon09,Jean09,Prantzos11}.  Observations of CR propagation in Galactic molecular clouds, especially those in the GCCMZ, could prove helpful.  

\item Although I discuss turbulence in separate ``phases'' of the ISM, the reality is probably a continuum of physical conditions \citep[e.g.,][]{Norman96}.  How does turbulence work in the chaotically varying conditions of the ISM?  Are CRs transported between regions of differing physical conditions, and if so, how does that mixing between ``phases'' work?  \citet{Crocker11-Wild} argued that CRs in the GCCMZ do not generally penetrate into dense molecular clouds, based on the gamma-ray faintness of the region.  Yet modeling of M82 and NGC 253's gamma-ray brightness is consistent with CRs experiencing average density material \citep{Lacki11-Obs}.  What microphysics is behind this difference, and how does the relationships between the ``phases'' come in?

\item How analogous is turbulence in $z \approx 0$ starbursts to turbulence in high-redshift disk galaxies?  Is turbulence in high-redshift galaxies powered by supernovae?  Is the ISM in those galaxies hot or cold?  What is the outer scale of turbulence in those galaxies?   
\end{itemize}

Turbulence shapes nearly everything about starburst ISMs, and understanding its role will require synthesizing observations of very different kinds and theories of different processes.  Radio observations of synchrotron emission tell us about the turbulent magnetic fields in starbursts, and with high enough resolution, the propagation of CRs.  VLBI monitoring of radio supernovae illuminate the sources of the turbulence, the ISM they go off in, and the CRs they accelerate.  Observations in molecular lines trace the distribution and kinematics of the stirred molecular gas.  With infrared emission, we see the sites of star-formation.  The superwind glows in hard X-rays, and X-ray spectroscopy may tell us about its kinematics.  Finally gamma rays (and eventually neutrinos) tell us about the CR protons and their interaction with dense gas.  On the theory side, a complete theory of starburst turbulence will draw upon fluid dynamics, plasma physics, magnetic dynamo theory, theories of interstellar feedback, and the physics of CRs.  Many years of discoveries and further questions await in this area.

\acknowledgments
I was supported by a Jansky Fellowship for this work from the National Radio Astronomy Observatory.  The National Radio Astronomy Observatory is operated by Associated Universities, Inc., under cooperative agreement with the National Science Foundation.  The motivation for this paper largely arose from discussions with Todd Thompson while I was at Ohio State University.  An early version of this work was delivered at the 2012 Sant Cugat Forum for Astrophysics; I wish to acknowledge the participants.  I also thank Jim Condon, Jean Eilek, Josh Marvil, Mark Krumholz, Roger Chevalier, Casey Law, and Chiara Ferrari for discussions and comments.

\begin{appendix}
\section{Bulk properties of the hot superwind in CC85}
\label{sec:CC85Properties}
The theory of supernovae-driven superwinds has been thoroughly developed \citep[see][]{Chevalier85,Silich04,Strickland09}.  The characteristics of the superwind are given by the rate of hot matter being injected into it ($\dot{M}$) and the energy pumped into the wind ($\dot{E}_{\rm mech}$) by stellar winds and supernovae.  \citet{Strickland09} gives the mass injection rate from supernovae and stellar winds as 
\begin{equation}
%Exact number: 0.117
\dot{M} = 0.12\ \Msun\ \yr^{-1}\ \left(\frac{\rm SFR}{\Msun\ \yr^{-1}}\right)
\end{equation}
and the energy injection rate as 
\begin{equation}
\label{eqn:EDot}
%Exact number: 2.535
\dot{E}_{\rm mech} = 2.5 \times 10^{41} \ergps\ \left(\frac{\rm SFR}{\Msun\ \yr^{-1}}\right).
\end{equation}
For these equations, I have converted to a Salpeter IMF from 0.1 to 100~$\Msun$, as used by \citet{Kennicutt98}, using ${\rm SFR} (\ge 1\ \Msun) = 0.39 \times {\rm SFR}$.  While starbursts may have a non-Salpeter IMF \citep[e.g.,][]{Papadopoulos11-SF,vanDokkum10}, both $\dot{M}$ and $\dot{E}_{\rm mech}$ are set by the formation rate of massive stars, for which the IMF should not vary; supernovae in particular are only generated by stars with masses above $\sim 8\ \Msun$.  These massive stars are also the power sources of the radiation used as star-formation indicators (such as the bolometric luminosity, ultraviolet light, radio continuum, and radio recombination lines), so for star-formation rates derived from these tracers, the derived $\dot{E}_{\rm mech}$ and $\dot{M}$ of a starburst should not depend on IMF, as long as it consistent throughout \citep[c.f.,][]{Condon92}.

For the physical conditions in the superwind, I use the simple model of CC85, which assumes a spherical geometry with no gravity from the starburst and no radiative cooling from the wind.  The star-formation is assumed to be entirely contained and uniform within a radius $R$.  Elaborations on this model have been given to include cooling \citep{Silich04}, but the basic arguments should remain the same.  After adjusting for geometry \citep{Strickland09}, the central density of a CC85 superwind is parameterized by an effective radius $R_{\rm eff} = f_{\rm geom} R / \sqrt{2}$, where $f_{\rm geom} = \sqrt{1 + 2 h / R}$ for a thin disk (with total surface area $A = 2 \pi R^2 + 4 \pi R h$) and $\sqrt{2}$ for a sphere (with total surface area $4 \pi R^2$).  The central density of the superwind in CC85 is
\begin{equation}
\label{eqn:rhoC}
%Exact value: 1.85982 = 0.296 * 2 * pi
\rho_c = 1.860 \frac{\zeta \beta^{3/2}}{f_{\rm geom}^2 \epsilon_{\rm therm}^{1/2}} \frac{\dot{M}^{3/2}}{\dot{E}_{\rm mech}^{1/2} A_{\rm proj}}
\end{equation}
where $\beta$ is the mass loading, $\zeta$ is the participation fraction, $\epsilon_{\rm therm}$ is the supernova thermalization efficiency, and $A_{\rm proj} = \pi R^2$ is the projected area of the star-forming region.  This is for an ideal gas with $\gamma = 5/3$.  Note that the density is directly proportional to the mean surface density of star-formation, $\Sigma_{\rm SFR} = {\rm SFR} / A_{\rm proj}$.  From observations of diffuse, hard X-ray emitting gas in M82, \citet{Strickland09} find that $\beta \approx 2$, $\zeta \approx 1$, and $\epsilon_{\rm therm} \approx 0.75$.  The central number density of electrons, $n_e = n_H + 2 n_{\rm He} = X \rho / m_H + Y \rho / m_{\rm He}$, is 
\begin{equation}
\label{eqn:nC}
%Exact value: 0.0132626
n_e = 0.013\ \cm^{-3} \frac{\zeta}{f_{\rm geom}^2}\ \left(\frac{\beta}{2}\right)^{3/2} \left(\frac{\epsilon_{\rm therm}}{0.75}\right)^{-1/2} \left(\frac{\Sigma_{\rm SFR}}{\Msun\ \yr^{-1}\ \kpc^{-2}}\right),
\end{equation}
for $X = 0.75$ and $Y = 0.25$.  The total particle density is $n = n_e + n_H + n_{\rm He}$:
\begin{equation}
\label{eqn:nAll}
%Exact value: 0.0255779
n = 0.026\ \cm^{-3} \frac{\zeta}{f_{\rm geom}^2}\ \left(\frac{\beta}{2}\right)^{3/2} \left(\frac{\epsilon_{\rm therm}}{0.75}\right)^{-1/2} \left(\frac{\Sigma_{\rm SFR}}{\Msun\ \yr^{-1}\ \kpc^{-2}}\right).
\end{equation}

In the CC85 model, the central temperature of the superwind is given by the ratio of the energy and mass injection rates, which should be the same for all starbursts:
\begin{equation}
\label{eqn:SuperwindTc}
%Exact value: 37.0393 MK
T_c = \frac{(\gamma - 1) \mu m_H \epsilon_{\rm therm} \dot{E}}{\gamma k_B \beta \dot{M}} = 3.7 \times 10^7\ \Kelv \left(\frac{\epsilon_{\rm therm}}{0.75}\right) \left(\frac{\beta}{2}\right)^{-1}
\end{equation}
I assume $\mu = 1/(2 X + 3 Y / 4) = 0.59$, appropriate for a completely ionized plasma, since the superwind is extremely hot.  The associated sound speed is also universal in the CC85 model:
\begin{equation}
\label{eqn:SoundSpeed}
%Exact value: 927.036 km/s
c_s = \sqrt{\frac{(\gamma - 1) \epsilon_{\rm therm} \dot{E}}{\beta \dot{M}}} = 930\ \kms\ \left(\frac{\epsilon_{\rm therm}}{0.75}\right)^{1/2} \left(\frac{\beta}{2}\right)^{-1/2}.
\end{equation}

Finally, the central thermal pressure in the CC85 model is $P_c = n k_B T$:
\begin{eqnarray}
%Exact value: 0.947387 MK / cm^3
P_c / k_B & = & 9.5 \times 10^{5}\ \Kelv\ \cm^{-3}\ \frac{\zeta}{f_{\rm geom}^2}\ \left(\frac{\epsilon_{\rm therm}}{0.75}\right)^{1/2} \left(\frac{\beta}{2}\right)^{1/2} \left(\frac{\Sigma_{\rm SFR}}{\Msun\ \yr^{-1}\ \kpc^{-2}}\right).
\end{eqnarray}

\section{Addressing objections to the existence of a volume-filling hot phase}
\label{sec:HotObjections}

\subsection{Do X-rays rule out a wind in the GCCMZ?}
The Galactic Ridge shines in seemingly-diffuse hard X-ray (2 - 10 keV) emission, and is especially bright in the GCCMZ \citep{Worrall82,Koyama89}.  While a diffuse X-ray hot plasma was considered a possible source, the required to keep the plasma hot enough is implausibly high \citep{Muno04}.  A deep Chandra image presented in \citet{Revnivtsev09} largely settled the matter by demonstrating that much of the X-ray emission, especially at higher energies ($6 - 8\ \keV$) actually comes from discrete sources.  But do these observations rule out a superwind from the CMZ \citep[c.f.,][]{Crocker12}?

If the basic theory of CC85 is correct, and the temperature of the superwind is $3.7 \times 10^7\ \Kelv$ (eqn.~\ref{eqn:SuperwindTc}), the free-free emission per unit volume in 2 - 10 keV X-rays is 
\begin{equation}
\label{eqn:HardXRayEmissivity}
%Exact value: 8.51329e-28, includes H and He
\varepsilon_{2-10} = 8.5 \times 10^{28} \erg\ \cm^{-3}\ \sec^{-1} f_{\rm geom}^{-4} \left(\frac{\Sigma_{\rm SFR}}{\Msun\ \yr^{-1}\ \kpc^{-2}}\right)^2.
\end{equation}
The surface density of star-formation in the GCCMZ is $2.2\ \Msun\ \yr^{-1}\ \kpc^{-2}$, and with a radius of $\sim 100\ \pc$, the typical sightline has length $s \sim 100\ \pc$.  That means that, if the superwind exists with a CC85-like density, the 2 -- 10 keV X-ray intensity observed at Earth from its free-free emission should be $s \varepsilon_{2-10} / (4 \pi) \approx 7.9 \times 10^{-12}\ \erg\ \cm^{-2}\ \sec^{-1}\ \deg^{-2}$.  This is just $9\%$ of the total 2 -- 10 keV emission from the region \citep{Revnivtsev09}.  For comparison, \citet{Revnivtsev09} calculate that about half of the total emission is resolved at $2 - 3\ \keV$, increasing to $\sim 90 \%$ at $6 - 8\ \keV$.  Note that the temperature of the superwind is 3.2 keV, so if the remaining unresolved emission is from the superwind we expect it to fall off at higher energies.

The observations of \citet{Revnivtsev09} therefore are consistent with the existence of a CC85 superwind, if the unresolved emission is partly free-free emission from the superwind: we would not expect the superwind to be as bright as the Galactic X-ray Ridge.  The weak constraints on the superwind go back to the energetics issue.  The Galactic X-ray Ridge is too bright to be powered by known sources, but a CC85 superwind necessarily can be powered by the supernovae in the CMZ \citep[c.f.,][]{Crocker12}.  

\subsection{Do X-ray observations rule out a volume-filling hot phase in Arp 220 and other ULIRGs?}
Although the huge pressures in dense starbursts likely prevent hot winds from filling their volumes, \citet{Murray10} went a step further and argued that the existence of the correlation between hard X-ray emission and star-formation rate ($L_{2-10} \approx 10^{-4} L_{\star}$; \citealt{David92,Lehmer10}) actually rules out hot superwinds in starbursts with gas surface densities $\Sigma_g \ga 0.15\ \gcm2$ (taking a temperature of $3.7 \times 10^7\ \Kelv$), comparable to M82.  However, they assumed that the thermal pressure of the wind was $\pi G \Sigma_g^2$; as I showed in Section~\ref{sec:UStarburst}, that estimate gives problematically high pressures for dense starbursts.

Using equation~\ref{eqn:HardXRayEmissivity} instead, I calculate a 2 -- 10 keV luminosity of $1.0 \times 10^8\ \Lsun$ from free-free emission of both nuclei of Arp 220, or $6 \times 10^{-5}\ L_{\rm TIR}$.  Since Arp 220 actually is X-ray faint compared to the X-ray--SFR correlation \citep{Iwasawa05,Lehmer10}, and since X-ray binaries should contribute an additional $10^{-4}\ L_{\rm TIR}$ \citep{Grimm03,Persic04}, this is problematic.  On the other hand, Arp 220's nuclei are Compton thick, reducing the X-ray emission by a factor of a few \citep{Downes98}.  I conclude there is some tension between the X-ray emission of Arp 220 and the prediction for a CC85, but not enough to actually rule out the superwind observationally.  For M82, where diffuse hard X-ray emission is observed directly, there is no problem with the X-ray emission.  I predict its superwind's diffuse 2 - 10 keV X-ray luminosity is $\sim 1 \times 10^5\ \Lsun$, only $2 \times 10^{-6} L_{\rm TIR}$ and just one tenth of the amount of diffuse, unresolved X-ray emission observed from M82's starburst \citep{Strickland07}. (The observed continuum may be brighter because (1) the wind's transonic turbulence makes it clumpy, and/or (2) the electron temperature is greater than $T_c$ because the ion-electron equilibrium time is greater than the advection time; I argued that both are plausible previously in the paper.)

The X-ray-SFR correlation \emph{does} become constraining for extreme starbursts that are very big, like those in some submillimeter galaxies (SMGs).  Whereas the free-free emission from the wind grows as $L_{\rm ff} \propto n_e^2 R^3 \propto \Sigma_{\rm SFR}^2 R^3$, the bolometric emission only grows as $L_{\rm TIR} \propto {\rm SFR} \propto \Sigma_{\rm SFR} R^2$.  Thus, free-free emission becomes more dominant for starbursts with high $\Sigma_{\rm SFR}$ and large radius: $L_{\rm ff} / L_{\rm TIR} \propto \Sigma_{\rm SFR} R$.  SMGs have typical radii of about two kiloparsecs \citep{Tacconi06} as opposed to the $\sim 100\ \pc$ scales of compact ULIRGs observed at $z \approx 0$, and SMG surface densities can reach those of Arp 220 \citep{Walter09}.  Furthermore, those without an AGN lie on the X-ray-SFR correlation, demonstrating that they do not have a CC85-like superwind \citep{Alexander05}.  Finally, the winds in such galaxies would emit so much free-free radiation that they cool radiatively before escaping, stalling the wind and invalidating the CC85 theory \citep{Silich10}.

To summarize, the X-ray emission observed from starbursts is perfectly consistent with a superwind existing in M82-like starbursts.  In compact ULIRGs observed at $z \approx 0$, the X-ray-SFR correlation is not a conclusive disproof of the superwind phase, although the superwind would have to be a major contributor to the hard X-ray luminosity.  Extended ULIRGs as at $z \approx 2$ are indeed too X-ray dim to host a hot superwind.

\end{appendix}

\end{document}